\documentclass[fleqn,twoside,11pt]{article}
\usepackage{latexsym} 

\font\msbmten=msbm10
\def\kappa{\mbox{\msbmten\char'173}}
\def\R{\mbox{\msbmten\char'122}}
\def\N{\mbox{\msbmten\char'116}}
\newcommand{\beg}{\begin{eqnarray}}
\newcommand{\eeq}{\end{eqnarray}}
\newcommand{\bg}{\begin{eqnarray*}}
\newcommand{\ed}{\end{eqnarray*}}

\newtheorem{theorem}{Theorem}[section]
\newtheorem{note}{Note}[section]
\newtheorem{lemma}{Lemma}[section]
\begin{document}   

\noindent
{\LARGE\bf
{Symmetry reduction and exact solutions
\vskip 2mm
\noindent
of the Navier-Stokes equations}}
\vskip 8mm
\noindent
{{\sl
Wilhelm FUSHCHYCH and Roman POPOVYCH,\\
{\small\it Institute of  Mathematics of
the  National  Ukrainian Academy of Sciences, \\
Tereshchenkivska Street 3, 01601 Kyiv,  Ukraina}
}}

\bigskip

\noindent  {\small URL:
{\tt http://www.imath.kiev.ua/\~{}appmath/wif.html} and {\tt http://www.imath.kiev.ua/\~{}rop/} 
\\ E-mail: {\tt rop@imath.kiev.ua}}

\begin{abstract}
\noindent
Ansatzes for the Navier-Stokes field are described. These
ansatzes reduce the Navier-Stokes equations to system of
differential equations in three, two, and one independent
variables. The large sets of exact solutions of
the Navier-Stokes equations are constructed.
\end{abstract}

\vspace{1ex}


\section{
Introduction}
\label{sec1}

The Navier-Stokes equations (NSEs)
\begin{equation}\label{e1.1}
\begin{array}{l}
\vec u_{t}+(\vec u\cdot\vec\nabla)\vec u-\triangle\vec u+
\vec\nabla p=\vec 0,
\\[1ex]
\mbox{div}\vec u=0
\end{array}
\end{equation}
which describe the motion of an incompressible viscous fluid are the
basic equations of modern hydrodynamics. In (\ref{e1.1}) and
below  $\:\vec u=\{u^{a}(t,\vec x)\}\:$ denotes the velocity field of a
fluid, $\:p=p(t,\vec x)\:$ denotes the pressure,
\(\:\vec x=\{x_{a}\},\:\)
$\:\partial_{t}=\partial/\partial t,\:$
$\:\partial_{a}=\partial/\partial x_{a},\:$
$\:\vec\nabla=\{\partial_{a}\},\:$
$\:\triangle=\vec \nabla\cdot\vec \nabla\:$ is the Laplacian, the
kinematic coefficient of viscosity and fluid density are set
equal to unity. Repead indices denote summation whereby we consider
the indices $a$, $b$ to take on values in $\{1,2,3\}$ and
the indices $i$, $j$ to take on values in $\{1,2\}$.

The problem of finding exact solutions of non-linear equations
(\ref{e1.1}) is an important but rather complicated one. There are
some ways to solve it. Considerable progress in this field can be
achieved by means of making use of a symmetry approach.
Equations (\ref{e1.1}) have non-trivial symmetry properties.
It was known long ago \cite{wilzynski,birkgoff} that they are
invariant under the eleven-parametric extended Galilei group. Let
us denote it by $G_{1}(1,3)$. This group includes the Galilei
group and scale transformations. The Lie algebra $AG_{1}(1,3)$ of
$G_{1}(1,3)$
is generated by the operators
\begin{displaymath}
P_{0},\quad J_{ab},\quad D,\quad P_{a},\quad G_{a},
\end{displaymath}
where
\begin{displaymath}
P_{0}=\partial_{t}, \quad
D=2t\partial_{t}+x_{a}\partial_{a}-u^{a}\partial_{u^{a}}-
2p\partial_{p},
\end{displaymath}
\begin{displaymath}
J_{ab}=x_{a}\partial_{b}-x_{b}\partial_{a}+
u^{a}\partial_{u^{b}}-u^{b}\partial_{u^{a}}, \quad a\not=b,
\end{displaymath}
\begin{displaymath}
G_{a}=t\partial_{a}+\partial_{u^{a}}, \quad P_{a}=\partial_{a}.
\end{displaymath}

Relatively recently it was found by means of the Lie method
\cite{danilov,bytev.a,lloyd} that the maximal Lie
invariance algebra (MIA) of the NSEs (\ref{e1.1}) is the
infinite-dimensional algebra $A(NS)$ with the basis elements
\begin{equation}\label{e1.2}
\partial_{t}, \quad D, \quad J_{ab}, \quad R(\vec m),
\quad Z(\chi),
\end{equation}
where
\begin{equation}\label{e1.3}
R(\vec m)=R(\vec m(t))=m^{a}(t)\partial_{a}+
m^{a}_{t}(t)\partial_{u^{a}}-m^{a}_{tt}(t)x_{a}\partial_{p},
\end{equation}
\begin{equation}\label{e1.4}
Z(\chi)=Z(\chi(t))=\chi(t)\partial_{p},
\end{equation}
$m^{a}=m^{a}(t)$ and $\chi=\chi(t)$ are arbitrary smooth functions
of $t$ (degree of their smoothness is discussed in Note
\ref{na.1}).

The algebra $AG_{1}(1,3)$ is a subalgebra of $A(NS)$. Indeed,
setting \mbox{$\:m^{a}=\delta_{ab},\:$} where $b$ is fixed,
we obtain $\:R(\vec m)=\partial_{b},\:$
and if $\:m^{a}=\delta_{ab}t\:$ then $\:R(\vec m)=G_{b}\:$. Here
$\delta_{ab}$ is the Kronecker symbol ($\:\delta_{ab}=1\:$ if $\:a=b,\:$
$\:\delta_{ab}=0\:$ if $\:a\not=b\:$).

Operators (\ref{e1.2}) generate the following invariance
transformations of system (\ref{e1.1}):
\begin{equation}\label{e1.5}
\begin{array}{ll}
\partial_{t}:&
\vec{\tilde u}(t,\vec x)=\vec u(t+\varepsilon,\vec x),
\quad \tilde p(t,\vec x)=p(t+\varepsilon,\vec x)
\\[1ex]
&(\mbox{translations with respect to} \: t),
\\[2ex]
J_{ab}:&
\vec{\tilde u}(t,\vec x)=B\vec u(t,B^{T}\vec x),
\quad \tilde p(t,\vec x)=p(t,B^{T}\vec x)
\\[1ex]
&(\mbox{space rotations}),
\\[2ex]
D:&
\vec{\tilde u}(t,\vec x)=
e^{\varepsilon}\vec u(e^{2\varepsilon}t,e^{\varepsilon}\vec x),
\quad \tilde p(t,\vec x)=
e^{2\varepsilon}p(e^{2\varepsilon}t,e^{\varepsilon}\vec x)
\\[1ex]
&(\mbox{scale transformations}),
\\[2ex]
R(\vec m):&
\vec{\tilde u}(t,\vec x)=\vec u(t,\vec x-\vec m(t))+\vec m_{t}(t),
\\[1ex]
&\tilde p(t,\vec x)=p(t,\vec x-\vec m(t))-\vec m_{tt}\cdot\vec x-
\frac{1}{2}\vec m\cdot\vec m_{tt}
\\[1ex]
&(\mbox{these transformations include the space translations}
\\[1ex]
&\mbox{and the Galilei transformations}),
\\[2ex]
Z(\chi):&
\vec{\tilde u}(t,\vec x)=\vec u(t,\vec x),
\quad \tilde p(t,\vec x)=p(t,\vec x)+\chi(t).
\end{array}
\end{equation}
Here $\:\varepsilon\in{\R},\:$
\mbox{$\:B=\{\beta_{ab}\}\in O(3),\:$} i.e. $BB^{T}=\{\delta_{ab}\},\:$
$\:B^{T}$ is the transposed matrix.

Besides continuous transformations (\ref{e1.5}) the NSEs
admit discrete transformations of the form
\begin{equation}\label{e1.6}
\begin{array}{l}
\tilde t=t,\quad \tilde x_{a}=x_{a},\:a\not=b, \quad
\tilde x_{b}=-x_{b},
\\[1ex]
\tilde p=p,\quad \tilde u^{a}=u^{a},\:a\not=b, \quad
\tilde u^{b}=-u^{b},
\end{array}
\end{equation}
where $b$ is fixed.
% Using formulas (\ref{e1.5}), (\ref{e1.6})
%one can construct new solutions of the NSEs (\ref{e1.1}),
%starting from known ones.
Invariance under transformations (\ref{e1.5}) and (\ref{e1.6})
means that $(\vec{\tilde u}, \tilde p)$ is a solution of
(\ref{e1.1}) if $(\vec u, p)$ is a solution of (\ref{e1.1}).

A complete review of exact solutions found for the NSEs before
1963 is contained in \cite{berker}. We should like also to mark
more modern reviews \cite{goldshtik,craik&criminal,wang} despite
their subjects slightly differ from subjects of our
investigations. To find exact solutions of (\ref{e1.1}), symmetry
approach in explicit form was used in
\cite{birkgoff,pukhnachev.a,pukhnachev.b,bytev.b,kapitanskiy.a,kapitanskiy.b,ames&other,grauel&steeb,fshsl,fshp,nses_to_lin,prepr93,my_phdthesis}.
This article is a continuation and a extention of our works
\cite{fshsl,fshp,nses_to_lin,prepr93,my_phdthesis}.
In it we make symmetry reduction of the NSEs to systems of PDEs
in three and two independent variables and to systems of ODEs,
using subalgebraic structure of $A(NS)$. We investigate symmetry
properties of the reduced systems of PDEs and construct exact
solutions of the reduced systems of ODEs when it is possible. As
a result, large classes of exact solutions of the NSEs are
obtained.

The reduction problem for the NSEs is to describe ansatzes of the
form \cite{f_ansatz}:
\begin{equation}\label{e1.7}
u^{a}=f^{ab}(t,\vec x)v^{b}(\omega)+g^{a}(t,\vec x), \quad
p=f^{0}(t,\vec x)q(\omega)+g^{0}(t,\vec x)
\end{equation}
that reduce system (\ref{e1.1}) in four independent variables
to systems of differential equations in the functions $v^{a}$ and $q$
depending on the variables $\omega=\{\omega_{n}\}$
$(n=\overline{1,N})$, where $N$ takes on a fixed value from the set
$\{1,2,3\}$. In formulas (\ref{e1.7}) $f^{ab}$,
$g^{a}$, $f^{0}$, $g^{0}$, and $\omega_{n}$ are smooth functions to
be described. In such a general formulation the reduction problem
is too complex to solve. But using Lie symmetry, some ansatzes
(\ref{e1.7}) reducing the NSEs can be obtained. According to the
Lie method, first a complete set of $A(NS)$-inequivalent
subalgebras of dimension $M=4-N$ is to be constructed. For $N=3$,
$N=2$, and $N=1$ such sets are given in Subsections \ref{subsec_a.2},
\ref{subsec_a.3}, and \ref{subsec_a.4}, correspondingly. Knowing
subalgebraic structure of $A(NS)$, one can find explicit forms
for the functions $f^{ab}$, $g^{a}$, $f^{0}$, $g^{0}$, and
$\omega_{n}$ and obtain reduced systems in the functions $v^{k}$ and
$q$. This is made in Sec.\  \ref{sec2} ($N=3$), Sec.\ \ref{sec3} ($N=2$)
and Sec.\ \ref{sec4} ($N=1$). Moreover, in Subsec.\ \ref{subsec2.3}
symmetry properties of all reduced systems of PDEs in three
independent variables are investigated, and in Subsec.
\ref{subsec4.3} exact solutions of the reduced systems of ODEs are
constructed. Symmetry properties and exact solutions of some
reduced systems of PDEs in two independent variables are
discussed in Sections \ref{sec5} and \ref{sec6}. In Sec.\
\ref{sec7} we make symmetry reduction of a some reduced system of
PDEs in three independent variables.

In conclusion of the section, for convenience, we give some
abbreviations, notations, and default rules used in this article.

\vspace{1ex}

Abbreviations:
\begin{description}
\item[the NSEs:]
     the Navier-Stokes equations
\item[the MIA:]
     the maximal Lie invariance algebra (of either
     a some equation or a some system of equations)
\item[a ODE:]
     a ordinary differential equation
\item[a PDE:]
     a partial differential equation
\end{description}

\vspace{1ex}

Notations:
\begin{description}
\item[$ C^{\infty}((t_{0},t_{1}),{\R})$:]
     the set of infinite-differentiable functions from
     $(t_{0},t_{1})$ into ${\R}$, where
     $-\infty\le t_{0}<t_{1}\le+\infty$
\item[$ C^{\infty}((t_{0},t_{1}),{\R}^{3})$:]
     the set of infinite-differentiable vector-functions from
     $(t_{0},t_{1})$ into ${\R}^{3}$, where
     $-\infty\le t_{0}<t_{1}\le+\infty$
\end{description}
$\partial_{t}=\partial/\partial_{t}$,
$\partial_{a}=\partial/\partial_{x_{a}}$,
$\partial_{u^{a}}=\partial/\partial_{u^{a}}$, \dots

\vspace{1ex}

Default rules:

Repead indices denote summation whereby we consider
the indices $a$, $b$ to take on values in $\{1,2,3\}$ and
the indices $i$, $j$ to take on values in $\{1,2\}$.

All theorems on the MIAs of PDEs are proved by means of the
standard Lie algorithm.

Subscripts of functions denote differentiation.

\setcounter{equation}{0}

\section
{Reduction of the Navier-Stokes equations
to systems of PDEs in three independent variables }
\label{sec2}

\subsection
{Ansatzes of codimension one }\label{subsec2.1}

In this subsection we give ansatzes that reduce the NSEs
to systems of PDEs in three independent variables. The ansatzes
are constructed with the subalgebraic analysis of $A(NS)$ ( see
Subsec.\@\ref{subsec_a.2} ) by means of the method discribed in
Sec.\@\ref{sec_b} .

\begin{equation}\label{e2.1}
\!\!\!\!\!\!\!\!\!\!\!
\begin{array}{llcl}
\mbox{1.} & u^{1} & = & \vert t \vert^{-1/2}(v^{1}\cos  \tau  -  v^{2}\sin
\tau ) + \frac{1}{2}x_{1}t^{-1} - \kappa x_{2}t^{-1},\\
\\
& u^{2} & = & \vert t \vert^{-1/2}(v^{1}\sin  \tau  +  v^{2}\cos
\tau ) + \frac{1}{2}x_{2}t^{-1} + \kappa x_{1}t^{-1},\\
\\
& u^{3} & = & \vert t \vert^{-1/2}v^{3} + \frac{1}{2}x_{3}t^{-1},\\
\\
& p & = & \vert t \vert^{-1}q + \frac{1}{2}\kappa^{2}t^{-2}r^{2} +
\frac{1}{8}t^{-2}x_{a}x_{a},
\end{array}
\end{equation}
where
\begin{displaymath}
y_{1}=\vert t \vert^{-1/2}(x_{1}\cos \tau + x_{2}\sin \tau ),
 \quad
y_{2}=\vert t \vert^{-1/2}(-x_{1}\sin  \tau  +  x_{2}\cos  \tau  ),
\end{displaymath}
\begin{displaymath}
y_{3}=\vert t \vert^{-1/2}x_{3},\quad
\kappa\geq 0, \quad \tau = \kappa\ln\vert t \vert.
\end{displaymath}
Here and below \quad $v^{a} = v^{a}(y_{1},y_{2},y_{3})$,
\quad $q = q(y_{1},y_{2},y_{3})$,
\quad $r = ( x^{2}_{1} + x^{2}_{2} )^{1/2}$.

\begin{equation}\label{e2.2}
\!\!\!\!\!\!\!\!\!\!\!
\begin{array}{llcl}
\mbox{2.} & u^{1} & = & v^{1}\cos\kappa t - v^{2}\sin\kappa t
 - \kappa x_{2},\\
\\
 & u^{2} & = & v^{1}\sin\kappa t + v^{2}\cos\kappa t
 + \kappa x_{1},\\
\\
& u^{3} & = & v^{3},\\
\\
& p & = & q + \frac{1}{2}\kappa^{2}r^{2},
\end{array}
\end{equation}
where
\begin{displaymath}
y_{1}=x_{1}\cos\kappa t + x_{2}\sin\kappa t,
 \quad
y_{2}=-x_{1}\sin\kappa t + x_{2}\cos\kappa t,
\end{displaymath}
\begin{displaymath}
y_{3}=x_{3},\quad \kappa\in \{ 0;1 \}.
\end{displaymath}

\begin{equation}\label{e2.3}
\!\!\!\!\!\!\!\!\!\!\!
\begin{array}{llcl}
\mbox{3.} & u^{1} & = & x_{1}r^{-1}v^{1} - x_{2}r^{-1}v^{2}+x_{1}r^{-2},\\
\\
& u^{2} & = & x_{2}r^{-1}v^{1} + x_{1}r^{-1}v^{2}+x_{2}r^{-2},\\
\\
& u^{3} & = & v^{3}+\eta(t)r^{-1}v^{2}+\eta_{t}(t)\arctan x_{2}/x_{1},\\
\\
& p & = & q - \frac{1}{2}\eta_{tt}(t)(\eta(t))^{-1}x_{3}^{2}-
\frac{1}{2}r^{-2}+\chi(t)\arctan x_{2}/x_{1},
\end{array}
\end{equation}
where
\begin{displaymath}
y_{1}=t,\quad\! y_{2}=r,\quad\! y_{3}=x_{3}-\eta(t)\arctan x_{2}/x_{1},
\quad \eta,\chi\!\in\! C^{\infty}((t_{0},t_{1}),{\R}).
\end{displaymath}

\begin{note}\label{n2.1}
The expression for the pressure $p$ from ansatz (\ref{e2.3})
is indeterminate in the points $t\in(t_{0},t_{1})$ where $\eta(t)=0$.
If there are such points $t$, we will consider ansatz (\ref{e2.3})
on the intervals $(t_{0}^{n},t_{1}^{n})$ that are contained in the
interval $(t_{0},t_{1})$ and that satisfy one of the conditions:
\begin{displaymath}
\mbox{a)}\ \eta(t)\not= 0 \quad\forall t\in (t_{0}^{n},t_{1}^{n});
\end{displaymath}
\begin{displaymath}
\mbox{b)}\ \eta(t) = 0 \quad\forall t\in (t_{0}^{n},t_{1}^{n}).
\end{displaymath}
In the last case we consider $\eta_{tt}/\eta:=0$.
\end{note}

\begin{equation}\label{e2.4}
\!\!\!\!\!\!\!\!\!\!\!
\begin{array}{llcl}
\mbox{4.} & \vec u & = & v^{i}\vec n^{i} +
(\vec m\cdot\vec m)^{-1}v^{3}\vec m +
(\vec m\cdot\vec m)^{-1}(\vec m\cdot\vec x)\vec m_{t} -
 y_{i}\vec n^{i}_{t} ,\\
\\
& p & = & q - \frac{3}{2}(\vec m\cdot\vec m)^{-1}
((\vec m_{t}\cdot\vec n^{i})y_{i})^{2}-
(\vec m\cdot\vec m)^{-1}(\vec m_{tt}\cdot\vec x )
(\vec m\cdot\vec x)+\\
\\
&&&+\frac{1}{2}(\vec m_{tt}\cdot\vec m)(\vec m\cdot\vec m)^{-2}
(\vec m\cdot\vec x)^{2},
\end{array}
\end{equation}
where
\begin{displaymath}
y_{i}=\vec n^{i}\cdot\vec x, \quad y_{3}=t, \quad
\vec m,\vec n^{i}\in C^{\infty}((t_{0},t_{1}),{\R}^{3}).
\end{displaymath}
\begin{equation}\label{e2.5}
\vec n^{i}\cdot\vec m = \vec n^{1}\cdot\vec n^{2} =
\vec n^{1}_{t}\cdot\vec n^{2}= 0,\quad
\vert \vec n^{i} \vert = 1.
\end{equation}

\begin{note}\label{n2.2}
There exist vector-functions $\vec n^{i}$ which satisfy conditions
(\ref{e2.5}). They can be constructed in the following way: let us fix
the vector-functions \mbox{$\:\vec k^{i}=\vec k^{i}(t)\:$} such that
$\:\vec k^{i}\cdot\vec m = \vec k^{1}\cdot\vec k^{2}=0,\:$
 $\:\vert\vec k^{i}\vert=1,\:$ and set
\begin{equation}\label{e2.6}
\begin{array}{l}
\vec n^{1} = \vec k^{1}\cos \psi(t) - \vec k^{2}\sin\psi(t),\\[1ex]
\vec n^{2} = \vec k^{1}\sin \psi(t) + \vec k^{2}\cos\psi(t).
\end{array}
\end{equation}
Then \quad $\vec n^{1}_{t}\cdot\vec n^{2}=
\vec k^{1}_{t}\cdot\vec k^{2}-\psi_{t} = 0$\quad
if\quad $\psi=\int(\vec k^{1}_{t}\cdot\vec k^{2})dt$.
\end{note}

\subsection
{Reduced systems}\label{subsec2.2}

1--2. Substituting ansatzes (\ref{e2.1}) and (\ref{e2.2}) into the NSEs
(\ref{e1.1}), we obtain reduced systems of PDEs with the same
general form
\begin{equation}\label{e2.7}
\begin{array}{l}
v^{a}v^{1}_{a}-v^{1}_{aa}+q_{1}+\gamma_{1}v^{2}=0,\\[1ex]
v^{a}v^{2}_{a}-v^{2}_{aa}+q_{2}-\gamma_{1}v^{1}=0,\\[1ex]
v^{a}v^{3}_{a}-v^{3}_{aa}+q_{3}=0,\\[1ex]
v^{a}_{a}=\gamma_{2}.
\end{array}
\end{equation}
Hereafter subscripts 1, 2, and 3 of functions denote
differentiation with respect to $y_{1}$, $y_{2}$, and $y_{3}$,
accordingly. The constants $\gamma_{i}$ take the values
\begin{displaymath}
\!\!\!\!\!\!\!\!\!\!\!
\begin{array}{l}
\mbox{1.}\quad     \gamma_{1}=-2\kappa,\quad  \gamma_{2}=-\frac{3}{2}\quad
\mbox{if}\quad t>0,
\quad     \gamma_{1}=2\kappa,\quad   \gamma_{2}=\frac{3}{2}\quad
\mbox{if}\quad t<0.
\\[1ex]
\mbox{2.}\quad     \gamma_{1}=-2\kappa,\quad     \gamma_{2}=0.
\end{array}
\end{displaymath}

For ansatzes (\ref{e2.3}) and (\ref{e2.4}) the  reduced  equations
have the form

\begin{equation}\label{e2.8}
\begin{array}{l}
\!\!\!\!\!\!\!\!\!\!\!
\mbox{3.} \quad v^{1}_{1}+v^{1}v^{1}_{2}+v^{3}v^{1}_{3}-y^{-1}_{2}v^{2}v^{2}-
\bigl(v^{1}_{22}+(1+\eta^{2}y^{-2}_{2})v^{1}_{33}\bigr)-
\\[1ex]
-2\eta y^{-2}_{2}v^{2}_{3}+q_{2}=0,\\
\\
  v^{2}_{1}+v^{1}v^{2}_{2}+v^{3}v^{2}_{3}+y^{-1}_{2}v^{1}v^{2}-
\bigl(v^{2}_{22}+(1+\eta^{2}y^{-2}_{2})v^{2}_{33}\bigr)+
\\[1ex]
+2\eta y^{-2}_{2}v^{1}_{3}+2y^{-2}_{2}v^{2}-\eta y^{-1}_{2}q_{3}+
\chi y^{-1}_{2}=0,\\
\\
  v^{3}_{1}+v^{1}v^{3}_{2}+v^{3}v^{3}_{3}-
\bigl(v^{3}_{22}+(1+\eta^{2}y^{-2}_{2})v^{3}_{33}\bigr)-
2\eta^{2} y^{-3}_{2}v^{1}_{3}+
\\[1ex]
+2\eta_{1}y^{-1}_{2}v^{2}+2\eta y^{-1}_{2}(y^{-1}_{2}v^{2})_{2}+
(1+\eta^{2}y^{-2}_{2})q_{3}-
\\[1ex]
-\eta_{11}\eta^{-1}y_{3}-\chi\eta y^{-2}_{2}=0,\\
\\
y^{-1}_{2}v^{1}+v^{1}_{2}+v^{3}_{3}=0.
\end{array}
\end{equation}

\begin{equation}\label{e2.9}
\begin{array}{l}
\!\!\!\!\!\!\!\!\!\!\!
\mbox{4.}\quad v^{i}_{3}+v^{j}v^{i}_{j}-v^{i}_{jj}+q_{i}+
\rho^{i}(y_{3})v^{3}=0,\\
\\
 v^{3}_{3}+v^{j}v^{3}_{j}-v^{3}_{jj}=0,\\
\\
 v^{i}_{i}+\rho^{3}(y_{3})=0,
\end{array}
\end{equation}
where
\begin{equation}\label{e2.10}
\begin{array}{l}
\rho^{i}=\rho^{i}(y_{3})=2(\vec m\cdot\vec m)^{-1}(\vec m_{t}\cdot
\vec n^{i}),
\\[1ex]
\rho^{3}=\rho^{3}(y_{3})=(\vec m\cdot\vec m)^{-1}(\vec m_{t}\cdot
\vec m).
\end{array}
\end{equation}

\subsection
{Symmetry of reduced systems}\label{subsec2.3}

Let us study symmetry properties of systems (\ref{e2.7}),
(\ref{e2.8}), and (\ref{e2.9}). All results of this subsection
are obtained by means of the standard Lie algorithm
\cite{ovsiannikov,olver}. First, let us consider system
(\ref{e2.7}).

\begin{theorem}\label{t2.1}
The MIA of system (\ref{e2.7}) is the algebra

\vspace{1ex}

a) $ < \partial_{a}, \partial_{q}, J^{1}_{12} >$ \quad if
\quad  $ \gamma_{1}\not= 0$;

\vspace{1ex}

b) $ < \partial_{a}, \partial_{q}, J^{1}_{ab} > $ \quad if
\quad  $ \gamma_{1}= 0$, $\gamma_{2}\not= 0$;

\vspace{1ex}

c) $< \partial_{a}, \partial_{q}, J^{1}_{ab}, D^{1}_{1} >$
\quad  if \quad  $ \gamma_{1}= \gamma_{2}= 0$.
\begin{tabbing}
Here\quad \= $ J^{1}_{ab}=y_{a}\partial_{b}-y_{b}\partial_{a}+
v^{a}\partial_{v^{b}}-v^{b}\partial_{v^{a}} $ ,\\[1ex]
\> $ D^{1}_{1}=y_{a}\partial_{a}-
v^{a}\partial_{v^{a}}-2q\partial_{q} $ .
\end{tabbing}
\end{theorem}

\begin{note}\label{n2.3}
All Lie symmetry operators of (\ref{e2.7}) are induced by
operators from $A(NS)$: The operators $J^{1}_{ab}$ and
$D^{1}_{1}$ are induced by $J_{ab}$ and $D$. The operators
$\:c_{a}\partial_{a}\:$ $\:(c_{a}={\rm const})\:$ and $\partial_{q}$
are induced by either
\begin{displaymath}
R(\vert t \vert^{1/2}(c_{1}\cos\tau-c_{2}\sin\tau,
c_{1}\sin\tau+c_{2}\cos\tau,c_{3})),\quad Z(\vert t \vert^{-1}),
\end{displaymath}
where $\tau=\kappa\ln\vert t \vert$, for ansats (\ref{e2.1}) or
\begin{displaymath}
R(c_{1}\cos\kappa t-c_{2}\sin\kappa t,
c_{1}\sin\kappa t+c_{2}\cos\kappa t,c_{3}),\quad Z(1)
\end{displaymath}
for ansatz (\ref{e2.2}), respectively. Therefore, Lie
reductions of system (\ref{e2.7}) give only solutions that
can be obtained by reducing the NSEs with two- and
three-dimensional subalgebras of $A(NS)$.
\end{note}

Let us continue to system (\ref{e2.8}). We denote $A^{max}$ as
the MIA of (\ref{e2.8}). Studying symmetry properties of
(\ref{e2.8}), one has to consider the following cases:

A. $\eta,\chi\equiv 0$. Then
\begin{displaymath}
A^{max}  =   <   \partial^{1},   D^{1}_{2},   R_{1}(\psi(y_{1})),
Z^{1}(\lambda(y_{1})) >,
\end{displaymath}
\begin{tabbing}
where\quad \= $  D^{1}_{2}=2y_{1}\partial_{1}+y_{2}\partial_{2}+
y_{3}\partial_{3}-v^{a}\partial_{v^{a}}-2q\partial_{q} $ ,\\[1ex]
\> $
R_{1}(\psi(y_{1}))=\psi\partial_{3}+\psi_{1}\partial_{v^{3}}-
\psi_{11}y_{3}\partial_{q}$,\quad
 $ Z^{1}(\lambda(y_{1}))=\lambda(y_{1})\partial_{q} $.
\end{tabbing}
Here and below $\psi=\psi(y_{1})$ and $\lambda=\lambda(y_{1})$  are
arbitrary smooth functions of $y_{1}=t$.

B. $\eta\equiv 0$, $\chi\not\equiv 0$. In this case an extension  of
$A^{max}$ exists for \mbox{$\chi=(C_{1}y_{1}+C_{2})^{-1}$}, where
$C_{1},C_{2}=\mbox{const}$. Let $C_{1}\not=0$. We can make $C_{2}$
vanish by means of equivalence transformation (\ref{ea.9}),
i.e., $\chi=Cy^{-1}_{1}$, where $C=\mbox{const}$. Then
\begin{displaymath}
A^{max}  =   <   D^{1}_{2},   R_{1}(\psi(y_{1})),
Z^{1}(\lambda(y_{1})) >.
\end{displaymath}
If $C_{1}=0$, $\chi=C=\mbox{const}$ and
\begin{displaymath}
A^{max} = < \partial_{1}, R_{1}(\psi(y_{1})),
Z^{1}(\lambda(y_{1})) >.
\end{displaymath}
For other values of $\chi$, i.e., when
$\chi_{11}\chi\not=\chi_{1}\chi_{1}$,
\begin{displaymath}
A^{max} = < R_{1}(\psi(y_{1})), Z^{1}(\lambda(y_{1})) >.
\end{displaymath}

C. $\eta\not=0$. By means of equivalence transformation
(\ref{ea.9}) we make \mbox{$\chi=0$}. In this case an extension of $A^{max}$
exists for $\eta=\pm\vert C_{1}y_{1}+C_{2}\vert^{1/2}$, where
$C_{1},C_{2}=\mbox{const}$. Let $C_{1}\not=0$. We can make $C_{2}$
vanish by means of equivalence transformation (\ref{ea.9}),
i.e., $\eta=C\vert y_{1}\vert^{1/2}$, where $C=\mbox{const}$. Then
\begin{displaymath}
A^{max}  =   <   D^{1}_{2},   R_{2}(\vert y_{1}\vert^{1/2}),
R_{2}(\vert y_{1}\vert^{1/2}\ln\vert y_{1}\vert ),
 Z^{1}(\lambda(y_{1})) >,
\end{displaymath}
where
$R_{2}(\psi(y_{1}))=\psi\partial_{3}+\psi_{1}\partial_{v^{3}}$.
If $C_{1}=0$, i.e., $\eta=C=\mbox{const}$,
\begin{displaymath}
A^{max} = < \partial^{1}, \partial_{3},
y_{1}\partial_{3}+\partial_{v^{3}}
Z^{1}(\lambda(y_{1})) >.
\end{displaymath}
For other values of $\eta$, i.e., when $(\eta^{2})_{11}\not=0$,
\begin{displaymath}
\begin{array}{l}
A^{max} = < R_{2}(\eta(y_{1})),
R_{2}(\eta(y_{1})\int(\eta(y_{1}))^{-2}dy_{1}),
Z^{1}(\lambda(y_{1})) >.
\end{array}
\end{displaymath}

\begin{note}\label{n2.4}
In all cases considered above the Lie symmetry operators of
(\ref{e2.8}) are induced by operators from $A(NS)$: The
operators $\partial_{1}$, $D^{1}_{2}$, and $Z^{1}(\lambda(y_{1}))$
are induced by $\partial_{t}$, $D$, and $Z(\lambda(t))$,
respectively. The operator $R(0,0,\psi(t))$ induces the operator
$R_{1}(\psi(y_{1}))$ for $\eta\equiv 0$ and the operator
$R_{2}(\psi(y_{1}))$ (if \mbox{$\:\psi_{11}\eta-\psi\eta_{11}=0$}) for
$\eta\not=0$. Therefore, the Lie reduction of system (\ref{e2.8})
gives only solutions that can be obtained by reducing the NSEs
with two- and three-dimentional subalgebras of $A(NS)$.
\end{note}

When $\eta=\chi=0$, system (\ref{e2.8}) describes axially
symmetric motion of a fluid and can be transformed into a  system
of two equations for a stream function $\Psi^{1}$ and  a  function
$\Psi^{2}$ that are determined by
\begin{displaymath}
\Psi^{1}_{3}=y_{2}v^{1},\quad \Psi^{1}_{2}=-y_{2}v^{3}, \quad
\Psi^{2}=y_{2}v^{2}.
\end{displaymath}
The transformed system was studied by L.V. Kapitanskiy
\cite{kapitanskiy.a,kapitanskiy.b}.

Consider system (\ref{e2.9}). Let us introduce the notations
\begin{displaymath}
\begin{array}{l}
t=y_{3},\quad \rho=\rho(t)=\int\rho^{3}(t)dt,
\end{array}
\end{displaymath}
\begin{displaymath}
R_{3}(\psi^{1}(t),\psi^{2}(t))=\psi^{i}\partial_{y_{i}}+
\psi^{i}_{t}\partial_{v^{i}}-\psi^{i}_{tt}y_{i}\partial_{q},
\end{displaymath}
\begin{displaymath}
Z^{1}(\lambda(t))=\lambda(t)\partial_{q},\quad
S=\partial_{v^{3}}-\rho^{i}(t)y_{i}\partial_{q},
\end{displaymath}
\begin{displaymath}
\begin{array}{l}
E(\chi(t))=2\chi\partial_{t}+\chi_{t}y_{i}\partial_{y_{i}}+
(\chi_{tt}y_{i}-\chi_{t}v^{i})\partial_{v^{i}}-
(2\chi_{t}q+\frac{1}{2}\chi_{ttt}y_{j}y_{j})\partial_{q},
\end{array}
\end{displaymath}
\begin{displaymath}
J^{1}_{12}=y_{1}\partial_{2}-y_{2}\partial_{1}+
v^{1}\partial_{v^{2}}-v^{2}\partial_{v^{1}}.
\end{displaymath}

\begin{theorem}\label{t2.3}
The MIA of(\ref{e2.9}) is the algebra

\vspace{2ex}

1) $< R_{3}(\psi^{1}(t),\:\psi^{2}(t)),\: Z^{1}(\lambda(t)),\: S,\:
E(\chi^{1}(t)),\: E(\chi^{2}(t)),\: v^{3}\partial_{v^{3}},\:
J^{1}_{12} >$,
\\[1ex] where
$\:\chi^{1}=e^{-\rho(t)}\int e^{\rho(t)}dt\:$ and
$\:\chi^{2}=e^{-\rho(t)},\:$  if  $\:\rho^{i}=0$;

\vspace{2ex}

2) $< R_{3}(\psi^{1}(t),\: \psi^{2}(t)),\: Z^{1}(\lambda(t)),\: S,\:
E(\chi(t))+2a_{1}v^{3}\partial_{v^{3}}+2a_{2}J^{1}_{12} >$,
\quad where \\[1ex] $a_{1}$,  $a_{2}$, and  $a_{3}$
are fixed constants,
$\:\chi=e^{-\rho(t)}\Bigl( \int e^{\rho(t)}dt+a_{3} \Bigr),\:$
if
\begin{displaymath}
\rho^{1}=e^{\frac{3}{2}\rho}\hat\rho^{-\frac{3}{2}-a_{1}}
\Bigl(
C_{1}\cos(a_{2}\ln\hat\rho)- C_{2}\sin(a_{2}\ln\hat\rho)
\Bigr),
\end{displaymath}
\begin{displaymath}
\rho^{2}=e^{\frac{3}{2}\rho}\hat\rho^{-\frac{3}{2}-a_{1}}
\Bigl(
C_{1}\sin(a_{2}\ln\hat\rho)+C_{2}\cos(a_{2}\ln\hat\rho)
\Bigr)
\end{displaymath}
with
$\:\hat\rho=\hat\rho(t)=\vert\int e^{\rho(t)}dt+a_{3}\vert,\:$
 $\:C_{1},C_{2}={\rm const},\:$ $\:(C_{1},C_{2})\not=(0,0)$;

\vspace{2ex}

3) $< R_{3}(\psi^{1}(t),\psi^{2}(t)),\: Z^{1}(\lambda(t)),\: S,\:
E(\chi(t))+2a_{1}v^{3}\partial_{v^{3}}+
2a_{2}J^{1}_{12} >$, \quad where \\[1ex] $a_{1}$ and $a_{2}$
are fixed constants,
$\:\chi=e^{-\rho(t)},\:$  if
\begin{displaymath}
\rho^{1}=e^{\frac{3}{2}\rho-a_{1}\hat\rho}
\Bigl( C_{1}\cos(a_{2}\hat\rho)-C_{2}\sin(a_{2}\hat\rho) \Bigr),
\end{displaymath}
\begin{displaymath}
\rho^{2}=e^{\frac{3}{2}\rho-a_{1}\hat\rho}
\Bigl( C_{1}\sin(a_{2}\hat\rho)+C_{2}\cos(a_{2}\hat\rho) \Bigr)
\end{displaymath}
with
$\:\hat\rho=\hat\rho(t)=\int e^{\rho(t)}dt,\:$
 $\:C_{1},C_{2}={\rm const},\:$ $\:(C_{1},C_{2})\not=(0,0)$;

\vspace{2ex}

4) $< R_{3}(\psi^{1}(t),\: \psi^{2}(t)),\: Z^{1}(\lambda(t)),\: S >$
in all other cases.

\vspace{1ex}

Here $\:\psi^{i}=\psi^{i}(t)$, $\lambda=\lambda(t)$  are  arbitrary
smooth function of $\:t=y_{3}$.
\end{theorem}

\begin{note}\label{n2.5}
If functions $\:\rho^{b}\:$ are determined by
(\ref{e2.10}), then \mbox{$\:e^{\rho(t)}=C\vert\vec m(t)\vert,\:$} where
$\:C={\rm const},\:$ and the condition $\:\rho^{i}=0\:$ implies that
$\:\vec m=\vert\vec m(t)\vert\vec e,\:$ where
$\:\vec e={\rm const}\:$ and $\:\vert\vec e\vert=1$.
\end{note}

\begin{note}\label{n2.6}
The vector-functions $\vec n^{i}$ from Note \ref{n2.2} are  determined  up
to the transformation
\begin{displaymath}
\vec n^{1}=\vec n^{1}\cos\delta - \vec n^{2}\sin\delta, \quad
\vec n^{2}=\vec n^{1}\sin\delta + \vec n^{2}\cos\delta,
\end{displaymath}
where $\delta={\rm const}$. Therefore, $\delta$ can be chosen
such that $C_{2}=0$ (then \mbox{$C_{1}\not= 0$}).
\end{note}

\begin{note}\label{n2.7}
The operators $R_{3}(\psi^{1},\psi^{2})+\alpha S$ and
$Z^{1}(\lambda)$ are induced by $R(\vec l)+Z(\chi)$ and
$Z(\lambda)$, respectively. Here
$\:\vec l=\psi^{i}\vec n^{i}+\psi^{3}\vec m,\:$
$\:\psi^{3}_{t}(\vec m\cdot\vec m) + 2\psi^{i}(\vec
n^{i}_{t}\cdot\vec m)=\alpha$,
\begin{displaymath}
\begin{array}{l}
\chi-\frac{3}{2}(\vec m\cdot\vec m)^{-1}((\vec m_{t}\cdot\vec
n^{i})\psi^{i})^{2}-\frac{1}{2}(\vec m_{tt}\cdot\vec n^{i})
\psi^{3}\psi^{i} + \frac{1}{2}(\vec l_{tt}\cdot\vec n^{i})\psi^{i}
=0.
\end{array}
\end{displaymath}

If $\:\vec m=\vert\vec m\vert\vec e,\:$ where $\:\vec e={\rm const}\:$
and $\:\vert\vec e\vert=1,\:$ the operator $\:J^{1}_{12}\:$ is induced by
$e^{1}J_{23}+e^{2}J_{31}+e^{3}J_{12}$.

For
\begin{displaymath}
\vec m= \beta_{3}e^{\sigma t}(\beta_{2}\cos\tau,
\beta_{2}\sin\tau,\beta_{1})^{T}
\end{displaymath}
with $\tau=\kappa t+\delta$ and
$\beta_{a}={\rm const}$, where $\beta^{2}_{1}+\beta^{2}_{2}=1$,
the operator $\partial_{t}+\kappa J_{12}$ induces the operator
$\:\partial_{y_{3}}-\beta_{1}\kappa J_{12}^{1}+\sigma
v^{3}\partial_{v^{3}}\:$
if the following vector-functions $\vec n^{i}$ are chosen:
\begin{equation}\label{e2.11}
\vec n^{1}=\vec k^{1}\cos\beta_{1}\tau+
\vec k^{2}\sin\beta_{1}\tau, \quad
\vec n^{2}=-\vec k^{1}\sin\beta_{1}\tau+
\vec k^{2}\cos\beta_{1}\tau,
\end{equation}
where $\vec k^{1}=(-\sin\tau,\cos\tau,0)^{T}$ and
$\vec k^{2}=(\beta_{1}\cos\tau,\beta_{1}\sin\tau,-\beta_{2})^{T}$.

For
\begin{displaymath}
\vec m= \beta_{3}\vert t+\beta_{4}\vert^{\sigma+1/2}
(\beta_{2}\cos\tau,\beta_{2}\sin\tau,\beta_{1})^{T}
\end{displaymath}
with $\tau=\kappa\ln\vert t+\beta_{4}\vert+\delta$ and
$\beta_{a},\beta_{4}={\rm const}$, where
$\beta^{2}_{1}+\beta^{2}_{2}=1$,
the operator $D+2\beta_{4}\partial_{t}+2\kappa J_{12}$
induces the operator
\begin{displaymath}
D^{1}_{3}+2\beta_{4}\partial_{y_{3}}-2\beta_{1}\kappa J_{12}^{1}+
2\sigma v^{3}\partial_{v^{3}},
\end{displaymath}
where $\:D^{1}_{3}=y_{i}\partial_{y_{i}}+2y_{3}\partial_{y_{3}}-
v^{i}\partial_{v^{i}}-2q\partial_{q},\:$
if the vector-functions $\vec n^{i}$ are chosen in form
(\ref{e2.11}). In all other cases the basis elements of the MIA
of (\ref{e2.9}) are not induced by operators from $A(NS)$.
\end{note}

\begin{note}\label{n2.8}
The invariance algebras of systems of form (\ref{e2.9}) with
different parameter-functions $\rho^{3}=\rho^{3}(t)$ and
$\tilde\rho^{3}=\tilde\rho^{3}(t)$ are similar .
It suggests that there exists a local transformation of
variables which make $\rho^{3}$ vanish. So, let us transform
variables in the following way:
\begin{equation}\label{e2.12}
\begin{array}{l}
\tilde y_{i} = y_{i}e^{\frac{1}{2}\rho(t)}, \quad
\tilde y_{3} = \int e^{\rho(t)}dt,
\\ \\
\tilde v^{i} =\Bigl( v^{i}+\frac{1}{2}y_{i}\rho^{3}(t)\Bigr)
e^{-\frac{1}{2}\rho(t)}, \quad \tilde v^{3}=v^{3},
\\ \\
\tilde q = qe^{-\rho(t)}+\frac{1}{8}y_{i}y_{i}
\Bigl( (\rho^{3}(t)^{2})-2\rho^{3}_{t}(t) \Bigr)e^{-\rho(t)}.
\end{array}
\end{equation}
As a result, we obtain the system
\begin{displaymath}
\begin{array}{l}
 \tilde v^{i}_{3}+\tilde v^{j}\tilde v^{i}_{j}-\tilde v^{i}_{jj}+
\tilde q_{i}+\tilde\rho^{i}(\tilde y_{3})\tilde v^{3}=0,\\
\\
\tilde v^{3}_{3}+\tilde v^{j}v^{3}_{j}-\tilde v^{3}_{jj}=0,\\
\\
\tilde v^{i}_{i}=0
\end{array}
\end{displaymath}
for the functions $\:\tilde v^{a}= \tilde v^{a}(\tilde y_{1},\tilde
y_{2}, \tilde y_{3})\:$ and
$\:\tilde q= \tilde q(\tilde y_{1},\tilde y_{2}, \tilde y_{3})$.
Here subscripts 1, 2, and 3 denote differentiation with respect to
$\tilde y_{1}$, $\tilde y_{2}$, and $\tilde y_{3}$, \vspace{4pt}
 accordingly.
 Also
\mbox{$\:\tilde\rho^{i}(\tilde y_{3})= \rho^{i}(t)e^{-\frac{3}{2}\rho(t)}$}.
\end{note}

\setcounter{equation}{0}

\section
{Reduction of the Navier-Stokes equations
to systems of PDEs in two independent variables }
\label{sec3}

\subsection
{Ansatzes of codimension two }\label{subsec3.1}

In this subsection we give ansatzes that reduce the NSEs
to systems of PDEs in two independent variables. The ansatzes
are constructed with the subalgebrical analysis of $A(NS)$ ( see
Subsec.~\ref{subsec_a.3} ) by means of the method discribed in Sec.~
\ref{sec_b} .
\begin{equation}\label{e3.1}
\!\!\!\!\!\!\!\!\!\!\!
\begin{array}{llll}
\mbox{1.} & u^{1} & = & (rR)^{-1}((x_{1}-\kappa x_{2})w^{1}-x_{2}w^{2}+
x_{1}x_{3}r^{-1}w^{3}),\\
\\
& u^{2} & = & (rR)^{-1}((x_{2}+\kappa x_{1})w^{1}+x_{1}w^{2}+
x_{2}x_{3}r^{-1}w^{3}),\\
\\
& u^{3} & = & x_{3}(rR)^{-1}w^{1}-R^{-1}w^{3},\\
\\
& p & = & R^{-2}s,
\end{array}
\end{equation}
where\quad $z_{1}=\arctan x_{2}/x_{1}-\kappa\ln R$,\quad
$z_{2}=\arctan r/x_{3}$,\quad  $\kappa\geq 0$.

Here and below
\quad$w^{a}=w^{a}(z_{1},z_{2})$,\quad $s=s(z_{1},z_{2})$,\quad
$r=(x^{2}_{1}+x^{2}_{2})^{1/2}$,\quad
$R=(x^{2}_{1}+x^{2}_{2}+x^{2}_{3})^{1/2}$, \quad
$\kappa$, $\varepsilon$, $\sigma$, $\mu$, and $\nu$ are real
constants.

\begin{equation}\label{e3.2}
\!\!\!\!\!\!\!\!\!\!\!
\begin{array}{llll}
\mbox{2.} & u^{1} & = & \vert t\vert^{-1/2}r^{-1}(x_{1}w^{1}-x_{2}w^{2})
+\frac{1}{2}t^{-1}x_{1}+x_{1}r^{-2},\\
\\
   & u^{2} & = & \vert t\vert^{-1/2}r^{-1}(x_{2}w^{1}+x_{1}w^{2})
+\frac{1}{2}t^{-1}x_{2}+x_{2}r^{-2},\\
\\
   & u^{3} & = & \vert t\vert^{-1/2}w^{3}+\kappa r^{-1}w^{2}
+\frac{1}{2}t^{-1}x_{3},\\
\\
   & p     & = & \vert t\vert^{-1}s-\frac{1}{2}r^{-2}+\frac{1}{8}
t^{-2}R^{2}+\varepsilon\vert t\vert^{-1}\arctan x_{2}/x_{1},
\end{array}
\end{equation}
where \quad $z_{1}=\vert t\vert^{-1/2}r$,\quad
$z_{2}=\vert t\vert^{-1/2}x_{3}-\kappa\arctan x_{2}/x_{1}$,
\quad $\kappa\geq 0$,\quad $\varepsilon\geq 0$.

\begin{equation}\label{e3.3}
\!\!\!\!\!\!\!\!\!\!\!
\begin{array}{llll}
\mbox{3.} & u^{1} & = & r^{-1}(x_{1}w^{1}-x_{2}w^{2})+x_{1}r^{-2},\\
\\
   & u^{2} & = & r^{-1}(x_{2}w^{1}+x_{1}w^{2})+x_{2}r^{-2},\\
\\
   & u^{3} & = & w^{3}+\kappa r^{-1}w^{2},\\
\\
   & p     & = & s-\frac{1}{2}r^{-2}+\varepsilon\arctan
x_{2}/x_{1},
\end{array}
\end{equation}
where\quad $z_{1}=r$,\quad $z_{2}=x_{3}-\kappa\arctan x_{2}/x_{1}$,
\quad $\kappa\in\{0;1\}$,
\quad $\varepsilon\geq 0$ \enspace  if \enspace  $\kappa=1$ \enspace and
\enspace $\varepsilon\in\{0;1\}$ \enspace if \enspace $\kappa=0$.

\begin{equation}\label{e3.4}
\!\!\!\!\!\!\!\!\!\!\!
\begin{array}{llll}
\mbox{4.} & u^{1} & = & \vert t \vert^{-1/2}(\mu w^{1}+\nu w^{3})
\cos\tau - \vert t \vert^{-1/2}w^{2}\sin\tau+\\
\\
& & & +\nu\xi t^{-1}\cos\tau + \frac{1}{2}t^{-1}x_{1} -
\kappa t^{-1}x_{2},\\
\\
   & u^{2} & = & \vert t \vert^{-1/2}(\mu w^{1}+\nu w^{3})
\sin\tau + \vert t \vert^{-1/2}w^{2}\cos\tau+\\
\\
& & & +\nu\xi t^{-1}\sin\tau + \frac{1}{2}t^{-1}x_{2} +
\kappa t^{-1}x_{1},\\
\\
   & u^{3} & = & \vert t \vert^{-1/2}(-\nu w^{1}+\mu w^{3})
+\mu\xi t^{-1} + \frac{1}{2}t^{-1}x_{3},\\
\\
   &  p    & = & \vert t \vert^{-1}s-\frac{1}{2}t^{-2}\xi^{2}+
\frac{1}{8}t^{-2}R^{2}+\frac{1}{2}\kappa^{2}t^{-2}r^{2}+\\
\\
& & & +\varepsilon\vert t \vert^{-3/2}(\nu x_{1}\cos\tau+
\nu x_{2}\sin\tau+\mu x_{3}),
\end{array}
\end{equation}
where
\begin{displaymath}
\begin{array}{lll}
z_{1} & = & \vert t \vert^{-1/2}(\mu x_{1}\cos\tau+
\mu x_{2}\sin\tau-\nu x_{3}),\\
\\
z_{2} & = & \vert t \vert^{-1/2}( x_{2}\cos\tau-x_{1}\sin\tau),\\
\\
\xi   & = & \sigma(\nu x_{1}\cos\tau+\nu x_{2}\sin\tau+
\mu x_{3})+2\kappa\nu( x_{2}\cos\tau-x_{1}\sin\tau),
\end{array}
\end{displaymath}

\begin{displaymath}
\tau=\kappa\ln\vert t \vert,\quad \kappa>0,\quad \mu\geq0, \quad
\nu\geq0, \quad \mu^{2}+\nu^{2}=1,\quad
\sigma\varepsilon=0,\quad \varepsilon\geq0.
\end{displaymath}

\begin{equation}\label{e3.5}
\!\!\!\!\!\!\!\!\!\!\!
\begin{array}{llll}
\mbox{5.} & u^{1} & = & \vert t \vert^{-1/2}w^{1}
+ \frac{1}{2}t^{-1}x_{1},\\
\\
   & u^{2} & = & \vert t \vert^{-1/2}w^{2}
+ \frac{1}{2}t^{-1}x_{2},\\
\\
   & u^{3} & = & \vert t \vert^{-1/2}w^{3}
+ (\sigma+\frac{1}{2})t^{-1}x_{3},\\
\\
   &  p    & = & \vert t \vert^{-1}s-\frac{1}{2}\sigma^{2}t^{-2}
x^{2}_{3}+\frac{1}{8}t^{-2}R^{2} +\varepsilon\vert t \vert^{-3/2}
x_{3},
\end{array}
\end{equation}
where
\begin{displaymath}
z_{1}  =  \vert t \vert^{-1/2}x_{1},\quad
z_{2}  =  \vert t \vert^{-1/2}x_{2},\quad
\sigma\varepsilon=0,\quad \varepsilon\geq0.
\end{displaymath}

\begin{equation}\label{e3.6}
\!\!\!\!\!\!\!\!\!\!\!
\begin{array}{llll}
\mbox{6.} & u^{1} & = & (\mu w^{1}+\nu w^{3})
\cos t - w^{2}\sin t+\nu\xi\cos t -x_{2},\\
\\
   & u^{2} & = & (\mu w^{1}+\nu w^{3})
\sin t + w^{2}\cos t+\nu\xi\sin t +x_{1},\\
\\
   & u^{3} & = & (-\nu w^{1}+\mu w^{3})+\mu\xi,\\
\\
   &  p    & = & s-\frac{1}{2}\xi^{2}+\frac{1}{2}r^{2}+
\varepsilon (\nu x_{1}\cos t+\nu x_{2}\sin t+\mu x_{3}),
\end{array}
\end{equation}
where
\begin{displaymath}
\begin{array}{lll}
z_{1} & = & (\mu x_{1}\cos t+\mu x_{2}\sin t-\nu x_{3}),\\
\\
z_{2} & = & ( x_{2}\cos t-x_{1}\sin t),\\
\\
\xi   & = & \sigma(\nu x_{1}\cos t+\nu x_{2}\sin t+\mu x_{3})
+2\nu( x_{2}\cos t-x_{1}\sin t),
\end{array}
\end{displaymath}
\begin{displaymath}
\mu\geq0, \quad \nu\geq0, \quad \mu^{2}+\nu^{2}=1,\quad
\sigma\varepsilon=0,\quad \varepsilon\geq0.
\end{displaymath}

\begin{equation}\label{e3.7}
\!\!\!\!\!\!\!\!\!\!\!
\begin{array}{llll}
\mbox{7.} & u^{1} & = & w^{1},\quad u^{2} = w^{2},\quad
u^{3} = w^{3} +\sigma x_{3},\\
\\
   &   p   & = & s-\frac{1}{2}\sigma^{2}x^{2}_{3}+\varepsilon
x_{3},
\end{array}
\end{equation}
where
\begin{displaymath}
z_{1}=x_{1},\quad z_{2}=x_{2},\quad
\sigma\varepsilon=0,\quad \varepsilon\in\{0;1\}.
\end{displaymath}

\begin{equation}\label{e3.8}
\!\!\!\!\!\!\!\!\!\!\!
\begin{array}{llll}
\mbox{8.} & u^{1} & = & x_{1}w^{1}-x_{2}r^{-2}(w^{2}-\chi(t)),\\
\\
   & u^{2} & = & x_{2}w^{1}+x_{1}r^{-2}(w^{2}-\chi(t)),\\
\\
   & u^{3} & = & (\rho(t))^{-1}(w^{3}+\rho_{t}(t)x_{3}+
\varepsilon\arctan x_{2}/x_{1}),\\
\\
   &   p   & = & s-\frac{1}{2}\rho_{tt}(t)(\rho(t))^{-1}x^{2}_{3}+
\chi_{t}(t)\arctan x_{2}/x_{1},
\end{array}
\end{equation}
where
\begin{displaymath}
z_{1}=t,\quad z_{2}=r,\quad \varepsilon\in\{0;1\},\quad
\chi,\rho\in C^{\infty}((t_{0},t_{1}),{\R}).
\end{displaymath}

\begin{equation}\label{e3.9}
\!\!\!\!\!\!\!\!\!\!\!
\begin{array}{llll}
\mbox{9.} & \vec u & = & \vec w+ \lambda^{-1}(\vec n^{i}\cdot\vec x)
\vec m^{i}_{t}-\lambda^{-1}(\vec k\cdot\vec x)\vec k_{t},\\
\\
   &   p    & = & s -\frac{1}{2}\lambda^{-1}
(\vec m^{i}_{tt}\cdot\vec x)(\vec n^{i}\cdot\vec x)-
\frac{1}{2}\lambda^{-2}(m^{i}_{tt}\cdot\vec k)
(\vec n^{i}\cdot\vec x)(\vec k\cdot\vec x),
\end{array}
\end{equation}
where
\begin{displaymath}
z_{1}=t,\quad z_{2}=(\vec k\cdot\vec x),\quad
\vec m^{i}\in C^{\infty}((t_{0},t_{1}),{\R}^{3}),
\end{displaymath}
\begin{displaymath}
\vec m^{1}_{tt}\cdot\vec m^{2}-
\vec m^{1}\cdot\vec m^{2}_{tt}=0, \quad
\vec k=\vec m^{1}\times\vec m^{2},\quad
\vec n^{1}=\vec m^{2}\times\vec k,
\end{displaymath}
\begin{displaymath}
\vec n^{2}=\vec k\times\vec m^{1}, \quad
\lambda=\lambda(t)=\vec k\cdot\vec k\not= 0 \quad
\forall t\in(t_{0},t_{1}).
\end{displaymath}

\subsection
{Reduced systems}\label{subsec3.2}

Substituting ansatzes (\ref{e3.1})--(\ref{e3.9}) into the
NSEs (\ref{e1.1}), we obtain the following systems of reduced
equations:
\begin{equation}\label{e3.10}
\begin{array}{l}
\!\!\!\!\!\!\!\!\!\!\!
\mbox{1.}\quad w^{2}w^{1}_{1}+w^{3}w^{1}_{2}-w^{1}w^{3}\cot z_{2}-
(w^{1})^{2}-(w^{2}+\kappa w^{1})^{2}\sin^{2}z_{2}-
\\[1ex]
  -(w^{3})^{2}-\bigl( (\kappa^{2}+\sin^{-2}z_{2})w^{1}_{11}+w^{1}_{22}-
\kappa w^{1}_{1}-2w^{3}_{2}-2w^{2}_{1}-
\\[1ex]
  -2w^{1}\bigr)\sin z_{2}+w^{1}_{2}\cos z_{2}-w^{1}\sin^{-1}z_{2}-
  (2s+\kappa s_{1})\sin^{2}z_{2}=0,\\
\\
  w^{2}w^{2}_{1}+w^{3}w^{2}_{2}+
w^{3}(w^{2}+2\kappa w^{1})\cot z_{2}-
\\[1ex]
  -\kappa\bigl(
(w^{1})^{2}+(w^{3})^{2}+(w^{2}+\kappa w^{1})^{2}\sin^{2}z_{2}
\bigr)-
\\[1ex]
  -\bigl( (\kappa^{2}+\sin^{-2}z_{2})w^{2}_{11}+w^{2}_{22}+
3\kappa w^{2}_{1}+2\kappa(w^{3}_{2}+\kappa w^{1}_{1}+w^{1})
\bigr)\cdot
\\[1ex]
  \cdot\sin z_{2}+
(2w^{1}_{1}+2w^{3}_{1}\cot z_{2}-w^{2}-2\kappa w^{1})
\sin^{-1}z_{2}-
\\[1ex]
  -(w^{2}_{2}+2\kappa w^{1}_{2})\cos z_{2}
  +2\kappa s\sin^{2}z_{2}+(1+\kappa^{2}\sin^{2}z_{2})s_{1}=0,\\
\\
  w^{2}w^{3}_{1}+w^{3}w^{3}_{2}-(w^{3})^{2}\cot z_{2}-
(w^{2}+\kappa w^{1})^{2}\sin z_{2}\cos z_{2}-
\\[1ex]
  -\bigl( (\kappa^{2}+\sin^{-2}z_{2})w^{3}_{11}+w^{3}_{22}+
\kappa w^{3}_{1}+2w^{1}_{2}\bigr)\sin z_{2}+
\\[1ex]
  +(2w^{1}+w^{3}_{2}+w^{2}_{1}+\kappa w^{1}_{1})\cos z_{2}+
s_{2}\sin^{2}z_{2}=0,\\
\\
  w^{1}+w^{2}_{1}+w^{3}_{2}=0.
\end{array}
\end{equation}

Hereafter numeration of the reduced systems corresponds to that of the
ansatzes in Subsec. \ref{subsec3.1}. Subscripts $1$ and $2$
denote differentiation with respect to the variables $z_{1}$ and
$z_{2}$, accordingly.

\begin{equation}\label{e3.11}
\begin{array}{l}
\!\!\!\!\!\!\!\!\!\!\!\!\!\!\!\!
\mbox{2--3.}\quad
      w^{1}w^{1}_{1}+w^{3}w^{1}_{2}-z^{-1}_{1}w^{2}w^{2}-
      \bigl(w^{1}_{11}+(1+\kappa^{2}z^{-2}_{1})w^{1}_{22}
      \bigr)-
\\[1ex]
      -2\kappa z^{-2}_{1}w^{2}_{2}+s_{1}=0,\\
\\
      w^{1}w^{2}_{1}+w^{3}w^{2}_{2}+z^{-1}_{1}w^{1}w^{2}
      -\bigl(w^{2}_{11}+(1+\kappa^{2}z^{-2}_{1})w^{2}_{22}\bigr)+
\\[1ex]
      +2\kappa z^{-2}_{1}w^{1}_{2}
      +2z^{-2}_{1}w^{2}-\kappa z^{-1}_{1}s_{2} +
      \varepsilon z^{-1}_{1}=0,\\
\\
      w^{1}w^{3}_{1}+w^{3}w^{3}_{2}-2\kappa z^{-2}_{1}w^{1}w^{2}-
\bigl(w^{3}_{11}+(1+\kappa^{2}z^{-2}_{1})w^{3}_{22}\bigr)+
\\[1ex]
      +2\kappa (z^{-2}_{1}w^{2})_{1}
      -2\kappa^{2}z^{-3}_{1}w^{1}_{2}
      +(1+\kappa^{2}z^{-2}_{1})s_{2}-
      \varepsilon\kappa z^{-2}_{1}=0,\\
\\
      w^{1}_{1}+w^{3}_{2}+z^{-1}_{1}w^{1}+\gamma=0,
\end{array}
\end{equation}
where $\gamma=\pm 3/2$ for ansatz (\ref{e3.2}) and $\gamma=0$
for ansatz (\ref{e3.3}). Here and below the upper and lower sign
in the symbols "$\pm$" and "$\mp$" are associated with $t>0$ and
$t<0$, respectively.

4--7.For ansatzes (\ref{e3.4})--(\ref{e3.7}) the reduced equations
can be written in the form
\begin{equation}\label{e3.12}
\begin{array}{l}
w^{i}w^{1}_{i}-w^{1}_{ii}+s_{1}+\alpha_{2}w^{2}=0,
\\[1ex]
w^{i}w^{2}_{i}-w^{2}_{ii}+s_{2}-\alpha_{2}w^{1}+\alpha_{1}w^{3}=0,
\\[1ex]
w^{i}w^{3}_{i}-w^{3}_{ii}+\alpha_{4}w^{3}+\alpha_{5}=0,
\\[1ex]
w^{i}_{i}=\alpha_{3}
\end{array}
\end{equation}
where the constants $\alpha_{n}\: (n=\overline{1,5})$, take on the
values
\begin{displaymath}
\begin{array}{llllll}
\!\!\!\!\!\!\!\!\!\!\!
\mbox{4.}&\alpha_{1}=\pm 2\kappa\nu,&\alpha_{2}=\mp 2\kappa\mu,
&\alpha_{3}=\mp(\sigma+3/2),&\alpha_{4}=\pm\sigma,
&\alpha_{5}=\varepsilon.\\
\!\!\!\!\!\!\!\!\!\!\!
\mbox{5.}&\alpha_{1}=0,              &\alpha_{2}=0,
&\alpha_{3}=\mp(\sigma+3/2),&\alpha_{4}=\pm\sigma,
&\alpha_{5}=\varepsilon.\\
\!\!\!\!\!\!\!\!\!\!\!
\mbox{6.}&\alpha_{1}= 2\nu,          &\alpha_{2}=-2\mu,
&\alpha_{3}=-\sigma,        &\alpha_{4}=\sigma,
&\alpha_{5}=\varepsilon.\\
\!\!\!\!\!\!\!\!\!\!\!
\mbox{7.}&\alpha_{1}= 0,            &\alpha_{2}=0,
&\alpha_{3}=-\sigma,        &\alpha_{4}=\sigma,
&\alpha_{5}=\varepsilon.
\end{array}
\end{displaymath}
\begin{equation}\label{e3.13}
\begin{array}{l}
\!\!\!\!\!\!\!\!\!\!\!
\mbox{8.}\quad w^{1}_{1}+(w^{1})^{2}-z^{-4}_{2}(w^{2}-\chi)^{2}+
z_{2}w^{1}w^{1}_{2}-w^{1}_{22}-
\\[1ex]
-3z_{2}w^{1}_{2}+z^{-1}_{2}s_{2}=0,
\end{array}
\end{equation}
\begin{equation}\label{e3.14}
w^{2}_{1}+z_{2}w^{1}w^{2}_{2}-w^{2}_{22}+z^{-1}_{2}w^{2}_{2}=0,
\end{equation}
\begin{equation}\label{e3.15}
w^{3}_{1}+z_{2}w^{1}w^{3}_{2}-w^{3}_{22}-z^{-1}_{2}w^{3}_{2}+
z^{-2}_{2}(w^{2}-\chi)=0,
\end{equation}
\begin{equation}\label{e3.16}
2w^{1}+z_{2}w^{1}_{2}+\rho_{1}/\rho=0.
\end{equation}
\begin{equation}\label{e3.17}
\!\!\!\!\!\!\!\!\!\!\!
\mbox{9.}\quad \vec w_{1}-\lambda\vec w_{22}+s_{2}\vec k+
\lambda^{-1}(\vec n^{i}\cdot\vec w)\vec m^{i}_{t}+z_{2}\vec e=
\vec 0,
\end{equation}
\begin{equation}\label{e3.18}
\vec k\cdot\vec w_{2}=0,
\end{equation}
where $y_{1}=t$ and
\begin{displaymath}
\vec e = \vec e(t) = 2\lambda^{-2}(\vec m^{1}_{t}\cdot\vec m^{2}-
\vec m^{1}\cdot\vec m^{2}_{t})\vec k_{t}\times\vec k+
\lambda^{-2}(2\vec k_{t}\cdot\vec k_{t}-\vec k_{tt}\cdot\vec k).
\end{displaymath}

Let us study symmetry properties of reduced systems
(\ref{e3.10}) and (\ref{e3.11}).

\begin{theorem}\label{t3.1}
The MIA of (\ref{e3.10}) is given by the algebra
$<\partial_{1}>$.
\end{theorem}

\begin{theorem}\label{t3.2}
The MIA of (\ref{e3.11}) is given by the following algebras:

\vspace{1ex}

a) $<\partial_{2}, \partial_{s},
D^{2}_{1}=z_{i}\partial_{i}-w^{a}\partial_{w^{a}}-2s\partial_{s}
>$\quad if \quad $\gamma=\kappa=\varepsilon=0$;

\vspace{1ex}

b) $<\partial_{2}, \partial_{s}>$\quad if \quad
$(\gamma,\kappa,\varepsilon)\not=(0,0,0)$.
\end{theorem}

All the Lie symmetry operators of systems (\ref{e3.10}) and
(\ref{e3.11}) are induced by elements of $A(NS)$. So, for
system (\ref{e3.10}) the operator $\partial_{1}$ is induced by
$J_{12}$. For system (\ref{e3.11}), when $\gamma=0$
($\gamma=\pm3/2$), the operators $D^{2}_{1}$, $\partial_{2}$, and
$\partial_{s}$ ($\partial_{2}$ and $\partial_{s}$) are induced by
$D$, $R(0,0,1)$, and $Z(1)$ ($R(0,0,\vert t \vert^{-1/2})$ and
$Z(\vert t \vert^{-1})$), accordingly. Therefore, the Lie
reductions of systems (\ref{e3.10}) and (\ref{e3.11}) give only
solutions that can be obtained by reducing the NSEs with
three-dimensional subalgebras of $A(NS)$ immediately to ODEs.

Investigation of reduced systems (\ref{e3.13})--(\ref{e3.16}),
(\ref{e3.17})--(\ref{e3.18}), and (\ref{e3.12})
 is given in Sec. \ref{sec5} and \ref{sec6}.

\setcounter{equation}{0}

\section
{Reduction of the Navier-Stokes equations
to ordinary differential equations}
\label{sec4}

\subsection
{Ansatzes of codimension three }\label{subsec4.1}

By means  of  subalgebraic  analysis  of  $A(NS)$  (see  Subsec.~
\ref{subsec_a.3}) and the method described in Sec.~\ref{sec_b} one can
obtain the following ansatzes that reduce the NSEs to ODEs:

\begin{equation}\label{e4.1}
\!\!\!\!\!\!\!\!\!\!\!
\begin{array}{llll}
\mbox{1.}&u^{1}&=&x_{1}R^{-2}\varphi^{1}-x_{2}(Rr)^{-1}\varphi^{2}+
x_{1}x_{3}r^{-1}R^{-2}\varphi^{3},\\
\\
         &u^{2}&=&x_{2}R^{-2}\varphi^{1}+x_{1}(Rr)^{-1}\varphi^{2}+
x_{2}x_{3}r^{-1}R^{-2}\varphi^{3},\\
\\
          &u^{3}&=&x_{3}R^{-2}\varphi^{1}-
rR^{-2}\varphi^{3},\\
\\
          &p    &=&R^{-2}h,
\end{array}
\end{equation}
where\quad $\omega=\arctan r/x_{3}$. Here and below
\quad$\varphi^{a}=\varphi^{a}(\omega)$,\quad $h=h(\omega)$, \\
$r=(x^{2}_{1}+x^{2}_{2})^{1/2}$, \quad
$R=(x^{2}_{1}+x^{2}_{2}+x^{2}_{3})^{1/2}$.

\begin{equation}\label{e4.2}
\!\!\!\!\!\!\!\!\!\!\!
\begin{array}{ll}
\mbox{2.}&u^{1}=r^{-2}(x_{1}\varphi^{1}-x_{2}\varphi^{2}),\quad
u^{2}=r^{-2}(x_{2}\varphi^{1}+x_{1}\varphi^{2}),\\
\\
&u^{3}=r^{-1}\varphi^{3},\quad p=r^{-2}h,
\end{array}
\end{equation}
where \quad $\omega=\arctan x_{2}/x_{1}-\kappa\ln r$,\quad
$\kappa\geq0$.

\begin{equation}\label{e4.3}
\!\!\!\!\!\!\!\!\!\!\!
\begin{array}{llll}
\mbox{3.}&u^{1}&=&x_{1}\vert t \vert^{-1}\varphi^{1}-
x_{2}r^{-2}\varphi^{2}+\frac{1}{2}x_{1}t^{-1},\\
\\
         &u^{2}&=&x_{2}\vert t \vert^{-1}\varphi^{1}+
x_{1}r^{-2}\varphi^{2}+\frac{1}{2}x_{2}t^{-1},\\
\\
         &u^{3}&=&\vert t \vert^{-1/2}\varphi^{3}+
(\sigma+\frac{1}{2})x_{3}t^{-1}+
\nu\vert t \vert^{1/2}t^{-1}\arctan x_{2}/x_{1},\\
\\
         &p    &=&\vert t \vert^{-1}h+\frac{1}{8}t^{-2}R^{2}-
\frac{1}{2}\sigma^{2}x^{2}_{3}t^{-2}+\\
\\
&&&+\varepsilon_{1}\vert t \vert^{-1}\arctan x_{2}/x_{1}+
\varepsilon_{2}x_{3}\vert t \vert^{-3/2},
\end{array}
\end{equation}
where \quad $\omega=\vert t \vert^{-1/2}r$,\quad $\nu\sigma=0,\:$
$\:\varepsilon_{2}\sigma=0,\:$ $\:\varepsilon_{1}\geq0$, $\nu\geq0$.

\begin{equation}\label{e4.4}
\!\!\!\!\!\!\!\!\!\!\!
\begin{array}{llll}
\mbox{4.}&u^{1}&=&x_{1}\varphi^{1}-x_{2}r^{-2}\varphi^{2},\\
\\
         &u^{2}&=&x_{2}\varphi^{1}+x_{1}r^{-2}\varphi^{2},\\
\\
         &u^{3}&=&\varphi^{3}+\sigma x_{3}+\nu\arctan x_{2}/x_{1},\\
\\
         &p    &=&h-\frac{1}{2}\sigma^{2}x^{2}_{3}+
\varepsilon_{1}\arctan x_{2}/x_{1}+\varepsilon_{2}x_{3},
\end{array}
\end{equation}
where \quad $\omega=r$,\quad $\nu\sigma=0$, $\:\varepsilon_{2}\sigma=0,\:$
and for $\sigma=0$ one of the conditions
\begin{displaymath}
\nu=1,\: \varepsilon_{1}\geq0;\quad
\nu=0,\: \varepsilon_{1}=1,\: \varepsilon_{2}\geq0;\quad
\nu=\varepsilon_{1}=0,\: \varepsilon_{2}\in\{0;1\}
\end{displaymath}
is satisfied.

Two ansatzes are described better in the following way:

5. The expressions for $u^{a}$ and $p$ are determined by
(\ref{e2.1}), where
\begin{equation}\label{e4.5}
\begin{array}{lll}
v^{1}&=&a_{1}\varphi^{1}+a_{2}\varphi^{3}+b_{1i}\omega_{i},\\
\\
v^{2}&=&\varphi^{2}+b_{2i}\omega_{i},\\
\\
v^{3}&=&a_{2}\varphi^{1}-a_{1}\varphi^{3}+b_{3i}\omega_{i},\\
\\
p   &=&h+c_{1i}\omega_{i}+c_{2i}\omega\omega_{i}+
\frac{1}{2}d_{ij}\omega_{i}\omega_{j}.
\end{array}
\end{equation}
In formulas (\ref{e4.5}) we use the following definitions:
\begin{displaymath}
\omega_{1}=a_{1}y_{1}+a_{2}y_{3}, \quad \omega_{2}=y_{2}, \quad
\omega=\omega_{3}=a_{2}y_{1}-a_{1}y_{3};
\end{displaymath}
\begin{displaymath}
a_{i}=\mbox{const},\quad a^{2}_{1}+a^{2}_{2}=1; \quad
a_{2}=0 \:\:  \mbox{if} \:\: \gamma_{1}=0;
\end{displaymath}
\begin{displaymath}
\begin{array}{l}
\gamma_{1}=-2\kappa,\: \gamma_{2}=-\frac{3}{2}\quad \mbox{if}\quad t>0
\quad \mbox{and} \quad
\gamma_{1}=2\kappa,\: \gamma_{2}=\frac{3}{2}\quad \mbox{if}\quad t<0.
\end{array}
\end{displaymath}
$b_{ai}$, $B_{i}$, $c_{ij}$, and $d_{ij}$ are real constants that
satisfy the equations
\begin{equation}\label{e4.6}
\begin{array}{l}
b_{1i}=a_{1}B_{i},\quad b_{3i}=a_{2}B_{i},\quad
c_{2i}+a_{2}\gamma_{1}b_{2i}=0,\\
\\
b_{21}B_{i}+b_{22}b_{2i}-\gamma_{1}a_{1}B_{i}+d_{2i}=0,\\
\\
B_{1}B_{i}+B_{2}b_{2i}+\gamma_{1}a_{1}B_{i}+d_{1i}=0,\\
\\
(B_{1}+b_{22})(B_{2}+a_{1}\gamma_{1}-b_{21})=0.
\end{array}
\end{equation}

6. The expressions for $u^{a}$ and $p$ have form (\ref{e2.2}),
where $v^{a}$ and $q$ are determined by (\ref{e4.5}),
(\ref{e4.6}), and $\:\gamma_{1}=-2\kappa$, $\:\gamma_{2}=0$.

\begin{note}\label{n4.1}
Formulas (\ref{e4.5}) and (\ref{e4.6}) determine an ansatz for
system (\ref{e2.7}), where equations (\ref{e4.6}) are the
necessary and sufficient condition to reduce system
(\ref{e2.7}) by means of an ansatz of form (\ref{e4.5}).
\end{note}

\begin{equation}\label{e4.7}
\!\!\!\!\!\!\!\!\!\!\!
\begin{array}{llll}
\mbox{7.}&u^{1}&=&\varphi^{1}\cos x_{3}/\eta^{3}-
\varphi^{2}\sin x_{3}/\eta^{3}+
x_{1}\theta^{1}(t)+x_{2}\theta^{2}(t),\\
\\
         &u^{2}&=&\varphi^{1}\sin x_{3}/\eta^{3}+
\varphi^{2}\cos x_{3}/\eta^{3}-
x_{1}\theta^{2}(t)+x_{2}\theta^{1}(t),\\
\\
         &u^{3}&=&\varphi^{3}+\eta^{3}_{t}(\eta^{3})^{-1}x_{3},\\
\\
         &p    &=&h-\frac{1}{2}\eta^{3}_{tt}(\eta^{3})^{-1}x_{3}^{2}-
\frac{1}{2}\eta^{j}_{tt}\eta^{j}(\eta^{i}\eta^{i})^{-1}r^{2},
\end{array}
\end{equation}
where $\:\omega=t$,
\begin{displaymath}
\begin{array}{l}
\eta^{a}\in C^{\infty}((t_{0},t_{1}),{\R}),\quad \eta^{3}\not=0,
\quad \eta^{i}\eta^{i}\not=0,\quad
\eta^{1}_{t}\eta^{2}-\eta^{1}\eta^{2}_{t}\in\{ 0; \frac{1}{2}\},
\end{array}
\end{displaymath}
\begin{displaymath}
\theta^{1}=\eta^{i}_{t}\eta^{i}(\eta^{j}\eta^{j})^{-1}, \quad
\theta^{2}=(\eta^{1}_{t}\eta^{2}-\eta^{1}\eta^{2}_{t})
(\eta^{j}\eta^{j})^{-1}.
\end{displaymath}

\begin{equation}\label{e4.8}
\!\!\!\!\!\!\!\!\!\!\!
\begin{array}{llll}
\mbox{8.}&\vec u&=&\vec \varphi+\lambda^{-1}
(\vec n^{a}\cdot\vec x)\vec m^{a}_{t},\\
\\
         &p    &=&h-\lambda^{-1}(\vec m^{a}_{tt}\cdot\vec x)
(\vec n^{a}\cdot\vec x)+\\
\\
&&&\quad\quad\quad\quad\quad\quad+\frac{1}{2}\lambda^{-2}
(\vec m^{b}_{tt}\cdot\vec m^{a})(\vec n^{a}\cdot\vec x)
(\vec n^{b}\cdot\vec x),
\end{array}
\end{equation}
where $\:\omega=t$, \quad
$\vec m^{a}\in C^{\infty}((t_{0},t_{1}),{\R})$,\quad
$\vec m^{a}_{tt}\cdot\vec m^{b}-\vec m^{a}\cdot\vec m^{b}_{tt}=0$,
\begin{displaymath}
\lambda=\lambda(t)=(\vec m^{1}\times\vec m^{2})\cdot\vec m^{3}
\not=0\quad\forall t\in (t_{0},t_{1}),
\end{displaymath}
\begin{displaymath}
\vec n^{1}=\vec m^{2}\times\vec m^{3}, \quad
\vec n^{2}=\vec m^{3}\times\vec m^{1}, \quad
\vec n^{3}=\vec m^{1}\times\vec m^{2}.
\end{displaymath}

\subsection
{Reduced systems}\label{subsec4.2}

Substituting the ansatzes 1--8 into the NSEs (\ref{e1.1}), we
obtain the following systems of ODE in the functions
\quad $\varphi^{a}$ and $h$:

\begin{equation}\label{e4.9}
\!\!\!\!\!\!\!\!\!\!\!
\begin{array}{ll}
\mbox{1.}&\varphi^{3}\varphi^{1}_{\omega}-\varphi^{a}\varphi^{a}-
\varphi^{1}_{\omega\omega}-\varphi^{1}_{\omega}\cot\omega-2h=0,\\
\\
  &\varphi^{3}\varphi^{2}_{\omega}+\varphi^{2}\varphi^{3}
\cot\omega-
\varphi^{2}_{\omega\omega}-\varphi^{2}_{\omega}\cot\omega
+\varphi^{2}\sin^{-2}\omega=0,\\
\\
  &\varphi^{3}\varphi^{3}_{\omega}-\varphi^{2}\varphi^{2}
\cot\omega-
\varphi^{3}_{\omega\omega}-\varphi^{3}_{\omega}\cot\omega
+\varphi^{3}\sin^{-2}\omega+
\\[1ex]
&-2\varphi^{1}_{\omega}+h_{\omega}=0,\\
\\
  &\varphi^{1}+\varphi^{3}_{\omega}+\varphi^{3}\cot\omega=0.
\end{array}
\end{equation}

\begin{equation}\label{e4.10}
\!\!\!\!\!\!\!\!\!\!\!
\begin{array}{ll}
\mbox{2.}&(\varphi^{2}-\kappa\varphi^{1})\varphi^{1}_{\omega}-
(1+\kappa^{2})\varphi^{1}_{\omega\omega}-\varphi^{1}\varphi^{1}-
\varphi^{2}\varphi^{2}-\kappa h_{\omega}-2h=0,\\
\\
  &(\varphi^{2}-\kappa\varphi^{1})\varphi^{2}_{\omega}-
(1+\kappa^{2})\varphi^{2}_{\omega\omega}-
2(\kappa\varphi^{2}_{\omega}+\varphi^{1}_{\omega})+
h_{\omega}=0,\\
\\
  &(\varphi^{2}-\kappa\varphi^{1})\varphi^{3}_{\omega}-
(1+\kappa^{2})\varphi^{3}_{\omega\omega}-\varphi^{1}\varphi^{3}-
\varphi^{3}-2\kappa \varphi^{3}_{\omega}=0,\\
\\
  &\varphi^{2}_{\omega}-\kappa\varphi^{1}_{\omega}=0.
\end{array}
\end{equation}

\begin{equation}\label{e4.11}
\!\!\!\!\!\!\!\!\!\!\!\!\!\!\!
\begin{array}{ll}
\mbox{3--4.}&\varphi^{1}\varphi^{1}-\omega^{-4}\varphi^{2}
\varphi^{2}+\omega\varphi^{1}\varphi^{1}_{\omega}-
\varphi^{1}_{\omega\omega}-3\omega^{-1}\varphi^{1}_{\omega}+
\omega^{-1}h_{\omega}=0,\\
\\
  &\omega\varphi^{1}\varphi^{2}_{\omega}-
\varphi^{2}_{\omega\omega}+\omega^{-1}\varphi^{2}_{\omega}+
\varepsilon_{1}=0,\\
\\
  &\omega\varphi^{1}\varphi^{3}_{\omega}+\sigma_{1}\varphi^{3}+
\nu\omega^{-2}\varphi^{2}-\varphi^{3}_{\omega\omega}-
\omega^{-1}\varphi^{3}_{\omega}+\varepsilon_{2}=0,\\
\\
  &2\varphi^{1}+\omega\varphi^{1}_{\omega}+\sigma_{2}=0,
\end{array}
\end{equation}
where
\begin{displaymath}
\!\!\!\!\!\!\!\!\!\!\!
\begin{array}{llll}
\mbox{3.} &\sigma_{1}=\sigma,\quad &\sigma_{2}=(\sigma+\frac{3}{2})
\quad &\mbox{if}\quad t>0,\\
& \sigma_{1}=-\sigma,\quad &\sigma_{2}=-(\sigma+\frac{3}{2})
\quad &\mbox{if} \quad  t<0.
\end{array}
\end{displaymath}
\begin{displaymath}
\!\!\!\!\!\!\!\!\!
\mbox{4.}\quad \sigma_{1}=\sigma_{2}=\sigma.
\end{displaymath}

\begin{equation}\label{e4.12}
\!\!\!\!\!\!\!\!\!\!\!\!\!\!\!\!
\begin{array}{ll}
\mbox{5--6.}&\varphi^{3}\varphi^{1}_{\omega}-
\varphi^{1}_{\omega\omega}-\mu_{1i}\varphi^{i}+
c_{11}+c_{21}\omega=0,\\
\\
  &\varphi^{3}\varphi^{2}_{\omega}-
\varphi^{2}_{\omega\omega}-\mu_{2i}\varphi^{i}+
c_{12}+c_{22}\omega+\gamma_{2}a_{2}\varphi^{3}=0,\\
\\
  &\varphi^{3}\varphi^{3}_{\omega}-
\varphi^{3}_{\omega\omega}+\gamma_{1}a_{2}\varphi^{2}+h_{\omega}=
0,\\
\\
&\varphi^{3}_{\omega}=\sigma,
\end{array}
\end{equation}
where $\:\mu_{11}=-B_{1}$, $\:\mu_{12}=-B_{2}-\gamma_{1}a_{1}$,
$\:\mu_{21}=-b_{21}+\gamma_{1}a_{1}$, $\:\mu_{22}=-b_{22}$,
\mbox{$\sigma=\gamma_{1}-B_{1}-b_{22}$}.

\begin{equation}\label{e4.13}
\!\!\!\!\!\!\!\!\!\!\!
\begin{array}{ll}
\mbox{7.}&\varphi^{1}_{\omega}+\theta^{1}\varphi^{1}+
\theta^{2}\varphi^{2}-(\eta^{3})^{-1}\varphi^{3}\varphi^{2}+
(\eta^{3})^{-2}\varphi^{1}=0,\\
\\
  &\varphi^{2}_{\omega}-\theta^{2}\varphi^{1}+
\theta^{1}\varphi^{2}+(\eta^{3})^{-1}\varphi^{3}\varphi^{1}+
(\eta^{3})^{-2}\varphi^{2}=0,\\
\\
  &\varphi^{3}_{\omega}+\eta^{3}_{t}(\eta^{3})^{-1}\varphi^{3}=0,\\
\\
  &2\theta^{1}+\eta^{3}_{t}(\eta^{3})^{-1}=0.
\end{array}
\end{equation}

\begin{equation}\label{e4.14}
\!\!\!\!\!\!\!\!\!\!\!
\begin{array}{ll}
\mbox{8.}&\vec\varphi_{\omega}+\lambda^{-1}(\vec n^{b}\cdot\vec\varphi)
\vec m^{b}_{t}=0,\\
\\
  &\vec n^{a}\cdot\vec m^{a}_{t}=0.
\end{array}
\end{equation}

\subsection
{Exact solutions of the reduced systems}
\label{subsec4.3}

1. Ansatz (\ref{e4.1}) and system (\ref{e4.9}) determine
the class of solutions of the NSEs (\ref{e1.1}) that are called
the steady axially symmetric conically similar flows of a viscous
fluid in hydrodynamics. This class of solutions was studied in
a number of works (for example, see references in
\cite{goldshtik}). For $\varphi^{2}=0$ it was shown, by
N.A.Slezkin \cite{slezkin}, that system (\ref{e4.9})
is reduced to a Riccati equation. The general solution of this
equation was expressed in terms of hypergeometric functions.
Later similar calculations were made by V.I.Yatseev
\cite{yatseev} and H.B.Squire \cite{squire}. The particular case
in the class of solutions with $\varphi^{2}=0$ is formed by
the Landau jets \cite{landau}. For swirling flows, where
$\varphi^{2}\not=0$, the order of system (\ref{e4.9}) can be
reduced too. For example \cite{sedov}, an arbitrary solution of
(\ref{e4.9}) satisfies the equation
\begin{displaymath}
\varphi^{2}\varphi^{2}\sin^{2}\omega-\sin\omega
(\Phi_{\omega}\sin^{-1}\omega)_{\omega}+2\Phi_{\omega}\cot\omega+
2\Phi=\mbox{const},
\end{displaymath}
where $\:\Phi=(\varphi^{3}_{\omega}-\frac{1}{2}\varphi^{3}\varphi^{3}
)\sin^{2}\omega-\varphi^{3}\cos\omega\sin\omega,\:$ and the Yatseev
results \cite{yatseev} are completely extended to the case
$\:\varphi^{2}\sin\omega=\mbox{const}$.

2. System (\ref{e4.10}) implies that
\begin{equation}\label{e4.15}
\!\begin{array}{l}
\varphi^{2}=\kappa\varphi^{1}+C_{1},\\
\\
h=\kappa(1+\kappa^{2})\varphi^{1}_{\omega}+
(2\kappa^{2}+2-\kappa C_{1})\varphi^{1}+C_{2},\\
\\
(1+\kappa^{2})\varphi^{1}_{\omega\omega}+
(4\kappa-C_{1})\varphi^{1}_{\omega}+
\varphi^{1}\varphi^{1}+4\varphi^{1}+
\\[1ex]
+(1+\kappa^{2})^{-1}(C^{2}_{1}+2C_{2})=0,\\
\\
(1+\kappa^{2})\varphi^{3}_{\omega\omega}-
(C_{1}-2\kappa)\varphi^{3}_{\omega}+(1+\varphi^{1})\varphi^{3}=0.
\end{array}
\end{equation}
If $\:\varphi^{3}=0,\:$ the solution determined by ansatz
(\ref{e4.10}) and formulas (\ref{e4.15}) coincides with the
Hamel solution \cite{hamel,kochin}. In Sec. \ref{sec6} we
consider system (\ref{e6.15}) which is more general than
system (\ref{e4.10}).

3--4. Let us integrate the last equation of system
(\ref{e4.11}), i.e.,
\begin{equation}\label{e4.16}
\begin{array}{l}
\varphi^{1}=C_{1}\omega^{-2}-\frac{1}{2}\sigma_{2}.
\end{array}
\end{equation}
Taking into account the integration result, the other equations of
system (\ref{e4.11}) can be written in the form
\begin{displaymath}
\begin{array}{l}
h_{\omega}=\omega^{-3}\varphi^{2}\varphi^{2}+C^{2}_{1}\omega^{-3}
-\frac{1}{4}\sigma^{2}_{2}\omega,
\end{array}
\end{displaymath}
\begin{displaymath}
\begin{array}{l}
\varphi^{2}_{\omega\omega}-((C_{1}+1)\omega^{-1}-
\frac{1}{2}\sigma_{2}\omega)\varphi^{2}_{\omega}=\varepsilon_{1},
\end{array}
\end{displaymath}
\begin{equation}\label{e4.17}
\begin{array}{l}
\varphi^{3}_{\omega\omega}-((C_{1}-1)\omega^{-1}-
\frac{1}{2}\sigma_{2}\omega)\varphi^{3}_{\omega}-
\sigma_{1}\varphi^{3}=\nu\omega^{-2}\varphi^{2}+\varepsilon_{2}.
\end{array}
\end{equation}
Therefore,

\begin{equation}\label{e4.18}
\begin{array}{l}
h=\int\omega^{-3}\varphi^{2}\varphi^{2}d\omega-
\frac{1}{2}C^{2}_{1}\omega^{-2}-
\frac{1}{8}\sigma^{2}_{2}\omega^{2},
\end{array}
\end{equation}

\begin{equation}\label{e4.19}
\begin{array}{ll}
\varphi^{2}=\!\!\!\!&C_{2}+C_{3}\int\vert\omega\vert^{C_{1}+1}
e^{-\frac{1}{4}\sigma_{2}\omega^{2}}d\omega+\\
\\
&+\varepsilon_{1}\int\vert\omega\vert^{C_{1}+1}
e^{-\frac{1}{4}\sigma_{2}\omega^{2}}\Bigl( \int
\vert\omega\vert^{-C_{1}-1}e^{\frac{1}{4}\sigma_{2}\omega^{2}}
d\omega\Bigr)d\omega.
\end{array}
\end{equation}
If $\:\sigma_{1}=0,\:$ it follows that

\begin{equation}\label{e4.20}
\begin{array}{ll}
\varphi^{3}\!\!\!\!&=C_{4}+C_{5}\int\vert\omega\vert^{C_{1}-1}
e^{-\frac{1}{4}\sigma_{2}\omega^{2}}d\omega+\\
\\
&+\int\vert\omega\vert^{C_{1}-1}
e^{-\frac{1}{4}\sigma_{2}\omega^{2}}\Bigl( \int
\vert\omega\vert^{-C_{1}+1}e^{\frac{1}{4}\sigma_{2}\omega^{2}}
(\varepsilon_{2}+\nu\omega^{-2}\varphi^{2})d\omega\Bigr)d\omega.
\end{array}
\end{equation}
Let $\sigma_{1}\not=0$ (and, therefore, $\nu=0$). Then, if
$\sigma_{2}\not=0$, the general solution of equation
(\ref{e4.17}) is expressed in terms of Whittaker functions:
\begin{displaymath}
\begin{array}{l}
\varphi^{3}=\vert\omega\vert^{\frac{1}{2}C_{1}-1}
e^{-\frac{1}{8}\sigma_{2}\omega^{2}}
W(-\sigma_{1}\sigma^{-1}_{2}+\frac{1}{4}C_{1}-\frac{1}{2},
\frac{1}{4}C_{1},\frac{1}{4}\sigma_{2}\omega^{2}),
\end{array}
\end{displaymath}
where $W(\kappa,\mu,\tau)$ is the general solution of the
Whittaker equation

\begin{equation}\label{e4.21}
4\tau^{2}W_{\tau\tau}=(\tau^{2}-4\kappa\tau+4\mu^{2}-1)W.
\end{equation}
If $\:\sigma_{2}=0,\:$ the general solution of equation
(\ref{e4.16}) is expressed in terms of Bessel functions:
\begin{displaymath}
\varphi^{3}=\vert\omega\vert^{\frac{1}{2}C_{1}}Z_{\frac{1}{2}C_{1}}
\bigl( (-\sigma_{1})^{1/2}\omega\bigr),
\end{displaymath}
where $Z_{\nu}(\tau)$ is the general solution of the Bessel
equation

\begin{equation}\label{e4.22}
\tau^{2}Z_{\tau\tau}+\tau Z_{\tau}+(\tau^{2}-\nu^{2})Z=0.
\end{equation}

\begin{note}\label{n4.2}
If $\:\sigma_{2}=0,\:$ all quadratures in formulas
(\ref{e4.18})--(\ref{e4.20}) are easily integrated. For example,
\begin{displaymath}
\varphi^{2}=\left\{
\begin{array}{ll}
C_{2}+C_{3}\ln\vert\omega\vert+\frac{1}{4}\varepsilon_{1}
\omega^{2} & \mbox{if}\quad C_{1}=-2,
\\[1ex]
C_{2}+C_{3}\frac{1}{2}\omega^{2}+\frac{1}{2}\varepsilon_{1}
\omega^{2}(\ln\omega-\frac{1}{2}) & \mbox{if} \quad C_{1}=0,
\\[1ex]
C_{2}+C_{3}(C_{1}+2)^{-1}\vert\omega\vert^{C_{1}+2}-
\frac{1}{2}\varepsilon_{1}C_{1}^{-1}\omega^{2}
& \mbox{if}\quad C_{1}\not=-2,0.
\end{array} \right.
\end{displaymath}
\end{note}

5--6. Let $\sigma=0$. Then the last equation of system
(\ref{e4.12}) implies that $\varphi^{3}=C_{0}=\mbox{const}$. The other
equations of system (\ref{e4.12}) can be written in the
form
\begin{displaymath}
\begin{array}{l}
h=-\gamma_{1}a_{2}\int\varphi^{2}(\omega)d\omega,
\end{array}
\end{displaymath}
\begin{equation}\label{e4.23}
\varphi^{i}_{\omega\omega}-C_{0}\varphi^{i}_{\omega}+
\mu_{ij}\varphi^{j}=\nu_{1i}+\nu_{2i}\omega,
\end{equation}
where $\:\nu_{11}=c_{11},\:$ $\:\nu_{21}=c_{21},\:$
$\:\nu_{12}=c_{12}+\gamma_{2}a_{2}C_{0},\:$ $\:\nu_{22}=c_{22}$.
System (\ref{e4.23}) is a linear nonhomogeneous system of ODEs
with constant coefficients. The form of its general solution depends
on the Jordan form of the matrix ${\rm M}=\{\mu_{ij}\}$. Now let us
transform the dependent variables
\begin{displaymath}
\varphi^{i}=e_{ij}\psi^{j},
\end{displaymath}
where the constants $e_{ij}$ are determined by means of the
system of linear algebraic equations
\begin{displaymath}
e_{ij}\tilde\mu_{jk}=\mu_{ij}e_{jk}\quad (i,j,k=1,2)
\end{displaymath}
with the condition $\det\{e_{ij}\}\not=0$. Here
${\rm\tilde M}=\{\tilde\mu_{ij}\}$ is the real Jordan form of the
matrix ${\rm M}$. The new unknown functions $\psi^{i}$ have
to satisfy the following system
\begin{equation}\label{e4.24}
\psi^{i}_{\omega\omega}-C_{0}\psi^{i}_{\omega}+
\tilde\mu_{ij}\psi^{j}=\tilde\nu_{1i}+\tilde\nu_{2i}\omega,
\end{equation}
where $\:\nu_{1i}=e_{ij}\tilde\nu_{1j},\:$
$\:\nu_{2i}=e_{ij}\tilde\nu_{2j}$. Depending on the form of
${\rm\tilde M}$, we consider the following cases:

\vspace{2ex}
A. $\det{\rm\tilde M}=0$ (this is equivalent to the condition
$\det {\rm M}=0$ ).

\vspace{2ex}
\hspace*{1em}
i. $\tilde M=\left( \begin{array}{cc}0&\varepsilon\\0&0
\end{array} \right),\:$
where $\:\varepsilon\in\{0;1\}$. Then
\begin{equation}\label{e4.25}
\begin{array}{l}
\psi^{2}=C_{1}+C_{2}e^{C_{0}\omega}-\frac{1}{2}\tilde\nu_{22}
C_{0}^{-1}\omega^{2}-(\tilde\nu_{12}-\tilde\nu_{22}C_{0}^{-1})
C_{0}^{-1}\omega,
\end{array}
\end{equation}
\begin{displaymath}
\begin{array}{ll}
\psi^{1}=\!\!\!\!&C_{3}+C_{4}e^{C_{0}\omega}-\frac{1}{2}\tilde\nu_{21}
C_{0}^{-1}\omega^{2}-(\tilde\nu_{11}-\tilde\nu_{21}C_{0}^{-1})
C_{0}^{-1}\omega+\\
\\
&+\varepsilon\Bigl( -\frac{1}{6}\tilde\nu_{22}C_{0}^{-2}\omega^{3}-
\frac{1}{2}(\tilde\nu_{12}-2\tilde\nu_{22}C_{0}^{-1})C_{0}^{-2}
\omega^{2}+\\
\\
&+\bigl(
C_{1}+(\tilde\nu_{21}-2\tilde\nu_{22}C_{0}^{-1})C_{0}^{-2}\bigr)
C_{0}^{-1}\omega-C_{2}C_{0}^{-1}\omega e^{C_{0}\omega} \Bigr)
\end{array}
\end{displaymath}
for $C_{0}\not=0$, and
\begin{equation}\label{e4.26}
\begin{array}{l}
\psi^{2}=C_{1}+C_{2}\omega+\frac{1}{6}\tilde\nu_{22}\omega^{3}+
\frac{1}{2}\tilde\nu_{12}\omega^{2},
\end{array}
\end{equation}
\begin{displaymath}
\begin{array}{ll}
\psi^{1}=\!\!\!\!&C_{3}+C_{4}\omega+\frac{1}{6}(\tilde\nu_{21}-C_{2})
\omega^{3}+\frac{1}{2}(\tilde\nu_{11}-C_{1})\omega^{2}-\\
\\
&-\frac{1}{120}\tilde\nu_{22}\omega^{5}-\frac{1}{24}\tilde\nu_{12}
\omega^{4}
\end{array}
\end{displaymath}
for $C_{0}=0$.

\vspace{2ex}
\hspace*{1em}
ii. ${\rm\tilde M}=\left(\begin{array}{cc}\kappa_{1}&0\\0&0
\end{array}   \right),\:$
where $\:\kappa_{1}\in{\R}\backslash\{0\}$. Then the form of
$\psi^{2}$ is given either by formula (\ref{e4.25}) for
$C_{0}\not=0$ or by formula (\ref{e4.26}) for $C_{0}=0$. The form
of
$\psi^{1}$ is given by formula (\ref{e4.28}) (see below).

\vspace{2ex}
 B. $\det{\rm\tilde M}\not=0$ (this is equivalent to the condition
$\:\det {\rm M}\not=0$).

\vspace{2ex}
\hspace*{1em}
i. ${\rm\tilde M}=\left( \begin{array}{cc}\kappa_{1}&0\\0&\kappa_{2}
\end{array}\right),\:$
where $\:\kappa_{i}\in{\R}\backslash\{0\}$. Then
\begin{equation}\label{e4.27}
\psi^{2}=\tilde\nu_{22}\kappa_{2}^{-1}\omega+(\tilde\nu_{12}-
C_{0}\tilde\nu_{22}\kappa_{2}^{-1})\kappa_{2}^{-1}+C_{1}
\theta^{21}(\omega)+C_{2}\theta^{22}(\omega),
\end{equation}
\begin{equation}\label{e4.28}
\psi^{1}=\tilde\nu_{21}\kappa_{1}^{-1}\omega+(\tilde\nu_{11}-
C_{0}\tilde\nu_{21}\kappa_{1}^{-1})\kappa_{1}^{-1}+C_{3}
\theta^{11}(\omega)+C_{4}\theta^{12}(\omega),
\end{equation}
where

\begin{tabbing}
\quad\quad\quad\=
$\theta^{i1}(\omega)=\exp\bigl(\frac{1}{2}(C_{0}-\sqrt{D_{i}})
\omega\bigr)$,\quad
\=$\theta^{i2}(\omega)=\exp\bigl(\frac{1}{2}(C_{0}+\sqrt{D_{i}})
\omega\bigr)$
\\[1ex]
if\quad $D_{i}=C_{0}^{2}-4\kappa_{i}>0$,
\\[1ex]
\>
$\theta^{i1}(\omega)=e^{\frac{1}{2}C_{0}\omega}
\cos\bigl(\frac{1}{2}\sqrt{-D_{i}}\omega\bigr)$,
\>$\theta^{i2}(\omega)=e^{\frac{1}{2}C_{0}\omega}
\sin\bigl(\frac{1}{2}\sqrt{-D_{i}}\omega\bigr)$
\\[1ex]
if\quad $D_{i}<0$,
\\[1ex]
\>$\theta^{i1}(\omega)=e^{\frac{1}{2}C_{0}\omega}$,
\>$\theta^{i2}(\omega)=\omega e^{\frac{1}{2}C_{0}\omega}$
\\[1ex]
if\quad $D_{i}=0$.
\end{tabbing}

\vspace{2ex}
\hspace*{1em}
ii. ${\rm\tilde M}=\left( \begin{array}{cc}\kappa_{2}&1\\0&\kappa_{2}
\end{array}\right),\:$
where $\:\kappa_{2}\in{\R}\backslash\{0\}$. Then the form of
$\psi^{2}$ is given by formula (\ref{e4.27}), and
\begin{displaymath}
\begin{array}{ll}
\psi^{1}=\!\!\!\!&
\bigl(\tilde\nu_{11}-(\tilde\nu_{12}-C_{0}\tilde\nu_{22}
\kappa_{2}^{-1})\kappa_{2}^{-1}-C_{0}(\tilde\nu_{21}-
\tilde\nu_{22}\kappa_{2}^{-1})\kappa_{2}^{-1}\bigr)
\kappa_{2}^{-1}+\\
\\
&+(\tilde\nu_{21}-\tilde\nu_{22}
\kappa_{2}^{-1})\kappa_{2}^{-1}\omega+
C_{3}\theta^{21}(\omega)+C_{4}\theta^{22}(\omega)-
C_{i}\eta^{i}(\omega),
\end{array}
\end{displaymath}
where
\begin{displaymath}
\begin{array}{ll}
\eta^{j}(\omega)=D_{2}^{-1}\omega\bigl(
2\theta^{2j}_{\omega}(\omega)-C_{0}\theta^{2j}(\omega)\bigr)
&\mbox{if}\quad D_{2}\not=0,\\
\\
\eta^{1}(\omega)=\frac{1}{2}\omega^{2}e^{\frac{1}{2}C_{0}\omega},
\quad \eta^{2}(\omega)=\frac{1}{6}\omega^{3}e^{\frac{1}{2}C_{0}\omega}
&\mbox{if}\quad D_{2}=0.
\end{array}
\end{displaymath}

\vspace{2ex}
\hspace*{1em}
iii. ${\rm\tilde M}=\left( \begin{array}{cc}
\kappa_{1}&-\kappa_{2}\\\kappa_{2}&\kappa_{1}
\end{array}    \right),\:$
where $\:\kappa_{i}\in{\R}$, $\:\kappa_{2}\not=0$. Then
\begin{displaymath}
\begin{array}{ll}
\psi^{1}=\!\!\!\!&(\kappa_{i}\kappa_{i})^{-1}
(\tilde\nu_{21}\kappa_{1}+\tilde\nu_{22}\kappa_{2})\omega+
(\kappa_{i}\kappa_{i})^{-1}
(\tilde\nu_{11}\kappa_{1}+\tilde\nu_{12}\kappa_{2})-\\
\\
&-C_{0}(\kappa_{i}\kappa_{i})^{-2}\bigl(
\tilde\nu_{21}(\kappa_{2}^{2}-\kappa_{1}^{2})-
\tilde\nu_{22}2\kappa_{1}\kappa_{2} \bigr)
+C_{n}\theta^{1n}(\omega),\\
\\
\psi^{2}=\!\!\!\!&(\kappa_{i}\kappa_{i})^{-1}
(-\tilde\nu_{21}\kappa_{2}+\tilde\nu_{22}\kappa_{1})\omega+
(\kappa_{i}\kappa_{i})^{-1}
(-\tilde\nu_{11}\kappa_{2}+\tilde\nu_{12}\kappa_{1})-\\
\\
&-C_{0}(\kappa_{i}\kappa_{i})^{-2}\bigl(
\tilde\nu_{21}2\kappa_{1}\kappa_{2}+
\tilde\nu_{22}(\kappa_{2}^{2}-\kappa_{1}^{2}) \bigr)
+C_{n}\theta^{2n}(\omega),
\end{array}
\end{displaymath}
where $\:n=\overline{1,4}$,

\begin{displaymath}
\begin{array}{l}
\gamma=\sqrt{(C_{0}^{2}-4\kappa_{1})^{2}+(4\kappa_{2})^{2}},
\end{array}
\end{displaymath}
\begin{displaymath}
\begin{array}{l}
\beta_{1}=\frac{1}{4}
\sqrt{2(\gamma+C_{0}^{2}-4\kappa_{1})},\quad
\beta_{2}=\frac{1}{4}
\frac{\vert\kappa_{2}\vert}{\kappa_{2}}
\sqrt{2(\gamma-C_{0}^{2}+4\kappa_{1})},
\end{array}
\end{displaymath}

\begin{displaymath}
\begin{array}{lllllll}
\!\!&\theta^{11}(\omega)\!\!&=\!\!&\!\!&\theta^{22}(\omega)\!\!&=\!\!&
\exp\bigl((\frac{1}{2}C_{0}-\beta_{1})\omega\bigr)\cos\beta_{2}\omega,
\\
\\
-\!\!\!\!&\theta^{21}(\omega)\!\!&=\!\!&\!\!&\theta^{12}(\omega)\!\!&=\!\!&
\exp\bigl((\frac{1}{2}C_{0}-\beta_{1})\omega\bigr)\sin\beta_{2}\omega,
\\
\\
\!\!&\theta^{13}(\omega)\!\!&=\!\!&\!\!&\theta^{24}(\omega)\!\!&=\!\!&
\exp\bigl((\frac{1}{2}C_{0}+\beta_{1})\omega\bigr)\cos\beta_{2}\omega,
\\
\\
\!\!&\theta^{23}(\omega)\!\!&=\!\!&-\!\!\!\!&\theta^{14}(\omega)\!\!&=\!\!&
\exp\bigl((\frac{1}{2}C_{0}+\beta_{1})\omega\bigr)\sin\beta_{2}\omega.
\end{array}
\end{displaymath}

If $\:\sigma\not=0,\:$ the last equation of system (\ref{e4.12})
implies that $\:\psi^{3}=\sigma\omega\:$ (translating $\omega$,
the integration constant can be made to vanish). The other
equations of system (\ref{e4.12}) can be written in the
form
\begin{displaymath}
\begin{array}{l}
h=-\gamma_{1}a_{2}\int\varphi^{2}(\omega)d\omega-
\frac{1}{2}\sigma^{2}\omega^{2},
\end{array}
\end{displaymath}
\begin{equation}\label{e4.29}
\varphi^{i}_{\omega\omega}-\sigma\omega\varphi^{i}_{\omega}+
\mu_{ij}\varphi^{j}=\nu_{1i}+\nu_{2i}\omega,
\end{equation}
where $\:\nu_{11}=c_{11},\:$ $\:\nu_{21}=c_{21},\:$
$\:\nu_{12}=c_{12},\:$ $\:\nu_{22}=c_{22}+\gamma_{2}a_{2}\sigma$.
The form of the general solution of system (\ref{e4.29}) depends
on the Jordan form of the matrix ${\rm M}=\{\mu_{ij}\}$. Now,
let us transform the dependent variables
\begin{displaymath}
\varphi^{i}=e_{ij}\psi^{j},
\end{displaymath}
where the constants $e_{ij}$ are determined by means of the
system of linear algebraic equations
\begin{displaymath}
e_{ij}\tilde\mu_{jk}=\mu_{ij}e_{jk}\quad (i,j,k=1,2)
\end{displaymath}
with the condition $\det\{e_{ij}\}\not=0$. Here
${\rm\tilde M}=\{\tilde\mu_{ij}\}$ is the real Jordan form of the
matrix ${\rm M}$. The new unknown functions $\psi^{i}$ have to
satisfy the following system

\begin{equation}\label{e4.30}
\psi^{i}_{\omega\omega}-\sigma\omega\psi^{i}_{\omega}+
\tilde\mu_{ij}\psi^{j}=\tilde\nu_{1i}+\tilde\nu_{2i}\omega,
\end{equation}
where $\:\nu_{1i}=e_{ij}\tilde\nu_{1j},\:$
$\:\nu_{2i}=e_{ij}\tilde\nu_{2j}$. Depending on the form of $
{\rm\tilde M}$, we consider the following cases:

\vspace{2ex}

A. $\det{\rm\tilde M}=0$ (this is equivalent to the condition
$\det {\rm M}=0$ ).

\vspace{2ex}
\hspace*{1em}
i. $\tilde M=\left( \begin{array}{cc}0&\varepsilon\\0&0
\end{array} \right),\:$
where $\:\varepsilon\in\{0;1\}$. Then
\begin{equation}\label{e4.31}
\begin{array}{ll}
\psi^{2}=\!\!\!\!&C_{1}+C_{2}\int e^{\frac{1}{2}\sigma\omega^{2}}d\omega-
\sigma^{-1}\tilde\nu_{22}\omega+\\
\\
&\quad\quad\quad\quad\quad\quad\quad+\tilde\nu_{12}
\int e^{\frac{1}{2}\sigma\omega^{2}}\bigl(
\int e^{-\frac{1}{2}\sigma\omega^{2}}d\omega\bigl) d\omega,
\end{array}
\end{equation}
\begin{displaymath}
\begin{array}{ll}
\psi^{1}=\!\!\!\!&C_{3}+C_{4}\int e^{\frac{1}{2}\sigma\omega^{2}}d\omega-
\sigma^{-1}\tilde\nu_{21}\omega+\\
\\
&\quad\quad\quad\quad\quad\quad\quad
+\int e^{\frac{1}{2}\sigma\omega^{2}}\bigl(
\int e^{-\frac{1}{2}\sigma\omega^{2}}(\tilde\nu_{11}-\varepsilon
\psi^{2})d\omega\bigl) d\omega.
\end{array}
\end{displaymath}

\vspace{2ex}
\hspace*{1em}
ii. ${\rm\tilde M}=\left(\begin{array}{cc}
\sigma&0\\0&0
\end{array}   \right)$.
Then the form of $\psi^{2}$ is given by formula (\ref{e4.31}),
and
\begin{displaymath}
\begin{array}{ll}
\psi^{1}=\!\!\!\!& C_{3}\omega+C_{4}\bigl(\omega\int
e^{\frac{1}{2}\sigma\omega^{2}}d\omega-
\sigma^{-1}e^{\frac{1}{2}\sigma\omega^{2}}\bigr)+\sigma^{-1}\tilde
\nu_{11}+\\
\\
&+\sigma^{-1}\tilde\nu_{21}\bigl( \sigma\omega
\int e^{\frac{1}{2}\sigma\omega^{2}}\lambda^{1}(\omega)d\omega-
e^{\frac{1}{2}\sigma\omega^{2}}\lambda^{1}(\omega)\bigr),
\end{array}
\end{displaymath}
where $\:\lambda^{1}(\omega)=\int
e^{-\frac{1}{2}\sigma\omega^{2}}d\omega$.

\vspace{2ex}
\hspace*{1em}
iii. ${\rm\tilde M}=\left(
\begin{array}{cc}
\kappa_{1}&0\\0&0
\end{array}    \right),\:$
where $\:\kappa_{1}\in{\R}\backslash\{0;\sigma\}$. Then $\psi^{2}$
is determited by (\ref{e4.31}), and the form of $\psi^{1}$ is
given by (\ref{e4.33}) (see below).

\vspace{2ex}
B. $\det{\rm\tilde M}\not=0$,
$\det\{\tilde\mu_{ij}-\sigma\delta_{ij}\}=0$  (this is equivalent
to the conditions \mbox{$\det {\rm M}\not=0$,}
$\det\{\mu_{ij}-\sigma\delta_{ij}\}=0$; here $\delta_{ij}$ is the
Kronecker symbol).

\vspace{2ex}
\hspace*{1em}
i. ${\rm\tilde M}=\left( \begin{array}{cc}
\sigma&\varepsilon\\0&\sigma
\end{array}   \right),\:$
where $\:\varepsilon\in\{0;1\}$. Then
\begin{equation}\label{e4.32}
\begin{array}{ll}
\psi^{2}=\!\!\!\!&C_{1}\omega+C_{2}\bigl(\omega
\int e^{\frac{1}{2}\sigma\omega^{2}}d\omega-\sigma^{-1}
e^{\frac{1}{2}\sigma\omega^{2}}\bigr)+\sigma^{-1}\tilde\nu_{12}+\\
\\
&+\sigma^{-1}\tilde\nu_{22}\bigl(\sigma\omega
\int e^{\frac{1}{2}\sigma\omega^{2}}\lambda^{1}(\omega)d\omega-
e^{\frac{1}{2}\sigma\omega^{2}}\lambda^{1}(\omega)\bigr),
\end{array}
\end{equation}
\begin{displaymath}
\begin{array}{ll}
\psi^{1}=\!\!\!\!&C_{3}\omega+C_{4}\bigl(\omega
\int e^{\frac{1}{2}\sigma\omega^{2}}d\omega-\sigma^{-1}
e^{\frac{1}{2}\sigma\omega^{2}}\bigr)+\sigma^{-1}\tilde\nu_{11}+\\
\\
&+\sigma\omega\int
e^{\frac{1}{2}\sigma\omega^{2}}\lambda^{2}(\omega)d\omega-
e^{\frac{1}{2}\sigma\omega^{2}}\lambda^{2}(\omega)+
\sigma^{-1}(\tilde\nu_{21}\omega-\varepsilon\psi^{2}),
\end{array}
\end{displaymath}
where $\:\lambda^{1}(\omega)=
\int e^{-\frac{1}{2}\sigma\omega^{2}}d\omega,\:$
$\:\lambda^{2}(\omega)=\sigma^{-1}
\int e^{-\frac{1}{2}\sigma\omega^{2}}(\tilde\nu_{21}-\varepsilon
\psi^{2}_{\omega})d\omega$.

\vspace{2ex}
\hspace*{1em}
ii. ${\rm\tilde M}=\left(
\begin{array}{cc}
\kappa_{1}&0\\0&\sigma
\end{array}  \right),\:$
where $\:\kappa_{1}\in{\R}\backslash\{0;\sigma\}$.
In this case $\psi^{2}$ is determined by (\ref{e4.32}), and the
form of $\psi^{1}$ is given by (\ref{e4.33}) (see below).

\vspace{2ex}
C. $\det{\rm\tilde M}\not=0$,
$\det\{\tilde\mu_{ij}-\sigma\delta_{ij}\}\not=0$ (this is
equivalent to the condition $\det {\rm M}\not=0$,
$\det\{\mu_{ij}-\sigma\delta_{ij}\}\not=0$: here $\delta_{ij}$ is
the Kronecker symbol).

\vspace{2ex}
\hspace*{1em}
i. ${\rm\tilde M}=\left( \begin{array}{cc}
\kappa_{1}&0\\0&\kappa_{2}
\end{array}  \right),\:$
where $\:\kappa_{i}\in{\R}\backslash\{0;\sigma\}$. Then

\begin{equation}\label{e4.33}
\begin{array}{l}
\psi^{1}=\kappa_{1}^{-1}\tilde\nu_{11}+(\kappa_{1}-\sigma)^{-1}
\tilde\nu_{21}\omega+\vert\omega\vert^{-1/2}
e^{\frac{1}{4}\sigma\omega^{2}}\cdot
\\[2ex]
\cdot\Bigl(
C_{3}M\bigl(\frac{1}{2}\kappa_{1}\sigma^{-1}+\frac{1}{4},\frac{1}{4},
\frac{1}{2}\sigma\omega^{2}\bigr)+
C_{4}M\bigl(\frac{1}{2}\kappa_{1}\sigma^{-1}+\frac{1}{4},-\frac{1}{4},
\frac{1}{2}\sigma\omega^{2}\bigr)\!\Bigr),
\end{array}
\end{equation}

\begin{equation}\label{e4.34}
\begin{array}{l}
\psi^{2}=\kappa_{2}^{-1}\tilde\nu_{12}+(\kappa_{2}-\sigma)^{-1}
\tilde\nu_{22}\omega+\vert\omega\vert^{-1/2}
e^{\frac{1}{4}\sigma\omega^{2}}\cdot
\\[2ex]
\cdot\Bigl(
C_{1}M\bigl(\frac{1}{2}\kappa_{2}\sigma^{-1}+\frac{1}{4},\frac{1}{4},
\frac{1}{2}\sigma\omega^{2}\bigr)+
C_{2}M\bigl(\frac{1}{2}\kappa_{2}\sigma^{-1}+\frac{1}{4},-\frac{1}{4},
\frac{1}{2}\sigma\omega^{2}\bigr)\!\Bigr),
\end{array}
\end{equation}
where $M(\kappa,\mu,\tau)$ is the Whittaker function:
\begin{equation}\label{e4.35}
\begin{array}{l}
M(\kappa,\mu,\tau)=\tau^{\frac{1}{2}+\mu}e^{-\frac{1}{2}\tau}
{_{1}F_{1}}\bigl(\frac{1}{2}+\mu-\kappa,2\mu+1,\tau\bigr),
\end{array}
\end{equation}
and $_{1}F_{1}(a,b,\tau)$ is the degenerate hypergeometric
function defined by means of the series:
\begin{displaymath}
_{1}F_{1}(a,b,\tau)=1+\sum_{n=1}^{\infty}
\frac{a(a+1)\ldots(a+n-1)}{b(b+1)\ldots(b+n-1)}
\frac{\tau^{n}}{n!},
\end{displaymath}
$b\not=0,-1,-2,\ldots$.

\vspace{2ex}
\hspace*{1em}
ii. ${\rm\tilde M}=\left(\begin{array}{cc}
\kappa_{1}&-\kappa_{2}\\\kappa_{2}&\kappa_{1}
\end{array}  \right),\:$
where $\:\kappa_{i}\in{\R},\:$ $\:\kappa_{2}\not=0$. Then

\begin{displaymath}
\begin{array}{ll}
\psi^{1}=\!\!\!\!&(\kappa_{j}\kappa_{j})^{-1}
(\kappa_{1}\tilde\nu_{11}+\kappa_{2}\tilde\nu_{12})+\\
\\
&+((\kappa_{1}-\sigma)^{2}+\kappa_{2}^{2})^{-1}
((\kappa_{1}-\sigma)\tilde\nu_{21}+\kappa_{2}\tilde\nu_{22})\omega+\\
\\
&+C_{1}\mbox{Re}\eta^{1}(\omega)-C_{2}\mbox{Im}\eta^{1}(\omega)
 +C_{3}\mbox{Re}\eta^{2}(\omega)-C_{4}\mbox{Im}\eta^{2}(\omega),
\end{array}
\end{displaymath}

\begin{displaymath}
\begin{array}{ll}
\psi^{2}=\!\!\!\!&(\kappa_{j}\kappa_{j})^{-1}
(-\kappa_{2}\tilde\nu_{11}+\kappa_{1}\tilde\nu_{12})+\\
\\
&+((\kappa_{1}-\sigma)^{2}+\kappa_{2}^{2})^{-1}
(-\kappa_{2}\tilde\nu_{21}+(\kappa_{1}-\sigma)\tilde\nu_{22})\omega+\\
\\
&+C_{1}\mbox{Im}\eta^{1}(\omega)+C_{2}\mbox{Re}\eta^{1}(\omega)
 +C_{3}\mbox{Im}\eta^{2}(\omega)+C_{4}\mbox{Re}\eta^{2}(\omega),
\end{array}
\end{displaymath}
where

\begin{displaymath}
\begin{array}{l}
\eta^{1}(\omega)=M\bigl(
\frac{1}{2}(\kappa_{1}+\kappa_{2}{\rm i})\sigma^{-1}+\frac{1}{4},
\frac{1}{4},\frac{1}{2}\sigma\omega^{2}\bigr),
\end{array}
\end{displaymath}

\begin{displaymath}
\begin{array}{l}
\eta^{2}(\omega)=M\bigl(
\frac{1}{2}(\kappa_{1}+\kappa_{2}{\rm i})\sigma^{-1}+\frac{1}{4},
-\frac{1}{4},\frac{1}{2}\sigma\omega^{2}\bigr),
\quad {\rm i}^{2}=-1.
\end{array}
\end{displaymath}

\vspace{2ex}
\hspace*{1em}
iii. ${\rm\tilde M}=\left(\begin{array}{cc}
\kappa_{2}&1\\0&\kappa_{2}
\end{array}   \right),\:$
where $\:\kappa_{2}\in{\R}\backslash\{0;\sigma\}$. Here the
form of $\psi^{2}$ is given by (\ref{e4.34}), and
\begin{displaymath}
\begin{array}{l}
\psi^{1}=(\tilde\nu_{11}-\tilde\nu_{12}\kappa_{2}^{-1})\kappa_{2}^{-1}+
\bigl(\tilde\nu_{21}-\tilde\nu_{22}(\kappa_{2}-\sigma)^{-1}
\bigr) (\kappa_{2}-\sigma)^{-1}\omega+\\
\\
\: +\vert\omega\vert^{-1/2}e^{\frac{1}{4}\sigma\omega^{2}}\Bigl(
C_{3}\theta^{1}(\tau)+C_{4}\theta^{2}(\tau)
-\sigma^{-1}\theta^{1}(\tau)
\int\tau^{-1}\theta^{2}(\tau)C_{i}\theta^{i}(\tau)d\tau+\\
\\
\quad\quad\quad\quad\quad\quad\quad\quad+\sigma^{-1}\theta^{2}(\tau)
\int\tau^{-1}\theta^{1}(\tau)C_{i}\theta^{i}(\tau)d\tau\Bigr),
\end{array}
\end{displaymath}
where $\tau=\frac{1}{2}\sigma\omega^{2}$,
\begin{displaymath}
\begin{array}{l}
\theta^{1}(\tau)=M\bigl(\frac{1}{2}\kappa_{2}\sigma^{-1}+
\frac{1}{4},\frac{1}{4},\tau\bigl),
\quad
\theta^{2}(\tau)=M\bigl(\frac{1}{2}\kappa_{2}\sigma^{-1}+
\frac{1}{4},-\frac{1}{4},\tau\bigl).
\end{array}
\end{displaymath}

\begin{note}\label{n4.3}
The general solution of the equation
\begin{displaymath}
\psi_{\omega\omega}-\sigma\omega\psi_{\omega}-(n+1)\sigma\psi=0,
\end{displaymath}
where $n$ is an integer and $n\geq0$, is determined by the formula
\begin{displaymath}
\psi=\left(\frac{d^{n}}{d\omega^{n}}
e^{\frac{1}{2}\sigma\omega^{2}}\right)
\left(C_{1}+C_{2}\int e^{\frac{1}{2}\sigma\omega^{2}}
\left(\frac{d^{n}}{d\omega^{n}}
e^{\frac{1}{2}\sigma\omega^{2}}\right)^{-2}d\omega\right).
\end{displaymath}
\end{note}

\begin{note}\label{n4.4}
If function $\psi$ satisfies the equation
\begin{displaymath}
\psi_{\omega\omega}-\sigma\omega\psi_{\omega}+\kappa\psi=0 \quad
(\kappa\not=-\sigma),
\end{displaymath}
then
$\:\int\psi(\omega)d\omega=(\kappa+\sigma)^{-1}(\sigma\omega\psi-
\psi_{\omega})+C_{1}$.
\end{note}

7. The last equation of system (\ref{e4.13}) is the
compatibility condition of the NSEs (\ref{e1.1}) and ansatz
(\ref{e4.7}). Integrating this equation, we obtain that
\begin{displaymath}
\eta^{3}=C_{0}(\eta^{i}\eta^{i})^{-1}, \quad C_{0}\not=0.
\end{displaymath}
As $\:\varphi^{3}_{\omega}=-\eta^{3}_{\omega}
(\eta^{3})^{-1}\varphi^{3}=2\theta^{1}\varphi^{3},\:$
$\:\varphi^{3}=C_{3}\eta^{i}\eta^{i}$.
Then system (\ref{e4.13}) is reduced to the equations
\begin{equation}\label{e4.36}
\begin{array}{l}
\varphi^{1}_{\omega}=\chi^{1}(\omega)\varphi^{1}-
\chi^{2}(\omega)\varphi^{2},\\
\\
\varphi^{2}_{\omega}=\chi^{2}(\omega)\varphi^{1}+
\chi^{1}(\omega)\varphi^{2},
\end{array}
\end{equation}
where $\:\chi^{1}=-C_{0}^{-2}(\eta^{i}\eta^{i})^{2}-\theta^{1}\:$ and
$\:\chi^{2}=\theta^{2}-C_{3}C_{0}^{-1}(\eta^{i}\eta^{i})^{2}.\:$
System (\ref{e4.36}) implies that
\begin{displaymath}
\begin{array}{l}
\varphi^{1}=\exp\bigl(\int\chi^{1}(\omega)d\omega\bigr)
\Bigl(C_{1}\cos\bigl(\int\chi^{2}(\omega)d\omega\bigr)-
C_{2}\sin\bigl(\int\chi^{2}(\omega)d\omega\bigr)\Bigr),
\end{array}
\end{displaymath}
\begin{displaymath}
\begin{array}{l}
\varphi^{2}=\exp\bigl(\int\chi^{1}(\omega)d\omega\bigr)
\Bigl(C_{1}\sin\bigl(\int\chi^{2}(\omega)d\omega\bigr)+
C_{2}\cos\bigl(\int\chi^{2}(\omega)d\omega\bigr)\Bigr).
\end{array}
\end{displaymath}

8. Let us apply the trasformation generated by the operator
$R(\vec k(t))$, where
\begin{displaymath}
\vec k_{t}=\lambda^{-1}(\vec n^{b}\cdot\vec k)\vec m^{b}_{t}-
\vec\varphi,
\end{displaymath}
to ansatz (\ref{e4.8}). As a result we obtain an ansatz of
the same form, where the functions $\vec\varphi$ and $h$ are
replaced by the new functions $\vec{\tilde\varphi}$ and $\tilde h$:
\begin{displaymath}
\begin{array}{l}
\vec{\tilde\varphi}=\vec\varphi-\lambda^{-1}(\vec n^{a}\cdot\vec
k)\vec m^{a}_{t}+\vec k_{t}=0,\\
\\
\tilde h=h-\lambda^{-1}(\vec m^{a}_{tt}\cdot\vec k)
(\vec n^{a}\cdot\vec k)+
\frac{1}{2}\lambda^{-2}(\vec m^{b}_{tt}\cdot\vec m^{a})
(\vec n^{a}\cdot\vec k)(\vec n^{b}\cdot\vec k).
\end{array}
\end{displaymath}
Let us make $\tilde h$ vanish by means of the transformation
generated by the operator $Z(-\tilde h(t))$. Therefore, the
functions $\varphi^{a}$ and $h$ can be considered to vanish. The
equation $\:(\vec n^{a}\cdot\vec m^{a}_{t})=0\:$ is the compatibility
condition of ansatz (\ref{e4.8}) and the NSEs (\ref{e1.1}).

\begin{note}\label{n4.5}
The solutions of the NSEs obtained by means of ansatzes 5--8 are
equivalent to either solutions (\ref{e5.1}) or solutions
(\ref{e5.5}).
\end{note}

\setcounter{equation}{0}

\section
{Reduction of the Navier-Stokes equations to linear systems of
PDEs}\label{sec5}

Let us show that non-linear systems 8 and 9, from Subsec.
\ref{subsec3.2}, are reduced to linear systems of PDEs.

\protect\subsection
{Investigation of system (\protect\ref{e3.17})--(\protect\ref{e3.18})
}\label{subsec5.1}

Consider system 9 from Subsec. \ref{subsec3.2}, i.e., equations
(\ref{e3.17}) and (\ref{e3.18}). Equation (\ref{e3.18})
integrates with respect to $z_{2}$ to the following expression:
\begin{displaymath}
\vec k\cdot\vec w=\psi(t).
\end{displaymath}
Here $\psi=\psi(t)$ is an arbitrary smooth function of $z_{1}=t$.
Let us make the transformation from the symmetry group of the
NSEs:
\begin{displaymath}
\vec{\tilde u}(t,\vec x)=\vec u(t,\vec x-\vec l) +\vec l_{t}(t),
\end{displaymath}
\begin{displaymath}
\tilde p(t,\vec x)=p(t,\vec x-\vec l) -
\vec l_{tt}(t)\cdot\vec x,
\end{displaymath}
where $\:\vec l_{tt}\cdot\vec m^{i}-\vec l\cdot\vec m^{i}_{tt}=0\:$
and
\begin{displaymath}
\vec k\cdot\bigl(\vec l_{t}-\lambda^{-1}(\vec n^{i}\cdot\vec l)
m^{i}_{t}+\lambda^{-1}(\vec k\cdot\vec l)\vec k_{t}\bigr)+\psi=0.
\end{displaymath}
This transformation does not modify ansatz (\ref{e3.9}), but it
makes the function $\psi(t)$ vanish, i.e.,
$\vec k\cdot\vec{\tilde w}=0$. Therefore, without loss of
generality we may assume, at once, that $\:\vec k\cdot\vec w=0$.

Let $\:f^{i}=f^{i}(z_{1},z_{2})=\vec m^{i}\cdot\vec w$. Since
$\:\vec m^{1}_{tt}\cdot\vec m^{2}-\vec m^{1}\cdot\vec m^{2}_{tt}=0,\:$
 it follows that \quad
$\vec m^{1}_{t}\cdot\vec m^{2}-\vec m^{1}\cdot\vec m^{2}_{t}=C=
\mbox{\rm const}$. Let us multiply the scalar equation (\ref{e3.17}) by
$\vec m^{i}$ and $\vec k$. As a result we obtain the linear system
of PDEs with variable coefficients in the functions $f^{i}$ and $s$:
\begin{displaymath}
\begin{array}{l}
f^{i}_{1}-\lambda f^{i}_{22}+C\lambda^{-1}\bigl(
(\vec m^{i}\cdot\vec m^{2})f^{1}-(\vec m^{i}\cdot\vec m^{1})f^{2}
\bigr)-
\\[1ex]
-2C\lambda^{-2}\bigl(
(\vec k\times\vec k_{t})\cdot\vec m^{i}\bigr)z_{2}=0,
\end{array}
\end{displaymath}
\begin{displaymath}
s_{2}=2\lambda^{-2}(\vec n^{i}\cdot\vec k_{t})f^{i}+\lambda^{-2}(
\vec k_{tt}\cdot\vec k-2\vec k_{t}\cdot\vec k_{t})z_{2}.
\end{displaymath}
Consider two possible cases.

A. Let $C=0$. Then there exist functions
$g^{i}=g^{i}(\tau,\omega)$, where $\tau=\int\lambda(t)dt$ and
$\:\omega=z_{2},\:$ such that $\:f^{i}=g^{i}_{\omega}\:$ and
$\:g^{i}_{\tau}-g^{i}_{\omega\omega}=0$. Therefore,
\begin{equation}\label{e5.1}
\begin{array}{ll}
\vec u=\!\!\!\!&\lambda^{-1}(g^{i}_{\omega}(\tau,\omega)+
\vec m^{i}_{t}\cdot\vec x)\vec n^{i}-\lambda^{-1}
(\vec k_{t}\cdot\vec x)k,\\
\\
p=\!\!\!\!&2\lambda^{-2}(\vec n^{i}\cdot\vec k_{t})g^{i}(\tau,\omega)+
\frac{1}{2}\lambda^{-2}(\vec k_{tt}\cdot\vec k-
2\vec k_{t}\cdot\vec k_{t})\omega^{2}-
\\ \\
&-\frac{1}{2}\lambda^{-1}(\vec n^{i}\cdot\vec x)
(\vec m^{i}_{tt}\cdot\vec x)-
\frac{1}{2}\lambda^{-2}(\vec k\cdot\vec m^{i}_{tt})
(\vec n^{i}\cdot\vec x)(\vec k\cdot\vec x),
\end{array}
\end{equation}
where
$\:\vec m^{1}_{t}\cdot\vec m^{2}-\vec m^{1}\cdot\vec m^{2}_{t}=0,\:$
$\:\vec k=\vec m^{1}\times\vec m^{2},\:$
$\:\vec n^{1}=\vec m^{2}\times\vec k,\:$
$\:\vec n^{2}=\vec k\times\vec m^{1},\:$
\mbox{$\:\lambda=\vert\vec k\vert^{2},\:$}
$\:\omega=\vec k\cdot\vec x,\:$
$\:\tau=\int\lambda(t)dt,\:$ and
$\:g^{i}_{\tau}-g^{i}_{\omega\omega}=0$.

\vspace{1ex}

For example, if $\vec m=(\eta^{1}(t),0,0)$ and
$\vec n=(0,\eta^{2}(t),0)$ with
$\eta^{i}(t)\not=0$, it follows that
\begin{displaymath}
u^{1}=(\eta^{1})^{-1}(f^{1}+\eta^{1}_{t}x_{1}),\quad
u^{2}=(\eta^{2})^{-1}(f^{2}+\eta^{2}_{t}x_{2}),
\end{displaymath}
\begin{displaymath}
u^{3}=-(\eta^{1}\eta^{2})_{t}(\eta^{1}\eta^{2})^{-1}x_{3},
\end{displaymath}
\begin{displaymath}
\begin{array}{ll}
p=\!\!\!\!&-\frac{1}{2}\eta^{1}_{tt}(\eta^{1})^{-1}x_{1}^{2}
-\frac{1}{2}\eta^{2}_{tt}(\eta^{2})^{-1}x_{2}^{2}+
\\ \\
&+\Bigl(\frac{1}{2}
(\eta^{1}\eta^{2})_{tt}(\eta^{1}\eta^{2})^{-1}-\bigl(
(\eta^{1}\eta^{2})_{t}(\eta^{1}\eta^{2})^{-1}\bigr)^{2}\Bigr)
x_{3}^{2},
\end{array}
\end{displaymath}
where $f^{i}=f^{i}(\tau,\omega)$,
$f^{i}_{\tau}-f^{i}_{\omega\omega}=0$,
$\tau=\int(\eta^{1}\eta^{2})^{2}dt$, and
$\omega=\eta^{1}\eta^{2}x_{3}$.
If $\vec m^{1}=(\eta^{1}(t),\eta^{2}(t),0)$ and
$\vec m^{2}=(0,0,\eta^{3}(t))$ with $\eta^{3}(t)\not=0$ and
$\eta^{i}(t)\eta^{i}(t)\not=0$, we obtain that
\begin{displaymath}
\begin{array}{lll}
u^{1}&=&(\eta^{i}\eta^{i})^{-1}\Bigl\{
\eta^{1}(g_{\omega}+\eta^{i}_{t}x_{i})-\eta^{2}\bigl(
\eta^{3}_{t}(\eta^{3})^{-2}\omega+\eta^{2}_{t}x_{1}-
\eta^{1}_{t}x_{2}\bigr)\Bigr\},\\
\\
u^{2}&=&(\eta^{i}\eta^{i})^{-1}\Bigl\{
\eta^{2}(g_{\omega}+\eta^{i}_{t}x_{i})+\eta^{1}\bigl(
\eta^{3}_{t}(\eta^{3})^{-2}\omega+\eta^{2}_{t}x_{1}-
\eta^{1}_{t}x_{2}\bigr)\Bigr\},\\
\\
u^{3}&=&(\eta^{3})^{-1}(f+\eta^{3}_{t}x_{3}),\\
\\
p&=&2(\eta^{3})^{-1}(\eta^{1}\eta^{2}_{t}-\eta^{1}_{t}\eta^{2})
(\eta^{i}\eta^{i})^{-2}g+\frac{1}{2}\lambda^{-1}\cdot\\
\\
&&\cdot\Bigl\{\lambda^{-1}\bigl(
(\eta^{3}_{tt}\eta^{3}-
2\eta^{3}_{t}\eta^{3}_{t})\eta^{i}\eta^{i}-
2\eta^{3}\eta^{3}_{t}\eta^{i}\eta^{i}_{t}-
2(\eta^{3})^{2}\eta^{i}_{t}\eta^{i}_{t}\bigr)\omega^{2}+\\
\\
&&+(\eta^{3})^{2}\bigl(
(\eta^{2}\eta^{2}_{tt}-\eta^{1}\eta^{1}_{tt})
(x_{1}^{2}-x_{2}^{2})-
2(\eta^{1}_{tt}\eta^{2}+\eta^{1}\eta^{2}_{tt})x_{1}x_{2}\bigr)-\\
\\
&&-\eta^{i}\eta^{i}\eta^{3}\eta^{3}_{tt}x_{3}^{2}\Bigr\}.
\end{array}
\end{displaymath}
Here $f=f(\tau,\omega)$, $f_{\tau}-f_{\omega\omega}=0$,
$g=g(\tau,\omega)$, $g_{\tau}-g_{\omega\omega}=0$,
$\tau=\int(\eta^{3})^{2}\eta^{i}\eta^{i}dt$,
$\omega=\eta^{3}(\eta^{2}x_{1}-\eta^{1}x_{2}),\:$ and
$\:\lambda=(\eta^{3})^{2}\eta^{i}\eta^{i}$.

\begin{note}\label{n5.1}
The equation
\begin{equation}\label{e5.2}
\vec m^{1}_{t}\cdot\vec m^{2}-\vec m^{1}\cdot\vec m^{2}_{t}=0
\end{equation}
can easily be solved in the following way: Let us fix arbitrary smooth
vector-functions \quad
\mbox{$\vec m^{1},\vec l\in C^{\infty}((t_{0},t_{1}),{\R}^{3})$}
\quad such that $\vec m^{1}(t)\not=\vec 0$,
$\vec l(t)\not=\vec 0$, and \mbox{$\vec m^{1}(t)\cdot\vec l(t)=0$} for
all $t\in(t_{0},t_{1})$. Then the vector-function
$\vec m^{2}=\vec m^{2}(t)$ is taken in the form
\begin{equation}\label{e5.3}
\vec m^{2}(t)=\rho(t)\vec m^{1}+\vec l(t).
\end{equation}
Equation (\ref{e5.2}) implies
\begin{equation}\label{e5.4}
\begin{array}{l}
\rho(t)=\int(\vec m^{1}\cdot\vec m^{1})^{-1}
(\vec m^{1}_{t}\cdot\vec l-\vec m^{1}\cdot\vec l_{t})dt.
\end{array}
\end{equation}
\end{note}

B. Let $C\not=0$. By means of the transformation
$\vec m^{i}\rightarrow a_{ij}\vec m^{j}$, where
\mbox{$\:a_{ij}=\mbox{\rm const}\:$}
and $\:\det\{a_{ij}\}=C,\:$ we make $\:C=1$.
Then we obtain the following solution of the NSEs (\ref{e1.1})
\begin{equation}\label{e5.5}
\begin{array}{ll}
\vec u=&\!\!\!\lambda^{-1}\Bigl(
\theta^{ij}(t)g^{j}_{\omega}(\tau,\omega)+\theta^{i0}(t)\omega+
\vec m^{i}_{t}\!\cdot\vec x-
\lambda^{-1}((\vec k\!\times\!\vec m^{i})\!\cdot\!\vec x)
\Bigr)\vec n^{i}-\\
\\
&\!\!\!-\lambda^{-1}(\vec k_{t}\cdot\vec x)\vec k,\\
\\
p=&\!\!\!2\lambda^{-2}(\vec n^{i}\cdot\vec k_{t})
(\theta^{ij}(t)g^{i}(\tau,\omega)+\frac{1}{2}\theta^{i0}(t)
\omega^{2})+\\
\\
&\!\!\!+\frac{1}{2}\lambda^{-2}(\vec k_{tt}\cdot\vec k-
2\vec k_{t}\cdot\vec k_{t})\omega^{2}-
\frac{1}{2}\lambda^{-1}(\vec n^{i}\cdot\vec x)
(\vec m^{i}_{tt}\cdot\vec x)-\\
\\
&\!\!\!-\frac{1}{2}\lambda^{-2}(\vec k\cdot\vec m^{i}_{tt})
(\vec n^{i}\cdot\vec x)(\vec k\cdot\vec x).
\end{array}
\end{equation}
Here
$\:\vec m^{1}_{t}\cdot\vec m^{2}-\vec m^{1}\cdot\vec m^{2}_{t}=1,\:$
$\:\vec k=\vec m^{1}\times\vec m^{2},\:$
$\:\vec n^{1}=\vec m^{2}\times\vec k,\:$
$\:\vec n^{2}=\vec k\times\vec m^{1},\:$
\mbox{$\:\lambda=\vert\vec k\vert^{2},\:$}
$\:\omega=\vec k\cdot\vec x,\:$
$\:\tau=\int\lambda(t)dt,\:$ and
$\:g^{i}_{\tau}-g^{i}_{\omega\omega}=0$.
$\:(\theta^{1i}(t),\theta^{2i}(t)) \: (i=1,2)\:$ are linearly
independent solutions of the system
\begin{equation}\label{e5.6}
\theta^{i}_{t}+\lambda^{-1}(\vec m^{i}\cdot\vec m^{2})\theta^{1}-
\lambda^{-1}(\vec m^{i}\cdot\vec m^{1})\theta^{2}=0,
\end{equation}
and $(\theta^{10}(t),\theta^{20}(t))$ is a particular solution of
the nonhomogeneous system
\begin{equation}\label{e5.7}
\theta^{i}_{t}+\lambda^{-1}(\vec m^{i}\cdot\vec m^{2})\theta^{1}-
\lambda^{-1}(\vec m^{i}\cdot\vec m^{1})\theta^{2}=
2\lambda^{-2}\bigl((\vec k\times\vec k_{t})\cdot\vec m^{i}\bigr).
\end{equation}

For example, if $\:\vec m^{1}=(\eta\cos\psi,\eta\sin\psi,0)\:$
and $\:\vec m^{2}=(-\eta\sin\psi,\eta\cos\psi,0),\:$ where
$\eta=\eta(t)\not=0$ and $\psi=-\frac{1}{2}\int(\eta)^{-2}dt$
(therefore,
$\vec m^{1}_{t}\cdot\vec m^{2}-\vec m^{1}\cdot\vec m^{2}_{t}=1$),
we obtain
\begin{displaymath}
\begin{array}{lll}
u^{1}&=&\eta^{-1}\bigl(f^{1}\cos\psi-f^{2}\sin\psi+\eta_{t}x_{1}-
\frac{1}{2}\eta^{-1}x_{2}\bigr),\\
\\
u^{2}&=&\eta^{-1}\bigl(f^{1}\sin\psi+f^{2}\cos\psi+\eta_{t}x_{2}+
\frac{1}{2}\eta^{-1}x_{1}\bigr),\\
\\
u^{3}&=&-2\eta_{t}\eta^{-1}x_{3},\\
\\
p&=&(\eta_{tt}\eta-3\eta_{t}\eta_{t})\eta^{-2}x_{3}^{2}-
\frac{1}{2}(\eta_{tt}\eta^{-1}-\frac{1}{4}\eta^{-4})x_{i}x_{i}.
\end{array}
\end{displaymath}
Here $\:f^{i}=f^{i}(\tau,\omega),\:$
$\:f^{i}_{\tau}-f^{i}_{\omega\omega}=0,\:$ $\:\tau=\int(\eta)^{4}dt,\:$
and $\:\omega=(\eta)^{2}x_{3}$.

\begin{note}\label{n5.2}
As in the case $C=0$, the solutions of the equation
\begin{equation}\label{e5.8}
\vec m^{1}_{t}\cdot\vec m^{2}-\vec m^{1}\cdot\vec m^{2}_{t}=1
\end{equation}
can be sought in form (\ref{e5.3}). As a result we obtain
that
\begin{equation}\label{e5.9}
\begin{array}{l}
\rho(t)=\int\vert\vec m^{1}\vert^{-2}
(\vec m^{1}_{t}\cdot\vec l-\vec m^{1}\cdot\vec l_{t}-1)dt.
\end{array}
\end{equation}
\end{note}

\begin{note}\label{n5.3}
System (\ref{e5.6}) can be reduced to a second-order
homogeneous differential equation either in $\theta^{1}$, i.e.,
\begin{equation}\label{e5.10}
\Bigl(\lambda\vert\vec m^{1}\vert^{-2}\theta^{1}_{t}
\Bigr)_{t}+
\Bigl(\bigl((\vec m^{1}\cdot\vec m^{2})
\vert\vec m^{1}\vert^{-2}\bigr)_{t}+
\vert\vec m^{1}\vert^{-2}\Bigr)\theta^{1}=0
\end{equation}
(then $\theta^{2}=\vert\vec m^{1}\vert^{-2}
(\lambda\theta^{1}_{t}+(\vec m^{1}\cdot\vec m^{2})\theta^{1})$),
or in $\theta^{2}$, i.e.,
\begin{equation}\label{e5.11}
\Bigl(\lambda\vert\vec m^{2}\vert^{-2}\theta^{2}_{t}
\Bigr)_{t}+
\Bigl(-\bigl((\vec m^{1}\cdot\vec m^{2})
\vert\vec m^{2}\vert^{-2}\bigr)_{t}+
\vert\vec m^{2}\vert^{-2}\Bigr)\theta^{2}=0
\end{equation}
(then $\theta^{1}=\vert\vec m^{2}\vert^{-2}
(-\lambda\theta^{2}_{t}+(\vec m^{1}\cdot\vec m^{2})\theta^{2})$).
Under the notation of Note \ref{n5.1} equation (\ref{e5.10})
has the form:
\begin{equation}\label{e5.12}
\bigl((\vec l\cdot\vec l)\theta^{1}_{t}\bigr)_{t}+
\vert\vec m^{1}\vert^{-2}
(\vec m^{1}_{t}\cdot\vec l-\vec m^{1}\cdot\vec l_{t})
\theta^{1}=0.
\end{equation}
The vector-functions $\vec m^{1}$ and $\vec l$ are chosen in such
a way that one can find a fundamental set of solutions for
equation (\ref{e5.12}). For example, let
\mbox{$\vec m\times\vec m_{t}\not=0\: \forall t\in(t_{0},t_{1})$}. Let
us introduce the notation $\vec m:=\vec m^{1}$ and put \quad
$\vec l=\eta(t)\vec m\times\vec m_{t}$, where
$\:\eta\in C^{\infty}((t_{0},t_{1}),{\R})$,
$\:\eta(t)\not=0$ $\forall t\in(t_{0},t_{1})$. Then
\begin{displaymath}
\begin{array}{l}
\vec m\cdot\vec l=0,\quad
\vec m_{t}\cdot\vec l-\vec m\cdot\vec l_{t}=0,\quad
\vec m^{2}=-\bigl(\int\vert\vec m\vert^{-2}dt\bigr)\vec m+\eta
\vec m\times\vec m_{t},
\end{array}
\end{displaymath}
\begin{displaymath}
\begin{array}{l}
\vec k=\eta\vec m\times(\vec m\times\vec m_{t}),\quad
\lambda=(\eta)^{2}\vert\vec m\vert^{2}
\vert\vec m\times\vec m_{t}\vert^{-2},
\end{array}
\end{displaymath}
\begin{displaymath}
\begin{array}{l}
\vec n^{2}=\eta\vert\vec m\vert^{2}\vec m\times\vec m_{t}, \quad
\vec n^{1}=\bigl(\int\vert\vec m\vert^{-2}dt\bigr)\vec n^{2}+
(\eta)^{2}\vert\vec m\times\vec m_{t}\vert^{-2}\vec m,
\end{array}
\end{displaymath}
\begin{displaymath}
\begin{array}{l}
\theta^{11}(t)=\int(\eta)^{-2}
\vert\vec m\times\vec m_{t}\vert^{-2}dt, \quad
\theta^{21}(t)=1-\theta^{11}\int\vert\vec m\vert^{-2}dt,
\end{array}
\end{displaymath}
\begin{displaymath}
\begin{array}{l}
\theta^{12}(t)=1,
\quad
\theta^{22}(t)=-\int\vert\vec m\vert^{-2}dt,
\end{array}
\end{displaymath}
\begin{displaymath}
\begin{array}{ll}
\theta^{10}(t)=&2\int\bigl(
((\vec m\times\vec m_{t})\cdot\vec m_{tt})
\vert\vec m\times\vec m_{t}\vert^{-2}+
\int\eta^{-1}\vert\vec m\vert^{-4}dt\bigr)\cdot\\
\\
&\cdot\,\eta^{-2}\vert\vec m\times\vec m_{t}\vert^{-2}dt,
\end{array}
\end{displaymath}
\begin{displaymath}
\begin{array}{l}
\theta^{20}(t)=-\theta^{10}(t)\int\vert\vec m\vert^{-2}dt+
2\int\eta^{-1}\vert\vec m\vert^{-4}dt.
\end{array}
\end{displaymath}

Consider the following cases:
 $\vec m\times\vec m_{t}\equiv\vec 0$, i.e.,
$\vec m=\chi(t)\vec a$,
where \mbox{$\:\chi(t)\in C^{\infty}((t_{0},t_{1}),{\R})$},
$\:\chi(t)\not=0$ $\forall t\in(t_{0},t_{1})$,
$\:\vec a=\mbox{\rm const},\:$ and $\:\vert\vec a\vert=1$.
Let us put
\begin{displaymath}
\vec l(t)=\eta^{1}(t)\vec b +\eta^{2}(t)\vec c,
\end{displaymath}
where $\eta^{1},\eta^{2}\in C^{\infty}((t_{0},t_{1}),{\R})$,
$(\eta^{1}(t),\eta^{2}(t))\not=(0,0)$
$\forall t\in(t_{0},t_{1})$, $\vec b=\mbox{\rm const}$,
$\vert\vec b\vert=1$, $\:\vec a\cdot\vec b=0,\:$ and
$\:\vec c=\vec a\times\vec b$. Then

\begin{displaymath}
\begin{array}{l}
\vec m^{2}=-\bigl(\chi\int\chi^{-2}dt\bigr)\vec a+
\eta^{1}\vec b+\eta^{2}\vec c,
\quad
\vec k=\chi\eta^{1}\vec c-\chi\eta^{2}\vec b,
\end{array}
\end{displaymath}

\begin{displaymath}
\begin{array}{l}
\lambda=(\chi)^{2}\eta^{i}\eta^{i},
\quad
\vec n^{2}=(\chi)^{2}(\eta^{1}\vec b+\eta^{2}\vec c),
\quad
\vec n^{1}=\bigl(\int\chi^{-2}dt\bigr)\vec n^{2}
+\chi\eta^{i}\eta^{i}\vec a,
\end{array}
\end{displaymath}

\begin{displaymath}
\begin{array}{l}
\theta^{11}=\int(\eta^{i}\eta^{i})^{-1}dt,
\quad
\theta^{21}=1-\theta^{11}\int\chi^{-2}dt,
\end{array}
\end{displaymath}

\begin{displaymath}
\begin{array}{l}
\theta^{12}=1,
\quad
\theta^{22}=-\int\chi^{-2}dt,
\end{array}
\end{displaymath}

\begin{displaymath}
\begin{array}{l}
\theta^{10}=2\int(\eta^{2}_{t}\eta^{1}-\eta^{2}\eta^{1}_{t})
\chi^{-1}(\eta^{i}\eta^{i})^{-1}dt,
\quad
\theta^{20}=-\theta^{10}\int\chi^{-2}dt.
\end{array}
\end{displaymath}
\end{note}

\begin{note}\label{n5.4}
In formulas (\ref{e5.1}) and (\ref{e5.5}) solutions of the NSEs
(\ref{e1.1}) are expressed in terms of solutions of the
decomposed system of two linear one-dimensional heat equations
(LOHEs) that have the form:
\begin{equation}\label{e5.13}
g^{i}_{\tau}=g^{i}_{\omega\omega}.
\end{equation}
The Lie symmetry of the LOHE are known. Large sets of its exact
solutions were
constructed \cite{olver,bluman&cole}. The Q-conditional
symmetries of LOHE were investigated in \cite{fshsp}. Moreover,
being decomposed system (\ref{e5.13}) admits transformations
of the form
\begin{displaymath}
\begin{array}{llll}
\tilde
g^{1}(\tau',\omega')\!\!&=F^{1}(\tau,\omega,g^{1}(\tau,\omega)),
\quad \tau'\!\!&=G^{1}(\tau,\omega),
\quad \omega'\!\!&=H^{1}(\tau,\omega),\\
\\
\tilde
g^{2}(\tau'',\omega'')\!\!\!\!&=F^{2}(\tau,\omega,g^{2}(\tau,\omega)),
\quad \tau''\!\!\!\!&=G^{2}(\tau,\omega),
\quad \omega''\!\!\!\!&=H^{2}(\tau,\omega),
\end{array}
\end{displaymath}
where $(G^{1},H^{1})\not=(G^{2},H^{2})$, i.e. the independent
variables can be transformed in the functions $g^{1}$ and $g^{2}$
in different ways. A similar statement is true for system
(\ref{e5.19})--(\ref{e5.20}) (see below) if $\:\varepsilon=0$.
\end{note}

\begin{note}\label{n5.5}
It can be proved that an arbitrary Navier-Stokes field
$(\vec u,p)$, where
\begin{displaymath}
\vec u=\vec w(t,\omega)+(\vec k^{i}(t)\cdot\vec x)\vec l^{i}(t)
\end{displaymath}
with $\:\vec k^{i},\vec l^{i}\in
C^{\infty}((t_{0},t_{1}),{\R}^{3}),\:$
$\:\vec k^{1}\times\vec k^{2}\not=0,\:$ and
$\:\omega=(\vec k^{1}\times\vec k^{2})\cdot\vec x,\:$
is equivalent to either a solution from family (\ref{e5.1})
or a solution from family (\ref{e5.5}). The equivalence
transformation is generated by $R(\vec m)$ and $Z(\chi)$.
\end{note}

\protect\subsection
{Investigation of system
(\protect\ref{e3.13})--(\protect\ref{e3.16})}
\label{subsec5.2}

Consider system 8 from Subsec. \ref{subsec3.2}, i.e., equations
(\ref{e3.13})--(\ref{e3.16}). Equation (\ref{e3.16})
immediately gives
\begin{equation}\label{e5.14}
\begin{array}{l}
w^{1}=-\frac{1}{2}\rho_{t}\rho^{-1}+(\eta-1)z_{2}^{-2},
\end{array}
\end{equation}
where $\eta=\eta(t)$ is an arbitrary smooth function of
$z_{1}=t$. Substituting (\ref{e5.14}) into remaining
equations (\ref{e5.13})--(\ref{e5.15}), we get

\begin{equation}\label{e5.15}
\begin{array}{ll}
q_{2}=\!\!\!\!&\frac{1}{2}\bigl((\rho_{t}\rho^{-1})_{t}-
\frac{1}{2}(\rho_{t}\rho^{-1})^{2}\bigr)z_{2}-\eta_{t}z_{2}^{-1}-
(\eta-1)^{2}z_{2}^{-3}+
\\[1ex]
&+(w^{2}-\chi)^{2}z_{2}^{-3},
\end{array}
\end{equation}

\begin{equation}\label{e5.16}
\begin{array}{l}
w^{2}_{1}-w^{2}_{22}+\bigl(\eta z_{2}^{-1}-
\frac{1}{2}\rho_{t}\rho^{-1}z_{2}\bigr)w^{2}_{2}=0,
\end{array}
\end{equation}

\begin{equation}\label{e5.17}
\begin{array}{l}
w^{3}_{1}-w^{3}_{22}+\bigl(\eta z_{2}^{-1}-
\frac{1}{2}\rho_{t}\rho^{-1}z_{2}\bigr)w^{3}_{2}
+\varepsilon(w^{2}-\chi)z_{2}^{-2}=0.
\end{array}
\end{equation}

Recall that $\rho=\rho(t)$ and $\chi=\chi(t)$ are arbitrary smooth
functions of $t$; \mbox{$\varepsilon\in\{0;1\}$}.
After the change of the independent variables

\begin{equation}\label{e5.18}
\begin{array}{l}
\tau=\int\vert\rho(t)\vert dt,
\quad
z=\vert\rho(t)\vert^{1/2}z_{2}
\end{array}
\end{equation}
in equations (\ref{e5.16}) and (\ref{e5.17}), we obtain a linear
system of a simpler form:

\begin{equation}\label{e5.19}
w^{2}_{\tau}-w^{2}_{zz}+
\hat\eta(\tau)z^{-1}w^{2}_{z}=0,
\end{equation}

\begin{equation}\label{e5.20}
w^{3}_{\tau}-w^{3}_{zz}+
(\hat\eta(\tau)-2)z^{-1}w^{3}_{z}
+\varepsilon(w^{2}-\hat\chi(\tau))z^{-2}=0,
\end{equation}
where $\hat\eta(\tau)=\eta(t)$ and $\hat\chi(\tau)=\chi(t)$.
Equation (\ref{e5.15}) implies

\begin{equation}\label{e5.21}
\begin{array}{ll}
q=\!\!\!\!&\frac{1}{4}
\bigl((\rho_{t}\rho^{-1})_{t}-
\frac{1}{2}(\rho_{t}\rho^{-1})^{2}\bigr)z_{2}^{2}-
\eta_{t}\ln\vert z_{2}\vert-
\\[1ex]
&-\frac{1}{2}(\eta-1)^{2}z_{2}^{-2}+
\int(w^{2}(\tau,z)-\hat\chi(\tau))^{2}z_{2}^{-3}dz_{2}.
\end{array}
\end{equation}
Formulas (\ref{e5.14}), (\ref{e5.18})--(\ref{e5.21}), and
ansatz (\ref{e3.8}) determine a solution of the NSEs
(\ref{e1.1}).

If $\:\varepsilon=0\:$ system (\ref{e5.19})--(\ref{e5.20}) is
decomposed and consists of two translational linear equations of
the general form
\begin{equation}\label{e5.22}
f_{\tau}+\tilde\eta(\tau)z^{-1}f_{z}-f_{zz}=0,
\end{equation}
where $\tilde\eta=\hat\eta$ ($\tilde\eta=\hat\eta-2$) for
equation (\ref{e5.19}) ((\ref{e5.20})). Tilde over $\eta$ is
omitted below. Let us investigate symmetry properties of equation
(\ref{e5.22}) and construct some of its exact solutions.

\begin{theorem}\label{t5.1}
The MIA of (\ref{e5.22}) is given by the following algebras

\vspace{1ex}

a) $L_{1}=<f\partial_{f},\: g(\tau,z)\partial_{f}>$\quad if \quad
$\eta(\tau)\not=\mbox{\rm const}$;

\vspace{1ex}

b) $L_{2}=<\partial_{\tau},\: \hat D,\: \Pi,
f\partial_{f}, \: g(\tau,z)\partial_{f}>$\quad if \quad
$\eta(\tau)=\mbox{\rm const}$, $\eta\not\in\{0;-2\}$;

\vspace{1ex}

c) $L_{3}=<\partial_{\tau},\: \hat D,\: \Pi, \:
\partial_{z}+\frac{1}{2}\eta z^{-1}f\partial_{f}, \:
G=2\tau\partial_{\tau}-(z-\eta z^{-1}\tau)f\partial_{f}, \:
f\partial_{f},$
\\[1ex]
$g(\tau,z)\partial_{f}>$ \quad if \quad
$\eta\in\{0;-2\}$.
\\[1ex]
Here $\hat D=2\tau\partial_{\tau}+z\partial_{z}$,
$\Pi=4\tau^{2}\partial_{\tau}+4\tau z\partial_{z}-
(z^{2}+2(1-\eta)\tau)f\partial_{f}$;
$g=g(\tau,z)$ is an arbitrary solution of (\ref{e5.22}).
\end{theorem}

When $\eta=0$, equation (\ref{e5.22}) is the heat equation, and,
when $\eta=-2$, it is reduced to the heat equation by means of the
change $\tilde f=zf$.

For the case $\:\eta=\mbox{\rm const}\:$ equation (\ref{e5.22}) can be
reduced by inequivalent one-dimensional subalgebras of $L_{2}$.
We construct the following solutions:

\vspace{2ex}

For the subalgebra $<\partial_{\tau}+af\partial_{f}>$, where
$a\in\{-1;0;1\}$, it follows that

\begin{displaymath}
f=e^{-\tau}z^{\nu}(C_{1}J_{\nu}(z)+C_{2}Y_{\nu}(z)) \quad \mbox{if}
\quad a=-1,
\end{displaymath}

\begin{displaymath}
f=e^{\tau}z^{\nu}(C_{1}I_{\nu}(z)+C_{2}K_{\nu}(z)) \quad \mbox{if}
\quad a=1,
\end{displaymath}

\begin{displaymath}
f=C_{1}z^{\eta+1}+C_{2} \quad \mbox{if}
\quad a=0 \quad \mbox{and} \quad \eta\not=-1,
\end{displaymath}

\begin{displaymath}
f=C_{1}\ln z+C_{2} \quad \mbox{if}
\quad a=0 \quad \mbox{and} \quad \eta=-1.
\end{displaymath}
Here $J_{\nu}$ and $Y_{\nu}$ are the Bessel functions of a real
variable, whereas $I_{\nu}$ and $K_{\nu}$ are the Bessel
functions of an imaginary variable, and $\:\nu=\frac{1}{2}(\eta+1)$.

\vspace{2ex}

For the subalgebra $<\hat D+2af\partial_{f}>$, where
$a\in{\R}$, it follows that
\begin{displaymath}
\begin{array}{l}
f=\vert \tau \vert^{a}e^{-\frac{1}{2}\omega}
\vert\omega\vert^{\frac{1}{2}(\eta-1)}
W\bigl(\frac{1}{4}(\eta-1)-a,\frac{1}{4}(\eta+1),\omega\bigr)
\end{array}
\end{displaymath}
with $\:\omega=\frac{1}{4}z^{2}\tau^{-1}$. Here $W(\kappa,\mu,\omega)$ is
the general solution of the Whittaker equation
\begin{displaymath}
4\omega^{2}W_{\omega\omega}=(\omega^{2}-4\kappa\omega+4\mu^{2}-1)W.
\end{displaymath}

\vspace{1ex}

For the subalgebra $<\partial_{\tau}+\Pi+af\partial_{f}>$, where
$a\in{\R}$, it follows that
\begin{displaymath}
\begin{array}{l}
f=(4\tau^{2}+1)^{\frac{1}{4}(\eta-1)}
\exp(-\tau\omega+\frac{1}{2}a\arctan 2\tau)\varphi(\omega)
\end{array}
\end{displaymath}
with $\:\omega=z^{2}(4\tau^{2}+1)^{-1}$. The function $\varphi$ is
a solution of the equation
\begin{displaymath}
4\omega\varphi_{\omega\omega}+2(1-\eta)\varphi_{\omega}+
(\omega-a)\varphi=0.
\end{displaymath}
For example if $a=0$, then
$\varphi(\omega)=\omega^{\mu}\Bigl(
C_{1}J_{\mu}(\frac{1}{2}\omega)+C_{2}Y_{\mu}(\frac{1}{2}\omega)
\Bigr)$,
where \mbox{$\mu=\frac{1}{4}(\eta+1)$}.

Consider equation (\ref{e5.22}), where $\eta$ is an arbitrary
smooth function of $\tau$.

\begin{theorem}\label{t5.2}
Equation (\ref{e5.22}) is Q-conditional invariant under the
operators
\begin{equation}\label{e5.23a}
Q^{1}=\partial_{\tau}+g^{1}(\tau,z)\partial_{z}+\bigl(
g^{2}(\tau,z)f+g^{3}(\tau,z)\bigr)\partial_{f}
\end{equation}
if and only if
\begin{equation}\label{e5.23}
\begin{array}{l}
g^{1}_{\tau}-\eta z^{-1}g^{1}_{z}+\eta z^{-2}g^{1}-g^{1}_{zz}+
2g^{1}_{z}g^{1}-\eta_{\tau}z^{-1}+2g^{2}_{z}=0,\\
\\
g^{k}_{\tau}+\eta z^{-1}g^{k}_{z}-g^{k}_{zz}+2g^{1}_{z}g^{k}=0,
\quad k=2,3,
\end{array}
\end{equation}
and
\begin{equation}\label{e5.24a}
Q^{2}=\partial_{z}+B(\tau,z,f)\partial_{f}
\end{equation}
if and only if
\begin{equation}\label{e5.24}
B_{\tau}-\eta z^{-2}B+\eta z^{-1}B_{z}-B_{zz}-2BB_{zf}-B^{2}B_{ff}
=0.
\end{equation}
An arbitrary operator of Q-conditional symmetry of equation
(\ref{e5.22}) is equivalent to either an operator of form
(\ref{e5.23a}) or an operator of form (\ref{e5.24a}).
\end{theorem}

Theorem \ref{t5.2} is proved by means of the method
described in \cite{fshs}.

\begin{note}\label{n5.6}
It can be shown (in a way analogous to one in \cite{fshs}) that
system (\ref{e5.23}) is reduced
to the decomposed linear system
\begin{equation}\label{e5.26}
f^{a}_{\tau}+\eta z^{-1}f^{a}_{z}-f^{a}_{zz}=0
\end{equation}
by means of the following non-local transformation
\begin{equation}\label{e5.25}
\begin{array}{l}
{\displaystyle
g^{1}=-\frac{f^{1}_{zz}f^{2}-f^{1}f^{2}_{zz}}
{f^{1}_{z}f^{2}-f^{1}f^{2}_{z}}+\eta z^{-1},}\\
\\
{\displaystyle
g^{2}=-\frac{f^{1}_{zz}f^{2}_{z}-f^{1}_{z}f^{2}_{zz}}
{f^{1}_{z}f^{2}-f^{1}f^{2}_{z}},}\\
\\
g^{3}=f^{3}_{zz}-\eta z^{-1}f^{3}_{z}+g^{1}f^{3}_{z}-g^{2}f^{3}.
\end{array}
\end{equation}
Equation (\ref{e5.24}) is reduced, by means of the change
\begin{displaymath}
B=-\Phi_{\tau}/\Phi_{f}, \quad \Phi=\Phi(\tau,z,f)
\end{displaymath}
and the hodograph transformation
\begin{displaymath}
y_{0}=\tau,\quad y_{1}=z, \quad y_{2}=\Phi, \quad \Psi=f,
\end{displaymath}
to the following equation in the function
$\Psi=\Psi(y_{0},y_{1},y_{2})$:
\begin{displaymath}
\Psi_{y_{0}}+\eta(y_{0})y_{1}^{-1}\Psi_{y_{1}}-
\Psi_{y_{1}y_{1}}=0.
\end{displaymath}
\end{note}

Therefore, unlike Lie symmetries Q-conditional symmetries of
(\ref{e5.22}) are more extended for an arbitrary smooth function
$\eta=\eta(\tau)$. Thus, Theorem \ref{t5.2} implies that
equation (\ref{e5.22}) is Q-conditional invariant under the
operators
\begin{displaymath}
\partial_{z},\quad
X=\partial_{\tau}+(\eta-1)z^{-1}\partial_{z}, \quad
G=(2\tau+C)\partial_{z}-zf\partial_{f}
\end{displaymath}
with $C=\mbox{\rm const}$. Reducing equation (\ref{e5.22}) by means
of the operator $G$, we obtain the following solution:
\begin{equation}\label{e5.27}
\begin{array}{l}
f=C_{2}\bigl(z^{2}-2\int(\eta(\tau)-1)d\tau\bigr)+C_{1}.
\end{array}
\end{equation}
In generalizing this we can construct solutions of the form
\begin{equation}\label{e5.28}
f=\sum_{k=0}^{N}T^{k}(\tau)z^{2k},
\end{equation}
where the coefficients $\:T^{k}=T^{k}(\tau)$ $\:(k=\overline{0,N})\:$
satisfy the system of ODEs:
\begin{equation}\label{e5.29}
\begin{array}{l}
T^{k}_{\tau}+(2k+2)(\eta(\tau)-2k-1)T^{k+1}=0,
\\ \\
k=\overline{0,N-1}, \quad T^{N}_{\tau}=0.
\end{array}
\end{equation}
Equation (\ref{e5.29}) is easily integrated for arbitrary
$N\in{\N}$. For example if $N=2$, it follows that
\begin{displaymath}
\begin{array}{ll}
f=\!\!\!&C_{3}\Bigl\{z^{4}-4z^{2}\!\!\int(\eta(\tau)-3)d\tau\!+
8\int\!\Bigl(
(\eta(\tau)-1)\int\!(\eta(\tau)-3)d\tau\Bigr)\!d\tau\Bigr\}+\\
\\
&+C_{2}\Bigl\{z^{2}-2\int(\eta(\tau)-1)d\tau\Bigr\}+C_{1}.
\end{array}
\end{displaymath}
An explicit form for solution (\ref{e5.28}) with $N=1$ is
given by (\ref{e5.27}).

Generalizing the solution
\begin{equation}\label{e5.30}
\begin{array}{l}
f=C_{0}\exp\bigl\{-z^{2}(4\tau+2C)^{-1}+
\int(\eta(\tau)-1)(2\tau+C)^{-1}d\tau\bigr\}
\end{array}
\end{equation}
obtained by means of reduction of (\ref{e5.22}) by the operator
$G$, we can construct solutions of the general form
\begin{equation}\label{e5.31}
\begin{array}{ll}

f=&
\displaystyle{\sum_{k=0}^{N}}
S^{k}(\tau)\bigl(z(2\tau+C)^{-1}\bigr)^{2k}\cdot\\
\\
&\cdot\exp\Bigl\{-z^{2}(4\tau+2C)^{-1}+
\int(\eta(\tau)-1)(2\tau+C)^{-1}d\tau\Bigr\},
\end{array}
\end{equation}
where the coefficients $\:S^{k}=S^{k}(\tau)$ $(k=\overline{0,N})\:$
satisfy the system of ODEs:
\begin{equation}\label{e5.32}
\begin{array}{l}
S^{k}_{\tau}+(2k+2)(\eta(\tau)-2k-1)(2\tau+C)^{-2}S^{k+1}=0,
\\ \\
k=\overline{0,N-1}, \quad S^{N}_{\tau}=0.
\end{array}
\end{equation}
For example if $\:N=1,\:$ then
\begin{displaymath}
\begin{array}{ll}
f=&\Bigl\{C_{1}\Bigl(z^{2}(2\tau+C)^{-2}-
2\int(\eta(\tau)-1)(2\tau+C)^{-2}d\tau\Bigr)+C_{0}\Bigr\}\cdot\\
\\
&\cdot\exp\Bigl\{-z^{2}(4\tau+2C)^{-1}+
\int(\eta(\tau)-1)(2\tau+C)^{-1}d\tau\Bigr\}.
\end{array}
\end{displaymath}
Here we do not present results for arbitrary $N$ as they are very
cumbersome.

Putting $\:g^{2}=g^{3}=0\:$ in system (\ref{e5.23}), we obtain one
equation in the function $g^{1}$:
\begin{displaymath}
g^{1}_{\tau}-\eta z^{-1}g^{1}_{z}+\eta z^{-2}g^{1}-g^{1}_{zz}+
2g^{1}_{z}g^{1}-\eta_{\tau}z^{-1}=0.
\end{displaymath}
It follows that $\:g^{1}=-g_{z}/g+(\eta-1)/z,\:$ where $\:g=g(\tau,z)\:$ is a
solution of the equation
\begin{equation}\label{e5.33}
g_{\tau}+(\eta-2)z^{-1}g_{z}-g_{zz}=0.
\end{equation}
Q-conditional symmetry of (\ref{e5.22}) under the operator
\begin{equation}\label{e5.34}
Q=\partial_{\tau}+\bigl(-g_{z}/g+(\eta-1)/z\bigr)\partial_{z}
\end{equation}
gives rise to the following

\begin{theorem}\label{t5.3}
If $g$ is a solution of equation (\ref{e5.33}) and
\begin{equation}\label{e5.35}
\begin{array}{ll}
f(\tau,z)=&\int_{z_{0}}^{z}z'g(\tau,z')dz'+\\
\\
&+\int_{\tau_{0}}^{\tau}\Bigl(z_{0}g_{z}(\tau',z_{0})-
(\eta(\tau')-1)g(\tau',z_{0})\Bigr)d\tau',
\end{array}
\end{equation}
where $(\tau_{0},z_{0})$ is a fixed point, then $f$ is a solution of
equation (\ref{e5.22}).
\end{theorem}

Proof. Equation (\ref{e5.33}) implies
\begin{displaymath}
(zg)_{\tau}=(zg_{z}-(\eta-1)g)_{z}
\end{displaymath}
Therefore,\quad $f_{z}=zg$,\quad $f_{\tau}=zg_{z}-(\eta-1)g$ \quad
and
\begin{displaymath}
f_{\tau}+\eta z^{-1}f_{z}-f_{zz}=zg_{z}-(\eta-1)g+\eta g-(zg)_{z}=0.
\quad \mbox{QED.}
\end{displaymath}

The converse of Theorem \ref{t5.3} is the following obvious

\begin{theorem}\label{t5.4}
If $f$ is a solution of (\ref{e5.22}), the function
\begin{equation}\label{e5.36}
g=z^{-1}f_{z}
\end{equation}
satisfies (\ref{e5.33}).
\end{theorem}

Theorems \ref{t5.3} and \ref{t5.4} imply that, when $\eta=2n$
$(n\in{\bf Z})$, solutions of (\ref{e5.22}) can be constructed
from known solutions of the heat equation by means of applying
either formula (\ref{e5.35}) (for $n>0$) or formula (\ref{e5.36})
(for $n<0$)  $\:\vert n\vert\:$ times.

Let us investigate symmetry properties and construct some
exact solutions of system (\ref{e5.19})--(\ref{e5.20}) for
$\varepsilon=1$, i.e., the system

\begin{equation}\label{e5.37}
w^{1}_{\tau}-w^{1}_{zz}+\hat\eta(\tau)z^{-1}w^{1}_{z}=0,
\end{equation}
\begin{equation}\label{e5.38}
w^{2}_{\tau}-w^{2}_{zz}+(\hat\eta(\tau)-2)z^{-1}w^{2}_{z}
+(w^{1}-\hat\chi(\tau))z^{-2}=0.
\end{equation}

If $(w^{1},w^{2})$ is a solution of system
\mbox{(\ref{e5.37})--(\ref{e5.38})}, then $(w^{1},w^{2}+g)$ (where
$g=g(\tau,z)$) is also a solution of
(\ref{e5.37})--(\ref{e5.38}) if and only if the function
$g$ satisfies the following equation
\begin{equation}\label{e5.39}
g_{\tau}-g_{zz}+(\hat\eta(\tau)-2)z^{-1}g_{z}=0
\end{equation}

System (\ref{e5.37})--(\ref{e5.38}), for some
$\hat\chi=\hat\chi(\tau)$, has particular solutions of the form
\begin{displaymath}
w^{1}=\sum_{k=0}^{N}T^{k}(\tau)z^{2k}, \quad
w^{2}=\sum_{k=0}^{N-1}S^{k}(\tau)z^{2k},
\end{displaymath}
where $T^{0}(\tau)=\hat\chi(\tau)$. For example, if
$\hat\chi(\tau)=-2C_{1}\int(\hat\eta(\tau)-1)d\tau+C_{2}$
and $N=1$, then
\begin{displaymath}
\begin{array}{l}
w^{1}=C_{1}\bigl(z^{2}-2\int(\hat\eta(\tau)-1)d\tau\bigr)+C_{2},
\quad
w^{2}=-C_{1}\tau.
\end{array}
\end{displaymath}

Let $\hat\chi(\tau)=0$.

\begin{theorem}\label{t5.5}
The MIA of system (\ref{e5.37})--(\ref{e5.38}) with
$\hat\chi(\tau)=0$ is given by the following algebras

\vspace{1ex}

a) $<w^{i}\partial_{w^{i}}, \:
\tilde w^{i}(\tau,z)\partial_{w^{i}}>\:$ if
$\:\hat\eta(\tau)\not=\mbox{\rm const}$;

\vspace{1ex}

b) $<2\tau\partial_{\tau}+z\partial_{z}, \: \partial_{\tau}, \:
w^{i}\partial_{w^{i}}, \: \tilde w^{i}(\tau,z)\partial_{w^{i}}>\:$
if
$\:\hat\eta(\tau)=\mbox{\rm const},\:$ $\:\hat\eta\not=0$;

\vspace{1ex}

c) $<2\tau\partial_{\tau}+z\partial_{z}, \: \partial_{\tau}, \:
w^{1}z^{-1}\partial_{w^{2}}, \: w^{i}\partial_{w^{i}}, \:
\tilde w^{i}(\tau,z)\partial_{w^{i}}>\:$ if
$\:\hat\eta\equiv0$.
\\[1ex]
Here $(\tilde w^{1},\tilde w^{2})$ is an arbitrary solution of
(\ref{e5.37})--(\ref{e5.38}) with $\:\hat\chi(\tau)=0$.
\end{theorem}

For the case $\:\hat\chi(\tau)=0\:$ and
$\:\hat\eta(\tau)=\mbox{\rm const}\:$ system
(\ref{e5.37})--(\ref{e5.38}) can be reduced by inequivalent
one-dimensional subalgebras of its MIA. We obtain the following
solutions:

\vspace{2ex}

For the subalgebra $\:<\partial_{\tau}>\:$ it follows that
\begin{displaymath}
\begin{array}{l}
w^{1}=C_{1}\ln z+C_{2},
\\[1ex]
w^{2}=\frac{1}{4}C_{1}(\ln^{2}z-\ln z)+
\frac{1}{2}C_{2}\ln z+C_{3}z^{-2}+C_{4}
\end{array}
\end{displaymath}
if $\:\hat\eta=-1$;
\begin{displaymath}
\begin{array}{l}
w^{1}=C_{1}z^{2}+C_{2},
\\[1ex]
w^{2}=\frac{1}{4}C_{1}z^{2}+\frac{1}{2}C_{2}\ln^{2}z+
C_{3}\ln z+C_{4}
\end{array}
\end{displaymath}
if $\:\hat\eta=1$;
\begin{displaymath}
\begin{array}{l}
w^{1}=C_{1}z^{\hat\eta+1}+C_{2},
\\[1ex]
w^{2}=\frac{1}{2}C_{1}(\hat\eta+1)^{-1}z^{\hat\eta+1}+
C_{2}(\hat\eta-1)^{-1}\ln z+C_{3}z^{\hat\eta-1}+C_{4}
\end{array}
\end{displaymath}
if $\:\hat\eta\not\in\{-1;1\}$.

\vspace{2ex}

For the subalgebra
$\:<\partial_{\tau}-w^{i}\partial_{w^{i}}>\:$ it follows that
\begin{displaymath}
w^{1}=e^{-\tau}z^{\frac{1}{2}(\hat\eta+1)}\psi^{1}(z),
\quad
w^{2}=e^{-\tau}z^{\frac{1}{2}(\hat\eta-1)}\psi^{2}(z),
\end{displaymath}
where the functions $\psi^{1}$ and $\psi^{2}$ satisfy the system

\begin{equation}\label{e5.40}
\begin{array}{l}
z^{2}\psi^{1}_{zz}+z\psi^{1}_{z}+
\bigl(z^{2}-\frac{1}{4}(\hat\eta+1)^{2}\bigr)\psi^{1}=0,
\end{array}
\end{equation}
\begin{equation}\label{e5.41}
\begin{array}{l}
z^{2}\psi^{2}_{zz}+z\psi^{2}_{z}+
\bigl(z^{2}-\frac{1}{4}(\hat\eta-1)^{2}\bigr)\psi^{2}=
z\psi^{1}.
\end{array}
\end{equation}
The general solution of system (\ref{e5.40})--(\ref{e5.41})
can be expressed by quadratures in terms of the Bessel functions
of a real variable $J_{\nu}(z)$ and $Y_{\nu}(z)$:

\begin{displaymath}
\psi^{1}=C_{1}J_{\nu+1}(z)+C_{2}Y_{\nu+1}(z),
\end{displaymath}
\begin{displaymath}
\begin{array}{ll}
\!\!\!\psi^{2}=\!\!\!\!&C_{3}J_{\nu}(z)+C_{4}Y_{\nu}(z)+
\\[1ex]
&+\frac{\pi}{2}Y_{\nu}(z)
\int J_{\nu}(z)\psi^{1}(z)dz-
\frac{\pi}{2}J_{\nu}(z)
\int Y_{\nu}(z)\psi^{1}(z)dz
\end{array}
\end{displaymath}
with $\:\nu=\frac{1}{2}(\hat\eta-1)$;

\vspace{2ex}

For the subalgebra
$\:<\partial_{\tau}+w^{i}\partial_{w^{i}}>\:$ it follows that
\begin{displaymath}
w^{1}=e^{\tau}z^{\frac{1}{2}(\hat\eta+1)}\psi^{1}(z),
\quad
w^{2}=e^{\tau}z^{\frac{1}{2}(\hat\eta-1)}\psi^{2}(z),
\end{displaymath}
where the functions $\psi^{1}$ and $\psi^{2}$ satisfy the system
\begin{equation}\label{e5.42}
\begin{array}{l}
z^{2}\psi^{1}_{zz}+z\psi^{1}_{z}-
\bigl(z^{2}+\frac{1}{4}(\hat\eta+1)^{2}\bigr)\psi^{1}=0,
\end{array}
\end{equation}
\begin{equation}\label{e5.43}
\begin{array}{l}
z^{2}\psi^{2}_{zz}+z\psi^{2}_{z}-
\bigl(z^{2}+\frac{1}{4}(\hat\eta-1)^{2}\bigr)\psi^{2}=
z\psi^{1}.
\end{array}
\end{equation}
The general solution of system (\ref{e5.42})--(\ref{e5.43})
can be expressed by quadratures in terms of the Bessel functions
of an imaginary variable $I_{\nu}(z)$ and $K_{\nu}(z)$:
\begin{displaymath}
\psi^{1}=C_{1}I_{\nu+1}(z)+C_{2}K_{\nu+1}(z),
\end{displaymath}
\begin{displaymath}
\begin{array}{ll}
\!\!\!\psi^{2}=\!\!\!\!&C_{3}I_{\nu}(z)+C_{4}K_{\nu}(z)+
\\[1ex]
&K_{\nu}(z)\int I_{\nu}(z)\psi^{1}(z)dz-
I_{\nu}(z)\int K_{\nu}(z)\psi^{1}(z)dz
\end{array}
\end{displaymath}
with $\:\nu=\frac{1}{2}(\hat\eta-1)$.

\vspace{2ex}

For the subalgebra
$\:<2\tau\partial_{\tau}+z\partial_{z}+aw^{i}\partial_{w^{i}}>\:$
it follows that
\begin{displaymath}
w^{1}=\vert\tau\vert^{a}e^{-\frac{1}{2}\omega}
\vert\omega\vert^{\frac{1}{4}(\hat\eta-1)}\psi^{1}(\omega),
\quad
w^{2}=\vert\tau\vert^{a}e^{-\frac{1}{2}\omega}
\vert\omega\vert^{\frac{1}{4}(\hat\eta-3)}\psi^{2}(\omega)
\end{displaymath}
with $\:\omega=\frac{1}{4}z^{2}\tau^{-1},\:$ where the functions
$\psi^{1}$ and $\psi^{2}$ satisfy the system
\begin{equation}\label{e5.44}
\begin{array}{l}
4\omega^{2}\psi^{1}_{\omega\omega}=\Bigl(
\omega^{2}+\bigl(a-\frac{1}{4}(\hat\eta-1)\bigr)\omega+
\frac{1}{4}(\hat\eta+1)^{2}-1\Bigr)\psi^{1},
\end{array}
\end{equation}
\begin{equation}\label{e5.45}
\begin{array}{ll}
4\omega^{2}\psi^{2}_{\omega\omega}=\!\!\!\!&\Bigl(
\omega^{2}+\bigl(a-\frac{1}{4}(\hat\eta-3)\bigr)\omega+
\frac{1}{4}(\hat\eta-1)^{2}-1\Bigr)\psi^{2}+
\\[1ex]
&+2\vert\omega\vert^{1/2}\psi^{1}.
\end{array}
\end{equation}
The general solution of system (\ref{e5.44})--(\ref{e5.45})
can be expressed by quadratures in terms of the Whittaker
functions.

\setcounter{equation}{0}

\protect
\section{Symmetry properties and exact solutions of system\\
(\protect\ref{e3.12})}\label{sec6}

As was mentioned in Sec. \ref{sec3}, ansatzes
(\ref{e3.4})--(\ref{e3.7}) reduce the NSEs (\ref{e1.1}) to the
systems of PDEs of a similar structure that have the general
form (see (\ref{e3.12})):
\begin{equation}\label{e6.1}
\begin{array}{l}
w^{i}w^{1}_{i}-w^{1}_{ii}+s_{1}+\alpha_{2}w^{2}=0,
\\[1ex]
w^{i}w^{2}_{i}-w^{2}_{ii}+s_{2}-\alpha_{2}w^{1}+
\alpha_{1}w^{3}=0,
\\[1ex]
w^{i}w^{3}_{i}-w^{3}_{ii}+\alpha_{4}w^{3}+\alpha_{5}=0,
\\[1ex]
w^{i}_{i}=\alpha_{3},
\end{array}
\end{equation}
where $\alpha_{n} \: (n=\overline{1,5})$ are real parameters.

Setting $\alpha_{k}=0 \: (k=\overline{2,5})$ in (\ref{e6.1}), we
obtain equations describing a plane convective flow that is
brought about by nonhomogeneous heating of boudaries
\cite{landau&lifshits}. In this case $w^{i}$ are the coordinates
of the flow velocity vector, $w^{3}$ is the flow temperature, $s$
is the pressure, the Grasshoff number $\lambda$ is equal to
$-\alpha_{1}$, and the Prandtl number $\sigma$ is equal to 1. Some
similarity solutions of these equations were constructed in
\cite{katkov}. The particular case of system (\ref{e6.1})
for $\:\alpha_{1}=\alpha_{2}=\alpha_{4}=\alpha_{5}=0\:$ and
$\:\alpha_{3}=1\:$ was considered in \cite{pukhnachev.a}.

In this section we study symmetry properties of system
(\ref{e6.1}) and construct large sets of its exact solutions.

\begin{theorem}\label{t6.1}
The MIA of (\ref{e6.1}) is the algebra

1. $E_{1}=<\partial_{1},\: \partial_{2}, \: \partial_{s}>$ \quad
if \quad $\alpha_{1}\not=0$, $\alpha_{4}\not=0$.

\vspace{1ex}

2. $E_{2}=<\partial_{1}, \: \partial_{2}, \: \partial_{s}, \:
\partial_{w^{3}}-\alpha_{1}z_{2}\partial_{s}>$ \quad
if \quad $\alpha_{1}\not=0$, $\alpha_{4}=0$,
\mbox{$(\alpha_{1},\alpha_{2},\alpha_{5})\not=(0,0,0)$}.

3. $E_{3}=<\partial_{1}, \: \partial_{2}, \: \partial_{s},
\: \partial_{w^{3}}-\alpha_{1}z_{2}\partial_{s}, \:
\tilde D-3w^{3}\partial_{w^{3}}>$ \quad
if \quad $\alpha_{1}\not=0$, $\alpha_{k}=0$, $k=\overline{2,5}$.

\vspace{1ex}

4. $E_{4}=<\partial_{1}, \: \partial_{2}, \: \partial_{s}, \: J,
\: (w^{3}+\alpha_{5}/\alpha_{4})\partial_{w^{3}}>$ \quad
if \quad $\alpha_{1}=0$, $\alpha_{4}\not=0$.

\vspace{1ex}

5. $E_{5}=<\partial_{1}, \: \partial_{2}, \: \partial_{s}, \: J,
\: \partial_{w^{3}}>$ \quad
if \quad $\alpha_{1}=\alpha_{4}=0$,
\mbox{$(\alpha_{2},\alpha_{3})\not=(0,0)$}, $\alpha_{5}\not=0$.
\vspace{1ex}

6. $E_{6}=<\partial_{1}, \: \partial_{2}, \: \partial_{s}, \: J,
\: \partial_{w^{3}}, \: w^{3}\partial_{w^{3}}>$ \quad
if \quad \mbox{$\alpha_{1}=\alpha_{4}=\alpha_{5}=0$},
\mbox{$(\alpha_{2},\alpha_{3})\not=(0,0)$}.

\vspace{1ex}

7. $E_{7}=<\partial_{1}, \: \partial_{2}, \: \partial_{s}, \: J,
\: \partial_{w^{3}}, \: \tilde D+2w^{3}\partial_{w^{3}}>$ \quad
if \quad $\alpha_{5}\not=0$,
$\alpha_{l}=0$, $l=\overline{1,4}$.

\vspace{1ex}

8. $E_{8}=<\partial_{1}, \: \partial_{2}, \: \partial_{s}, \: J,
\: \partial_{w^{3}}, \: \tilde D, w^{3}\partial_{w^{3}}>$ \quad
if \quad $\alpha_{n}=0$, $n=\overline{1,5}$.

\vspace{1ex}

Here\quad $\tilde D=
z_{i}\partial_{i}-w^{i}\partial_{w^{i}}-2s\partial_{s}$,\quad
$J=z_{1}\partial_{2}-z_{2}\partial_{1}+
w^{1}\partial_{w^{2}}-w^{2}\partial_{w^{1}}$, \quad
$\partial_{i}=\partial_{z_{i}}$.
\end{theorem}

\begin{note}\label{n6.1}
The bases of the algebras $E_{6}$ and $E_{8}$ contain the operator
$w^{3}\partial_{w^{3}}$ that is not induced by elements of
$A(NS)$.
\end{note}

\begin{note}\label{n6.2}
If $\alpha_{4}\not=0$, the constant $\alpha_{5}$ can be made to
vanish by means of local transformation
\begin{equation}\label{e6.2}
\tilde w^{3}=w^{3}+\alpha_{5}/\alpha_{4}, \quad
\tilde s=s-\alpha_{1}\alpha_{5}\alpha_{4}^{-1}z_{2},
\end{equation}
where the independent variables and the functions $w^{i}$ are not
transformed. Therefore, we consider below that $\:\alpha_{5}=0\:$ if
$\:\alpha_{4}\not=0\:$.
\end{note}

\begin{note}\label{n6.3}
Making the non-local transformation
\begin{equation}\label{e6.3}
\tilde s=s+\alpha_{2}\Psi,
\end{equation}
where $\Psi_{1}=w^{2}$, $\Psi_{2}=-w^{1}$ (such a function $\Psi$
exists in view of the last equation of (\ref{e6.1})), in system
(\ref{e6.1}) with $\alpha_{3}=0$, we obtain a system of form
(\ref{e6.1}) with $\tilde\alpha_{3}=\tilde\alpha_{2}=0$. In some
cases ($\alpha_{1}\not=0$,
$\alpha_{3}=\alpha_{4}=\alpha_{5}=0$, $\alpha_{2}\not=0$;
$\alpha_{1}=\alpha_{3}=\alpha_{4}=0$, $\alpha_{2}\not=0$)
transformation (\ref{e6.3}) allows the symmetry of (\ref{e6.1}) to be
extended and non-Lie solutions to be constructed. Moreover, it
means that in the cases listed above system (\ref{e6.1}) is
invariant under the non-local transformation
\begin{displaymath}
\hat z_{i}=e^{\varepsilon}z_{i},\quad
\hat w^{i}=e^{-\varepsilon}w^{i},\quad
\hat w^{3}=e^{\delta\varepsilon}w^{3},\quad
\hat s=e^{-2\varepsilon}s+\alpha_{2}(e^{-2\varepsilon}-1)\Psi,
\end{displaymath}
\begin{tabbing}
where \quad \= $\delta=-3$\= \quad if \quad
$\alpha_{3}=\alpha_{4}=\alpha_{5}=0$,
\quad $\alpha_{1},\alpha_{2}\not=0$;\\
\> $\delta=2$\> \quad if \quad
$\alpha_{1}=\alpha_{3}=\alpha_{4}=0$,
\quad $\alpha_{2},\alpha_{5}\not=0$;\\
\> $\delta=0$\> \quad if \quad
$\alpha_{1}=\alpha_{3}=\alpha_{4}=\alpha_{5}=0$,
\quad $\alpha_{2}\not=0$.
\end{tabbing}
\end{note}

\begin{table}
\protect\caption{
Complete sets of inequivalent one-dimensional subalgebras of the
algebras \protect $E_{1}-E_{8}$ (\protect$a$ and
\protect$a_{l}\:(l=\overline{1,4})$ are real constants)
}\label{table6.2}
\begin{center}
{\normalsize
\begin{tabular}{|c|l|c|}
\hline
$\!\!\!\mbox{Algebra}\!\!\!$ &
$\qquad\qquad\qquad\qquad\mbox{Subalgebras}$ &
$\!\!\!\begin{array}{c}\mbox{Values of}\\ \mbox{parameters}
\end{array}\!\!\!$
\\
\hline
$E_{1}$&
$\!\!\!\begin{array}{l} \\
         <a_{1}\partial_{1}+a_{2}\partial_{2}+a_{3}\partial_{s}>,\:
         <\partial_{s}>\\[2ex]
\end{array}\!\!\!$&
$a_{1}^{2}+a_{2}^{2}=1$
\\
\hline
$E_{2}$&
$\!\!\!\begin{array}{l}\\
        <a_{1}\partial_{1}+a_{2}\partial_{2}+
        a_{3}(\partial_{w^{3}}-\alpha_{1}z_{2}\partial_{s})>,
        \\
        <\partial_{1}+a_{4}\partial_{s}>,
        <\partial_{w^{3}}-\alpha_{1}z_{2}\partial_{s}>,
        <\partial_{s}>\\[2ex]
\end{array}\!\!\!$&
$\!\!\!\begin{array}{c}
a_{1}^{2}+a_{2}^{2}=1,\\
a_{4}\not=0
\end{array}\!\!\!$
\\
\hline
$E_{3}$&
$\!\!\!\begin{array}{l}\\
        <a_{1}\partial_{1}+a_{2}\partial_{2}+
        a_{3}(\partial_{w^{3}}-\alpha_{1}z_{2}\partial_{s})>,
        <\partial_{1}+a_{4}\partial_{s}>,
\\[1ex]
        <\tilde D-3w^{3}\partial_{w^{3}}>,
        <\partial_{w^{3}}-\alpha_{1}z_{2}\partial_{s}>,
        <\partial_{s}>\\[2ex]
\end{array}\!\!\!$&
$\!\!\!\begin{array}{c}
        a_{1}^{2}+a_{2}^{2}=1,\\ a_{3}\in\{-1;0;1\},\\
        a_{4}\in\{-1;1\}
\end{array}\!\!\!$
\\
\hline
$E_{4}$&
$\!\!\!\begin{array}{l}\\
        <J+a_{1}\partial_{s}+a_{2}w^{3}\partial_{w^{3}}>,
        <\partial_{2}+a_{1}\partial_{s}+a_{2}w^{3}\partial_{w^{3}}>,
\\[1ex]
        <w^{3}\partial_{w^{3}}+a_{1}\partial_{s}>,
        <\partial_{s}>\\[2ex]
\end{array}\!\!\!$&
\\
\hline
$E_{5}$&
$\!\!\!\begin{array}{l}\\
        <J+a_{1}\partial_{s}+a_{2}\partial_{w^{3}}>,
        <\partial_{2}+a_{1}\partial_{s}+a_{2}\partial_{w^{3}}>,
\\[1ex]
        <\partial_{w^{3}}+a_{1}\partial_{s}>, <\partial_{s}>\\[2ex]
\end{array}\!\!\!$&
\\
\hline
$E_{6}$&
$\!\!\!\begin{array}{l}\\
        <J+a_{1}\partial_{s}+a_{2}w^{3}\partial_{w^{3}}>,
        <\partial_{2}+a_{1}\partial_{s}+a_{2}w^{3}\partial_{w^{3}}>
        ,
\\[1ex]
        <J+a_{1}\partial_{s}+a_{3}\partial_{w^{3}}>,
        <\partial_{2}+a_{1}\partial_{s}+a_{3}\partial_{w^{3}}>
        ,
\\[1ex]
        <w^{3}\partial_{w^{3}}+a_{1}\partial_{s}>,
        <\partial_{w^{3}}+a_{1}\partial_{s}>, <\partial_{s}>\\[2ex]
\end{array}\!\!\!$&
$\!\!\!\begin{array}{c}
       a_{2}\not=0,\\ a_{3}\in\{-1;0;1\}
\end{array}\!\!\!$
\\
\hline
$E_{7}$&
$\!\!\!\begin{array}{l}\\
        <\tilde D+aJ+2w^{3}\partial_{w^{3}}>,
        <J+a_{1}\partial_{s}+a_{2}\partial_{w^{3}}>,
\\[1ex]
        <\partial_{2}+a_{1}\partial_{s}+a_{2}\partial_{w^{3}}>
        ,
        <\partial_{w^{3}}+a_{2}\partial_{s}>, <\partial_{s}>\\[2ex]
\end{array}\!\!\!$&
$\!\!\!\begin{array}{c}
        a_{2}\in\{-1;0;1\},\\ a_{1}\in\{-1;0;1\}\\
        \mbox{if}\: a_{2}=0
\end{array}\!\!\!$
\\
\hline
$E_{8}$&
$\!\!\!\begin{array}{l}\\
        <\tilde D+aJ+a_{3}w^{3}\partial_{w^{3}}>,
        <\tilde D+aJ+a_{3}\partial_{w^{3}}>,
\\[1ex]
        <J+a_{1}\partial_{s}+a_{4}w^{3}\partial_{w^{3}}>,
        <\partial_{2}+a_{1}\partial_{s}+a_{4}w^{3}\partial_{w^{3}}>
        ,
\\[1ex]
        <J+a_{1}\partial_{s}+a_{2}\partial_{w^{3}}>,
        <\partial_{2}+a_{1}\partial_{s}+a_{2}\partial_{w^{3}}>
        ,
\\[1ex]
        <w^{3}\partial_{w^{3}}+a_{1}\partial_{s}>,
        <\partial_{w^{3}}+a_{1}\partial_{s}>, <\partial_{s}>\\[2ex]
\end{array}\!\!\!$&
$\!\!\!\begin{array}{c}
        a_{i}\in\{-1;0;1\},\\ a_{4}\not=0
\end{array}\!\!\!$
\\
\hline
\end{tabular}
}
\end{center}
\end{table}

Let us consider an ansatz of the form:
\begin{equation}\label{e6.5}
\begin{array}{lll}
w^{1}&=&a_{1}\varphi^{1}-a_{2}\varphi^{3}+b_{1}\omega_{2},
\\[0.8em]
w^{2}&=&a_{2}\varphi^{1}+a_{1}\varphi^{3}+b_{2}\omega_{2},
\\[0.8em]
w^{3}&=&\varphi^{2}+b_{3}\omega_{2},
\\[0.8em]
s    &=&h+d_{1}\omega_{2}+d_{2}\omega_{1}\omega_{2}+\frac{1}{2}
       d_{3}\omega_{2}^{2},
\end{array}
\end{equation}
where $\:a_{1}^{2}+a_{2}^{2}=1$,
$\;\omega=\omega_{1}=a_{1}z_{2}-a_{2}z_{1}$,
$\;\omega_{2}=a_{1}z_{1}+a_{2}z_{2}$, $\;B,b_{a},d_{a}=\mbox{const}$,
\begin{equation}\label{e6.6}
\begin{array}{l}
b_{i}=Ba_{i},\quad b_{3}(B+\alpha_{4})=0,
\\[0.8em]
d_{2}=\alpha_{2}B-\alpha_{1}b_{3}a_{1},\quad
d_{3}=-B^{2}-\alpha_{1}b_{3}a_{2},
\end{array}
\end{equation}
Here and below $\varphi^{a}=\varphi^{a}(\omega)$ and $h=h(\omega)$.
Indeed, formulas (\ref{e6.5}) and (\ref{e6.6}) determine
a whole set of ansatzes for system (\ref{e6.1}). This set contains
both Lie ansatzes, constructed by means of subalgebras of the form
\begin{equation}\label{e6.7}
<a_{1}\partial_{1}+a_{2}\partial_{2}+a_{3}(\partial_{w^{3}}-
\alpha_{1}z_{2}\partial_{s})+a_{4}\partial_{s}>,
\end{equation}
and non-Lie ansatzes. Equation (\ref{e6.6}) is the necessary
and sufficient condition to reduce (\ref{e6.1}) by means of an
ansatz of form (\ref{e6.3}). As a result of reduction we obtain
the following system of ODEs:
\begin{equation}\label{e6.8}
\begin{array}{l}
\varphi^{3}\varphi^{1}_{\omega}-\varphi^{1}_{\omega\omega}+
\mu_{1j}\varphi^{j}+d_{1}+d_{2}\omega+\alpha_{2}\varphi^{3}=0,\\
\\
\varphi^{3}\varphi^{2}_{\omega}-\varphi^{2}_{\omega\omega}+
\mu_{2j}\varphi^{j}+\alpha_{5}=0,\\
\\
\varphi^{3}\varphi^{3}_{\omega}-\varphi^{3}_{\omega\omega}+
h_{\omega}-\alpha_{2}\varphi^{1}+\alpha_{1}a_{1}\varphi^{2}=0,\\
\\
\varphi^{3}_{\omega}=\sigma,
\end{array}
\end{equation}
where $\mu_{11}=-B$, $\mu_{12}=-\alpha_{1}a_{2}$,
$\mu_{21}=-b_{3}$, $\mu_{22}=-\alpha_{4}$, $\sigma=\alpha_{3}-B$.
If $\sigma=0$, system (\ref{e6.8}) implies that
\begin{displaymath}
\varphi^{3}=C_{0}=\mbox{const},
\end{displaymath}
\begin{displaymath}
\begin{array}{l}
h=\alpha_{2}\int\varphi^{1}(\omega)d\omega-
\alpha_{1}a_{1}\int\varphi^{2}(\omega)d\omega,
\end{array}
\end{displaymath}
and the functions $\varphi^{i}$ satisfy system (\ref{e4.23}),
where $\nu_{11}=d_{1}+\alpha_{2}C_{0}$, $\nu_{21}=d_{2}$,
$\nu_{12}=\alpha_{5}$, $\nu_{22}=0$. If $\sigma\not=0$, then
$\varphi^{3}=\sigma\omega$ (translating $\omega$, the integration
constant can be made to vanish),
\begin{displaymath}
\begin{array}{l}
h=-\frac{1}{2}\sigma^{2}\omega^{2}+
\alpha_{2}\int\varphi^{1}(\omega)d\omega-
\alpha_{1}a_{1}\int\varphi^{2}(\omega)d\omega,
\end{array}
\end{displaymath}
and the functions satisfy system (\ref{e4.29}), where
$\nu_{11}=d_{1}$, $\nu_{21}=d_{2}+\alpha_{2}\sigma$,
\mbox{$\nu_{12}=\alpha_{5}$}, $\nu_{22}=0$.

\begin{note}\label{n6.4}
Step-by-step reduction of the NSEs (\ref{e1.1}) by means of
ansatzes \mbox{(\ref{e3.4})--(\ref{e3.7})} and (\ref{e6.5}) is
equivalent to a particular case of immediate reduction of the
NSEs (\ref{e1.1}) to ODEs by means of ansatzes 5 and 6 from
Subsec. \ref{subsec4.1}.
\end{note}

Now let us choose such algebras, among the algebras from Table
\ref{table6.2},
that can be used to reduce system (\ref{e6.1}) and do not
belong to the set of algebras (\ref{e6.7}). By means of the chosen
algebras we construct ansatzes that are tabulated in the form
of Table \ref{table6.1}.

\begin{table}
\protect\caption{Ansatzes reducing system (\protect\ref{e6.1})
\quad
\protect$(r=(z_{1}^{2}+z_{2}^{2})^{1/2})$
}\label{table6.1}
\begin{center}
{\normalsize
\begin{tabular}{|c|c|c|c|l|}
\hline
$\!$N$\!$ &
$\!\!\begin{array}{c}\mbox{Values}\\\mbox{of}\quad\alpha_{n}
\end{array}\!\!\!$&
Algebra & \begin{tabular}{c}Invariant\\variable   \end{tabular} &
\qquad\quad  Ansatz
\\
\hline
$\!$
1
$\!$ &
$\!\!\begin{array}{c}
\alpha_{1}\not=0,\\\alpha_{k}=0,\\k=\overline{2,5}
\end{array}\!\!\! $
&
$<\tilde D-3w^{3}\partial_{w^{3}}>$
&
$\omega=\arctan\!\frac{z_{2}}{z_{1}}$
&
$\!\!\begin{array}{l}\\
w^{1}=r^{-2}(z_{1}\varphi^{1}-z_{2}\varphi^{2}),\\
w^{2}=r^{-2}(z_{2}\varphi^{1}+z_{1}\varphi^{2}),\\
w^{3}=r^{-3}\varphi^{3}, \: s=r^{-2}h\\[2ex]
\end{array}\!\!\!$
\\
\hline
$\!$
2
$\!$ &
$\!\!\begin{array}{c}
\alpha_{1}=0,\\\alpha_{5}=0
\end{array}\!\!\!$
&
$\!\!\!\!\begin{array}{c}
<\partial_{2}+a_{1}\partial_{s}+a_{2}w^{3}\partial_{w^{3}}>,\\
a_{2}\not=0
\end{array}\!\!\!\!$
&
$\omega=z_{1}$
&
$\!\!\begin{array}{l}\\
w^{1}=\varphi^{1},\quad w^{2}=\varphi^{2},\\
w^{3}=\varphi^{3}e^{a_{2}z_{2}},\\
s=h+a_{1}z_{2}\\[2ex]
\end{array}\!\!\!$
\\
\hline
$\!$
3
$\!$ &
$\!\!\begin{array}{c}
\alpha_{1}=0,\\\alpha_{4}=0
\end{array}\!\!\!$
&
$\!\!\!\begin{array}{c}
<J+a_{1}\partial_{s}+a_{2}\partial_{w^{3}}>
\end{array}\!\!\!$
&
$\omega=r$
&
$\!\!\begin{array}{l}\\
w^{1}=z_{1}\varphi^{1}-z_{2}r^{-2}\varphi^{2}, \\
w^{2}=z_{2}\varphi^{1}+z_{1}r^{-2}\varphi^{2}, \\
w^{3}=\varphi^{3}+a_{2}\arctan\!\frac{z_{2}}{z_{1}},\\
s=h+a_{1}\arctan\!\frac{z_{2}}{z_{1}}\\[2ex]
\end{array}\!\!\!$
\\
\hline
$\!$
4
$\!$ &
$\!\!\begin{array}{c}
\alpha_{1}=0,\\\alpha_{5}=0
\end{array}\!\!\!$
&
$\!\!\!\!\begin{array}{c}
<J+a_{1}\partial_{s}+a_{2}w_{3}\partial_{w^{3}}> \\
a_{2}\not=0 \quad\mbox{if}\quad\alpha_{4}=0
\end{array}\!\!\!\!$
&
$\omega=r$
&
$\!\!\begin{array}{l}\\
w^{1}=z_{1}\varphi^{1}-z_{2}r^{-2}\varphi^{2}, \\
w^{2}=z_{2}\varphi^{1}+z_{1}r^{-2}\varphi^{2}, \\
w^{3}=\varphi^{3}e^{a_{2}\arctan\!\frac{z_{2}}{z_{1}}},\\
s=h+a_{1}\arctan\!\frac{z_{2}}{z_{1}}\\[2ex]
\end{array}\!\!\!$
\\[1ex]
\hline
$\!$
5
$\!$ &
$\!\!\begin{array}{c}
\alpha_{5}\not=0,\\\alpha_{l}=0,\\l=\overline{1,4}
\end{array}\!\!\! $
&
$\!\!\!\begin{array}{c}
<\tilde D+aJ+2w^{3}\partial_{w^{3}}>
\end{array}\!\!\! $
&
$\!\!\begin{array}{c}
\omega=\arctan\!\frac{z_{2}}{z_{1}}-\\
-a\ln r
\end{array}\!\!\! $
&
$\!\!\begin{array}{l}\\
w^{1}=r^{-2}(z_{1}\varphi^{1}-z_{2}\varphi^{2}),\\
w^{2}=r^{-2}(z_{2}\varphi^{1}+z_{1}\varphi^{2}),\\
w^{3}=r^{2}\varphi^{3}, \: s=r^{-2}h\\[2ex]
\end{array}\!\!\!$
\\
\hline
$\!$
6
$\!$ &
$\!\!\begin{array}{c}
\alpha_{n}=0,\\n=\overline{1,5}
\end{array}\!\!\! $
&
$\!\!\!\begin{array}{c}
<\tilde D+aJ+a_{1}\partial_{w^{3}}>
\end{array}\!\!\! $
&
$\!\!\begin{array}{c}
\omega=\arctan\!\frac{z_{2}}{z_{1}}-\\
-a\ln r
\end{array}\!\!\! $
&
$\!\!\begin{array}{l}\\
w^{1}=r^{-2}(z_{1}\varphi^{1}-z_{2}\varphi^{2}),\\
w^{2}=r^{-2}(z_{2}\varphi^{1}+z_{1}\varphi^{2}),\\
w^{3}=\varphi^{3}+a_{1}\ln r, \\
s=r^{-2}h\\[2ex]
\end{array}\!\!\!$
\\
\hline
7
$\!$ &
$\!\!\begin{array}{c}
\alpha_{n}=0,\\n=\overline{1,5}
\end{array}\!\!\! $
&
$\!\!\!\!\begin{array}{c}
<\tilde D+aJ+a_{1}w^{3}\partial_{w^{3}}>,  \\  a_{1}\not=0
\end{array}\!\!\!\! $
&
$\!\!\begin{array}{c}
\omega=\arctan\!\frac{z_{2}}{z_{1}}-\\
-a\ln r
\end{array}\!\!\! $
&
$\!\!\begin{array}{l}\\
w^{1}=r^{-2}(z_{1}\varphi^{1}-z_{2}\varphi^{2}),\\
w^{2}=r^{-2}(z_{2}\varphi^{1}+z_{1}\varphi^{2}),\\
w^{3}=r^{a_{1}}\varphi^{3}, \: s=r^{-2}h\\[2ex]
\end{array}\!\!\!$
\\
\hline
\end{tabular}
}
\end{center}
\end{table}

Substituting the ansatzes from Table \ref{table6.1} into
system (\ref{e6.1}), we obtain the reduced systems of ODEs in the
functions $\varphi^{a}$ and $h$:

\begin{equation}\label{e6.9}
\!\!\!\!\!\!\!\!\!\!\!
\begin{array}{ll}
\mbox{1.}
&\varphi^{2}\varphi^{1}_{\omega}-
\varphi^{1}_{\omega\omega}-\varphi^{1}\varphi^{1}-
\varphi^{2}\varphi^{2}-2h+\alpha_{1}\varphi^{3}\sin\omega+
2\varphi^{2}_{\omega}=0,\\
\\
&\varphi^{2}\varphi^{2}_{\omega}-
\varphi^{2}_{\omega\omega}+h_{\omega}-2\varphi^{1}_{\omega}+
\alpha_{1}\varphi^{3}\cos\omega=0,\\
\\
&\varphi^{2}\varphi^{3}_{\omega}-
\varphi^{3}_{\omega\omega}-3\varphi^{1}\varphi^{3}-9\varphi^{3}
=0,\\
\\
&\varphi^{2}_{\omega}=0.
\end{array}
\end{equation}

\begin{equation}\label{e6.10}
\!\!\!\!\!\!\!\!\!\!\!
\begin{array}{ll}
\mbox{2.}
&\varphi^{1}\varphi^{1}_{\omega}-\varphi^{1}_{\omega\omega}+
\alpha_{2}\varphi^{2}+h_{\omega}=0,\\
\\
&\varphi^{1}\varphi^{2}_{\omega}-\varphi^{2}_{\omega\omega}-
\alpha_{2}\varphi^{1}+a_{1}=0,\\
\\
&\varphi^{1}\varphi^{3}_{\omega}-\varphi^{3}_{\omega\omega}+
(a_{2}\varphi^{2}+\alpha_{4}-a_{2}^{2})\varphi^{3}=0,\\
\\
&\varphi^{1}_{\omega}=\alpha_{3}.
\end{array}
\end{equation}

\begin{equation}\label{e6.11}
\!\!\!\!\!\!\!\!\!\!\!
\begin{array}{ll}
\mbox{3.}
&\omega\varphi^{1}\varphi^{1}_{\omega}-\varphi^{1}_{\omega\omega}
+\varphi^{1}\varphi^{1}-\omega^{-4}\varphi^{2}\varphi^{2}-
3\omega^{-1}\varphi^{1}_{\omega}+
\\[1ex]
&\quad\alpha_{2}\omega^{-2}\varphi^{2}
+\omega^{-1}h_{\omega}=0,\\
\\
&\omega\varphi^{1}\varphi^{2}_{\omega}-\varphi^{2}_{\omega\omega}
+\omega^{-1}\varphi^{2}_{\omega}-\alpha_{2}\omega^{2}\varphi^{1}+
a_{1}=0,\\
\\
&\omega\varphi^{1}\varphi^{3}_{\omega}-\varphi^{3}_{\omega\omega}
+a_{2}\omega^{-2}\varphi^{2}-\omega^{-1}\varphi^{3}_{\omega}+
\alpha_{5}=0,\\
\\
&2\varphi^{1}+\omega\varphi^{1}_{\omega}=\alpha_{3}.
\end{array}
\end{equation}

\begin{equation}\label{e6.12}
\!\!\!\!\!\!\!\!\!\!\!
\begin{array}{ll}
\mbox{4.}
&\omega\varphi^{1}\varphi^{1}_{\omega}-\varphi^{1}_{\omega\omega}
+\varphi^{1}\varphi^{1}-\omega^{-4}\varphi^{2}\varphi^{2}-
3\omega^{-1}\varphi^{1}_{\omega}+
\\[1ex]
&\quad\alpha_{2}\omega^{-2}\varphi^{2}
+\omega^{-1}h_{\omega}=0,\\
\\
&\omega\varphi^{1}\varphi^{2}_{\omega}-\varphi^{2}_{\omega\omega}
+\omega^{-1}\varphi^{2}_{\omega}-\alpha_{2}\omega^{2}\varphi^{1}+
a_{1}=0,\\
\\
&\omega\varphi^{1}\varphi^{3}_{\omega}-\varphi^{3}_{\omega\omega}
+a_{2}\omega^{-2}\varphi^{2}\varphi^{3}-
\omega^{-1}\varphi^{3}_{\omega}+
(\alpha_{4}-a_{2}^{2}\omega^{-2})\varphi^{3}=0,\\
\\
&2\varphi^{1}+\omega\varphi^{1}_{\omega}=\alpha_{3}.
\end{array}
\end{equation}

\begin{equation}\label{e6.13}
\!\!\!\!\!\!\!\!\!\!\!
\begin{array}{ll}
\mbox{5.}
&(\varphi^{2}-a\varphi^{1})\varphi^{1}_{\omega}-
(1+a^{2})\varphi^{1}_{\omega\omega}-\varphi^{1}\varphi^{1}-
\varphi^{2}\varphi^{2}-ah_{\omega}-2h=0,\\
\\
&(\varphi^{2}-a\varphi^{1})\varphi^{2}_{\omega}-
(1+a^{2})\varphi^{2}_{\omega\omega}-2(a\varphi^{2}_{\omega}+
\varphi^{1}_{\omega})+h_{\omega}=0,\\
\\
&(\varphi^{2}-a\varphi^{1})\varphi^{3}_{\omega}-
(1+a^{2})\varphi^{3}_{\omega\omega}+2\varphi^{1}\varphi^{3}
-4\varphi^{3}+4a\varphi^{3}_{\omega}+\alpha_{5}=0,\\
\\
&\varphi^{2}_{\omega}-a\varphi^{1}_{\omega}=0.
\end{array}
\end{equation}

\begin{equation}\label{e6.14}
\!\!\!\!\!\!\!\!\!\!\!
\begin{array}{ll}
\mbox{6.}
&(\varphi^{2}-a\varphi^{1})\varphi^{1}_{\omega}-
(1+a^{2})\varphi^{1}_{\omega\omega}-\varphi^{1}\varphi^{1}-
\varphi^{2}\varphi^{2}-ah_{\omega}-2h=0,\\
\\
&(\varphi^{2}-a\varphi^{1})\varphi^{2}_{\omega}-
(1+a^{2})\varphi^{2}_{\omega\omega}-2(a\varphi^{2}_{\omega}+
\varphi^{1}_{\omega})+h_{\omega}=0,\\
\\
&(\varphi^{2}-a\varphi^{1})\varphi^{3}_{\omega}-
(1+a^{2})\varphi^{3}_{\omega\omega}+a_{1}\varphi^{1}=0,\\
\\
&\varphi^{2}_{\omega}-a\varphi^{1}_{\omega}=0.
\end{array}
\end{equation}

\begin{equation}\label{e6.15}
\!\!\!\!\!\!\!\!\!\!\!
\begin{array}{ll}
\mbox{7.}
&(\varphi^{2}-a\varphi^{1})\varphi^{1}_{\omega}-
(1+a^{2})\varphi^{1}_{\omega\omega}-\varphi^{1}\varphi^{1}-
\varphi^{2}\varphi^{2}-ah_{\omega}-2h=0,\\
\\
&(\varphi^{2}-a\varphi^{1})\varphi^{2}_{\omega}-
(1+a^{2})\varphi^{2}_{\omega\omega}-2(a\varphi^{2}_{\omega}+
\varphi^{1}_{\omega})+h_{\omega}=0,\\
\\
&(\varphi^{2}-a\varphi^{1})\varphi^{3}_{\omega}-
(1+a^{2})\varphi^{3}_{\omega\omega}+a_{1}\varphi^{1}\varphi^{3}
-a_{1}^{2}\varphi^{3}+2aa_{1}\varphi^{3}_{\omega}=0,\\
\\
&\varphi^{2}_{\omega}-a\varphi^{1}_{\omega}=0.
\end{array}
\end{equation}

Numeration of reduced systems (\ref{e6.9})--(\ref{e6.15})
corresponds to that of the ansatzes in Table \ref{table6.1}.
Let us integrate systems (\ref{e6.9})--(\ref{e6.15}) in such
cases when it is possible. Below, in this section,
$C_{k}=\mbox{const}$ $\:(k=\overline{1,6})$.

1. We failed to integrate system (\ref{e6.9}) in the general
case, but we managed to find the following particular solutions:
\begin{displaymath}
\!\!\!\!\!\!\!\!\!\!\!
\begin{array}{ll}
\mbox{a)}
&\varphi^{1}=-6\wp(\omega+C_{3},\frac{1}{3}(4-2C_{1}),C_{2})-2,\\
\\
&\varphi^{2}=\varphi^{3}=0,\quad h=2\varphi^{1}+C_{1};\\
\\
\mbox{b)}
&\varphi^{1}=-6C_{1}^{2}e^{2C_{1}\omega}
\wp(e^{C_{1}\omega}+C_{3},0,C_{2})+3C_{1}^{2}-2,\\
\\
&\varphi^{2}=5C_{1},\quad \varphi^{3}=0,\\
\\
&h=-12C_{1}^{2}e^{2C_{1}\omega}\wp(e^{C_{1}\omega}+C_{3},0,C_{2})-
2-\frac{13}{2}C_{1}^{2}-\frac{9}{4}C_{1}^{4};\\
\\
\mbox{c)}
&\varphi^{1}=C_{1},\quad \varphi^{2}=C_{2},\quad \varphi^{3}=0,
\quad h=-\frac{1}{2}(C_{1}^{2}+C_{2}^{2}).
\end{array}
\end{displaymath}
Here $\wp(\tau,\kappa_{1},\kappa_{2})$ is the Weierstrass function
that satisfies the equation (see \cite{kamke}):
\begin{equation}\label{e6.16}
(\wp_{\tau})^{2}=4\wp^{3}-\kappa_{1}\wp-\kappa_{2}.
\end{equation}

2. If $\alpha_{3}=0$, the last equation of (\ref{e6.10}) implies
that \mbox{$\varphi^{1}=C_{1}$}. It follows from the other
equations of (\ref{e6.10}) that

\begin{displaymath}
\begin{array}{l}
\varphi^{2}=C_{3}+C_{2}e^{C_{1}\omega}
-(a_{1}C_{1}^{-1}-\alpha_{2})\omega,
\end{array}
\end{displaymath}

\begin{displaymath}
\begin{array}{l}
h=C_{6}-\alpha_{2}C_{3}\omega-
\alpha_{2}C_{2}C_{1}^{-1}e^{C_{1}\omega}
+\frac{1}{2}\alpha_{2}(a_{1}C_{1}^{-1}-\alpha_{2})\omega^{2}
\end{array}
\end{displaymath}
if $\:C_{1}\not=0,\:$ and

\begin{displaymath}
\begin{array}{l}
\varphi^{2}=C_{3}+C_{2}\omega+\frac{1}{2}a_{1}\omega^{2},
\end{array}
\end{displaymath}

\begin{displaymath}
\begin{array}{l}
h=C_{6}-\alpha_{2}C_{3}\omega-\frac{1}{2}\alpha_{2}C_{2}
\omega^{2}-\frac{1}{6}\alpha_{2}a_{1}\omega^{3}
\end{array}
\end{displaymath}
if $\:C_{1}=0$.
The function $\varphi^{3}$ satisfies the equation

\begin{equation}\label{e6.17}
\varphi^{3}_{\omega\omega}-C_{1}\varphi^{3}_{\omega}+
(a_{2}^{2}-\alpha_{4}-a_{2}\varphi^{2})\varphi^{3}=0.
\end{equation}
We solve equation (\ref{e6.17}) for the following cases:

\vspace{1ex}

A. $C_{2}=a_{1}-\alpha_{2}C_{1}=0$:
\begin{displaymath}
\varphi^{3}=\left\{
\begin{array}{ll}
e^{\frac{1}{2}C_{1}\omega}\bigl(
C_{4}e^{\mu^{1/2}\omega}+C_{5}e^{-\mu^{1/2}\omega}\bigr),&
\mu>0,\\
\\
e^{\frac{1}{2}C_{1}\omega}\bigl(C_{4}+C_{5}\omega \bigr),&
\mu=0,\\
\\
e^{\frac{1}{2}C_{1}\omega}\bigl(
C_{4}\cos((-\mu)^{1/2}\omega)+C_{5}\sin((-\mu)^{1/2}\omega)\bigr),&
\mu<0,
\end{array} \right.
\end{displaymath}
where $\:\mu=\frac{1}{4}C_{1}^{2}-a_{2}^{2}+\alpha_{4}+a_{2}C_{3}$.

\vspace{1ex}

B. $C_{1}=a_{1}=0$, $\:C_{2}\not=0$ (\cite{kamke}):
\begin{displaymath}
\begin{array}{l}
\varphi^{3}=\xi^{1/2}Z_{1/3}\bigl(\frac{2}{3}(-a_{2}C_{2})^{1/2}
\xi^{3/2}\bigr),
\end{array}
\end{displaymath}
where
$\:\xi=\omega+(C_{3}a_{2}-a_{2}^{2}-\alpha_{4})/(a_{2}C_{2})$. Here
$Z_{\nu}(\tau)$ is the general solution of the Bessel equation
(\ref{e4.22}).

\vspace{1ex}

C. $C_{1}=0$, $\:a_{1}\not=0$ (\cite{kamke}):
\begin{displaymath}
\begin{array}{l}
\varphi^{3}=(\omega+C_{2}a_{1}^{-1})^{-1/2}W\bigl(\nu,
\frac{1}{4},(\frac{1}{2}a_{1}a_{2})^{-1/2}
(\omega+C_{2}a_{1}^{-1})^{2}\bigr),
\end{array}
\end{displaymath}
where $\:\nu=\frac{1}{4}(\frac{1}{2}a_{1}a_{2})^{-1/2}\bigl(
a_{2}^{2}-\alpha_{4}-a_{2}C_{3}+\frac{1}{2}a_{2}C_{3}^{2}
a_{1}^{-1}\bigr)$. Here
$W(\kappa,\mu,\tau)$ is the general solution of the Whittaker
equation (\ref{e4.21}).

\vspace{1ex}

D. $C_{1}\not=0$, $\:C_{2}\not=0$, $\:a_{1}-\alpha_{2}C_{1}=0$
(\cite{kamke}):
\begin{displaymath}
\varphi^{3}=e^{\frac{1}{2}C_{1}\omega}Z_{\nu}\bigl(
2C_{1}^{-1}(-a_{2}C_{2})^{1/2}e^{\frac{1}{2}C_{1}\omega}\bigr),
\end{displaymath}
where $\:\nu=C_{1}^{-1}\bigl(C_{1}^{2}+4(\alpha_{4}+a_{2}C_{3}-
a_{2}^{2})\bigr)^{1/2}$. Here
$Z_{\nu}(\tau)$ is the general solution of the Bessel equation
(\ref{e4.22}).

\vspace{1ex}

E. $C_{1}\not=0$, $\:a_{1}-\alpha_{2}C_{1}\not=0$, $\:C_{2}=0$
(\cite{kamke}):
\begin{displaymath}
\begin{array}{l}
\varphi^{3}=e^{\frac{1}{2}C_{1}\omega}\xi^{1/2}Z_{1/3}\Bigl(
\frac{2}{3}\bigl(a_{2}(a_{1}C_{1}^{-1}-\alpha_{2})\bigr)
^{1/2}\xi^{3/2}\Bigr),
\end{array}
\end{displaymath}
where $\:\xi=\omega+\bigl(a_{2}^{2}-\frac{1}{4}C_{1}^{2}-C_{3}a_{2}-
\alpha_{4}\bigr)/\bigl(a_{2}(a_{1}C_{1}^{-1}-\alpha_{2})\bigr)$.
Here $Z_{\nu}(\tau)$ is the general solution of the Bessel equation
(\ref{e4.22}).

If $\alpha_{3}\not=0$, then $\varphi^{1}=\alpha_{3}\omega$
(translating $\omega$, the integration constant can be made to
vanish),

\begin{displaymath}
\begin{array}{l}
\varphi^{2}=C_{1}+
C_{2}\int e^{\frac{1}{2}\alpha_{3}\omega^{2}}d\omega+
a_{1}\int e^{\frac{1}{2}\alpha_{3}\omega^{2}}\Bigl(
\int e^{-\frac{1}{2}\alpha_{3}\omega^{2}}d\omega\Bigr)d\omega+
\alpha_{2}\omega,
\end{array}
\end{displaymath}

\begin{displaymath}
\begin{array}{ll}
h=&\!\!\!
C_{3}-\frac{1}{2}(\alpha_{2}^{2}+\alpha_{3}^{2})\omega^{2}-
\alpha_{2}C_{1}\omega-\alpha_{2}C_{2}\Bigl(
\omega\int e^{\frac{1}{2}\alpha_{3}\omega^{2}}d\omega-
\alpha_{3}^{-1}e^{\frac{1}{2}\alpha_{3}\omega^{2}}\Bigr)-
\\ \\
&\alpha_{2}a_{1}\Bigl(
\omega\int e^{\frac{1}{2}\alpha_{3}\omega^{2}}
\bigl(\int
e^{-\frac{1}{2}\alpha_{3}\omega^{2}}d\omega\bigr)d\omega-
\alpha_{3}^{-1}e^{\frac{1}{2}\alpha_{3}\omega^{2}}
\int e^{-\frac{1}{2}\alpha_{3}\omega^{2}}d\omega+
\alpha_{3}^{-1}\omega\Bigr),
\end{array}
\end{displaymath}
and the function $\varphi^{3}$ satisfies the equation
\begin{equation}\label{e6.18}
\varphi^{3}_{\omega\omega}-\alpha_{3}\omega\varphi^{3}_{\omega}+
(a_{2}^{2}-\alpha_{4}-a_{2}\varphi^{2})\varphi^{3}=0.
\end{equation}
We managed to find a solution of (\ref{e6.18}) only for
the case $a_{1}=C_{2}=0$, i.e.,
\begin{displaymath}
\varphi^{3}=e^{\frac{1}{4}\alpha_{3}\omega^{2}}V\bigl(
\alpha_{3}^{1/2}(\omega+2a_{2}\alpha_{2}\alpha_{3}^{-2}),\nu
\bigr),
\end{displaymath}
where $\:\nu=4\alpha_{3}^{-1}\bigl(\alpha_{4}+a_{2}C_{1}-
a_{2}^{2}(\alpha_{2}^{2}\alpha_{3}^{-2}+1)\bigr)$.
Here $V(\tau,\nu)$ is the general solution of the Weber equation
\begin{equation}\label{e6.19}
4V_{\tau\tau}=(\tau^{2}+\nu)V.
\end{equation}

3. The general solution of system (\ref{e6.11}) has the form:
\begin{equation}\label{e6.20}
\begin{array}{l}
\varphi^{1}=C_{1}\omega^{-2}+\frac{1}{2}\alpha_{3},
\end{array}
\end{equation}
\begin{equation}\label{e6.21}
\begin{array}{ll}
\varphi^{2}=\!\!\!\!&C_{2}+C_{3}
\int \omega^{C_{1}+1}e^{\frac{1}{4}\alpha_{3}\omega^{2}}d\omega
-\frac{1}{2}\alpha_{2}\omega^{2}+
\\ \\
&\quad a_{1}\int \omega^{C_{1}+1}e^{\frac{1}{4}\alpha_{3}\omega^{2}}
\Bigl(\int \omega^{-C_{1}-1}e^{-\frac{1}{4}\alpha_{3}\omega^{2}}
d\omega\Bigr)d\omega,
\end{array}
\end{equation}

\begin{displaymath}
\begin{array}{ll}
\varphi^{3}=\!\!\!\!&C_{4}+C_{5}
\int \omega^{C_{1}-1}e^{\frac{1}{4}\alpha_{3}\omega^{2}}d\omega+
\\ \\
&\int \omega^{C_{1}-1}e^{\frac{1}{4}\alpha_{3}\omega^{2}}\Bigl(
\int \omega^{1-C_{1}}e^{-\frac{1}{4}\alpha_{3}\omega^{2}}
(\alpha_{5}+a_{2}\omega^{-2}\varphi^{2})d\omega\Bigr)d\omega,
\end{array}
\end{displaymath}

\begin{equation}\label{e6.22}
\begin{array}{ll}
h=\!\!\!\!&C_{6}-\frac{1}{8}\alpha_{3}^{2}\omega^{2}-
\frac{1}{2}C_{1}^{2}\omega^{-2}+
\int(\varphi^{2}(\omega))^{2}\omega^{-3}d\omega-
\alpha_{2}\int\omega^{-1}\varphi^{2}(\omega)d\omega.
\end{array}
\end{equation}

4. System (\ref{e6.12}) implies that the functions
$\varphi^{i}$ and $h$ are determined by
(\ref{e6.20})--(\ref{e6.22}),
and the function $\varphi^{3}$ satisfies the equation
\begin{equation}\label{e6.23}
\begin{array}{l}
\varphi^{3}_{\omega\omega}\!-\bigl((C_{1}\!-\!1)\omega^{-1}\!+
\frac{1}{2}\alpha_{3}\omega\bigr)\varphi^{3}_{\omega}+
\bigl(a_{2}\omega^{-2}(a_{2}\!-\!\varphi^{2})-\alpha_{4}\bigr)
\varphi^{3}=0.
\end{array}
\end{equation}
We managed to solve equation (\ref{e6.23}) in following cases:

\vspace{1ex}

A. $C_{3}=a_{1}=0$, $\:\alpha_{3}\not=0$:
\begin{displaymath}
\begin{array}{l}
\varphi^{3}=\omega^{\frac{1}{2}C_{1}-1}
e^{\frac{1}{8}\alpha_{3}\omega^{2}}
W(\kappa,\mu,\frac{1}{4}\alpha_{3}\omega^{2}),
\end{array}
\end{displaymath}
where
$\:\kappa=\frac{1}{4}\bigl(2-C_{1}-(4\alpha_{4}+2\alpha_{2}a_{2})
\alpha_{3}^{-1}\bigr),\:$
$\:\mu=\frac{1}{4}(C_{1}^{2}-4a_{2}^{2}+4a_{2}C_{2})^{1/2}$. Here
$W(\kappa,\mu,\tau)$ is the general solution of the Whittaker
equation (\ref{e4.21}).

Let $\alpha_{3}=0$, then
\begin{displaymath}
\varphi^{2}=\left\{
\begin{array}{ll}
C_{2}+C_{3}\ln\omega+\frac{1}{4}(a_{1}+2\alpha_{2})\omega^{2},&
C_{1}=-2,\\
\\
C_{2}+\frac{1}{2}C_{3}\omega^{2}+\frac{1}{2}a_{1}\omega^{2}
(\ln\omega-\frac{1}{2}),&
C_{1}=0,\\
\\
C_{2}+C_{3}(C_{1}+2)^{-1}\omega^{C_{1}+2}-
\frac{1}{2}C_{1}^{-1}(a_{1}-\alpha_{2}C_{1})\omega^{2},&
C_{1}\not=0,-2.
\end{array}  \right.
\end{displaymath}

\vspace{1ex}

B. $C_{3}=a_{1}-\alpha_{2}C_{1}=0$:
\begin{equation}\label{e6.24}
\varphi^{3}=\left\{
\begin{array}{ll}
\omega^{\frac{1}{2}C_{1}}Z_{\nu}(\mu^{1/2}\omega),&
\mu\not=0,\\
\\
\omega^{\frac{1}{2}C_{1}}(C_{5}\omega^{\nu}+C_{6}\omega^{-\nu}),&
\mu=0, \nu\not=0,\\
\\
\omega^{\frac{1}{2}C_{1}}(C_{5}+C_{6}\ln\omega),&
\mu=0, \nu=0,
\end{array} \right.
\end{equation}
where $\:\mu=-\alpha_{4},\:$
$\:\nu=\frac{1}{2}(C_{1}^{2}-4a_{2}^{2}+4a_{2}C_{2})^{1/2}$.
Here and below $Z_{\nu}(\tau)$ is the general solution of the
Bessel equation (\ref{e4.22}).

\vspace{1ex}

C. $C_{3}=0$, $\:C_{1}\not=0$:\quad $\varphi^{3}$ is determined by
(\ref{e6.24}), where
\begin{displaymath}
\begin{array}{l}
\mu=\frac{1}{2}a_{2}C_{1}^{-1}(a_{1}-\alpha_{2}C_{1})-\alpha_{4},
\quad \nu=\frac{1}{2}(C_{1}^{2}-4a_{2}^{2}+4a_{2}C_{2})^{1/2}.
\end{array}
\end{displaymath}

D. $C_{1}=a_{1}=0$:\quad $\varphi^{3}$ is determined by
(\ref{e6.24}), where
\begin{displaymath}
\begin{array}{l}
\mu=-\frac{1}{2}a_{2}C_{3}-\alpha_{4},
\quad \nu=(-a_{2}^{2}+a_{2}C_{2})^{1/2}.
\end{array}
\end{displaymath}

E. $C_{3}\not=0$, $\:C_{1}\not\in\{0;-2\}$,
$\:a_{2}(a_{1}-\alpha_{2}C_{1})-2\alpha_{4}C_{1}=0$:
\begin{displaymath}
\varphi^{3}=\omega^{\frac{1}{2}C_{1}}
Z_{\nu}(\mu\omega^{1+\frac{1}{2}C_{1}}),
\end{displaymath}
where $\:\mu=2C_{3}^{1/2}(C_{1}+2)^{-3/2},\:$
$\:\nu=(C_{1}+2)^{-1}(C_{1}^{2}-4a_{2}^{2}+4a_{2}C_{2})^{1/2}$.

\vspace{1ex}

F. $C_{1}=-2$, $\:C_{3}\not=0$,
$\:a_{2}(a_{1}+2\alpha_{2})+4\alpha_{4}=0$ (\cite{kamke}):
\begin{displaymath}
\begin{array}{l}
\varphi^{3}=\omega^{-1}\xi^{1/2}
Z_{1/3}(\frac{2}{3}C_{3}^{1/2}\xi^{3/2}),
\end{array}
\end{displaymath}
where $\:\xi=\ln\omega+C_{3}^{-1}(a_{2}^{2}-a_{2}C_{2}-1)$.

\vspace{1ex}

G. $C_{1}=2$, $\:C_{3}<0$, $\:1-a_{2}^{2}+a_{2}C_{2}\geq0$:
\begin{displaymath}
\begin{array}{l}
\varphi^{3}=W(\kappa,\mu,\frac{1}{2}(-C_{3})^{1/2}\omega^{2}),
\end{array}
\end{displaymath}
where \quad $\kappa=\frac{1}{8}(-C_{3})^{-1/2}(-4\alpha_{4}+a_{2}^{2}-
2\alpha_{2}a_{2})$,
\quad $\mu=\frac{1}{2}(1-a_{2}^{2}+a_{2}C_{2})^{1/2}$. Here
$W(\kappa,\mu,\tau)$ is the general solution of the Whittaker
equation (\ref{e4.21}).

\vspace{2ex}

5--7. Identical corollaries of system (\ref{e6.13}),
(\ref{e6.14}), and (\ref{e6.15}) are the equations
\begin{equation}\label{e6.24a}
\varphi^{2}=a\varphi^{1}+C_{1},
\end{equation}
\begin{equation}\label{e6.25}
h=a(1+a^{2})\varphi^{1}_{\omega}+(2+2a^{2}-aC_{1})\varphi^{1}+
C_{2},
\end{equation}
\begin{equation}\label{e6.26}
(1+a^{2})\varphi^{1}_{\omega\omega}\!+
\!(4a\!-\!C_{1})\varphi^{1}_{\omega}+\varphi^{1}\varphi^{1}\!\!+
\!4\varphi^{1}\!\!+\!(1+a^{2})^{-1}\!(C_{1}^{2}\!+\!2C_{2})\!=\!0.
\end{equation}
We found the following solutions of (\ref{e6.26}):

\vspace{2ex}

A. If $\:(1+a^{2})^{-1}(C_{1}^{2}+2C_{2})<4$:
\begin{equation}\label{e6.27}
 \varphi^{1}=\bigl(4-(1+a^{2})^{-1}(C_{1}^{2}+2C_{2})
\bigr)^{1/2}-2.
\end{equation}

\vspace{2ex}

B. If $\:C_{1}=4a$:
\begin{equation}\label{e6.28}
\varphi^{1}=-6\wp\left(\frac{\omega}{(1+a^{2})^{1/2}}+C_{4},\:
\frac{4}{3}-\frac{(C_{1}^{2}+2C_{2})}{3(1+a^{2})},\: C_{3}
\right)-2.
\end{equation}
Here and below $\wp(\tau,\kappa_{1},\kappa_{2})$
is the Weierstrass function satisfying equation
(\ref{e6.16}).
If \mbox{$\:C_{2}=2-6a^{2}\:$} and $\:C_{3}=0,\:$
a particular case of (\ref{e6.28}) is the function
\begin{equation}\label{e6.29}
\varphi^{1}=-6(1+a^{2})\omega^{2}-2
\end{equation}
(the constant $C_{4}$ is considered to vanish).

\vspace{2ex}

C. If $\:C_{1}\not=4a,\:$
$\:(1+a^{2})^{-1}(C_{1}^{2}+2C_{2})-4=-9\mu^{4}$:
\begin{equation}\label{e6.30}
\varphi^{1}=-6\mu^{2}e^{-2\xi}
\wp(e^{-\xi}+C_{4},0,C_{3})+3\mu^{2}-2,
\end{equation}
where $\:\xi=(1+a^{2})^{-1/2}\mu\omega,\:$
$\:\mu=\frac{1}{5}(4a-C_{1})(1+a^{2})^{-1/2}.\:$
If $\:C_{3}=0,\:$ a paticular case of (\ref{e6.30}) is the function
\begin{equation}\label{e6.32}
\varphi^{1}=-6\mu^{2}e^{-2\xi}(e^{-\xi}+C_{4})^{-2}+3\mu^{2}-2,
\end{equation}
where the constant $C_{4}$ is considered not to vanish.

The function $\varphi^{3}$ has to be found for systems
(\ref{e6.13}), (\ref{e6.14}), and (\ref{e6.15}) individually.

\vspace{2ex}

5. The function $\varphi^{3}$ satisfy the equation
\begin{displaymath}
(1+a^{2})\varphi^{3}_{\omega\omega}-
(C_{1}+4a)\varphi^{3}_{\omega}-(2\varphi^{1}-4)\varphi^{3}-
\alpha_{5}=0.
\end{displaymath}
If $\varphi^{1}$ is determined by (\ref{e6.27}), we obtain
\begin{displaymath}
\begin{array}{ll}
\varphi^{3}=\!\!\!\!&
\exp\bigl(\frac{1}{2}(1+a^{2})^{-1}(C_{1}+4a)\omega\bigr)\cdot\\
\\
&\left\{
\begin{array}{ll}
C_{5}\exp(\nu^{1/2}\omega)+C_{6}\exp(-\nu^{1/2}\omega),&
\nu>0\\
C_{5}\cos((-\nu)^{1/2}\omega)+C_{6}\sin((-\nu)^{1/2}\omega),&
\nu<0\\
C_{5}+C_{6}\omega,&
\nu=0\\
\end{array}
\right\}+
\\[1ex] \\
&\left\{
\begin{array}{ll}
-\alpha_{5}(2\varphi^{1}-4)^{-1},& 2\varphi^{1}-4\not=0
\\
-\alpha_{5}(4a+C_{1})^{-1}\omega, &2\varphi^{1}-4=0,
\quad C_{1}+4a\not=0
\\
\frac{1}{2}\alpha_{5}(1+a^{2})^{-1}\omega^{2},& 2\varphi^{1}-4=0,
\quad C_{1}+4a=0
\end{array} \right\},
\end{array}
\end{displaymath}
where $\:\nu=\frac{1}{4}(1+a^{2})^{-2}(C_{1}+4a)^{2}-
(1+a^{2})^{-1}(4-2\varphi^{1})$.

\vspace{2ex}

6. In this case $\varphi^{3}$ satisfy the equation
\begin{displaymath}
(1+a^{2})\varphi^{3}_{\omega\omega}-C_{1}\varphi^{3}_{\omega}=
a_{1}\varphi^{1}.
\end{displaymath}
Therefore,
\begin{displaymath}
\begin{array}{ll}
\varphi^{3}=\!\!\!\!&C_{5}+C_{6}\exp\bigl((1+a^{2})^{-1}C_{1}\omega\bigr)
+a_{1}C_{1}^{-1}\Bigl(\int\varphi^{1}(\omega)d\omega+\\
\\
&\exp\bigl((1+a^{2})^{-1}C_{1}\omega\bigr)
\int\exp\bigl(-(1+a^{2})^{-1}C_{1}\omega\bigr)\varphi^{1}(\omega)
d\omega\Bigr)
\end{array}
\end{displaymath}
for $C_{1}\not=0$, and
\begin{displaymath}
\begin{array}{l}
\varphi^{3}=C_{5}+C_{6}\omega+a_{1}(1+a^{2})^{-1}\bigl(\omega
\int\varphi^{1}(\omega)d\omega-
\int\omega\varphi^{1}(\omega)d\omega\bigr)
\end{array}
\end{displaymath}
for $C_{1}=0$.

\vspace{2ex}

7. The function $\varphi^{3}$ satisfy the equation
\begin{equation}\label{e6.33}
(1+a^{2})\varphi^{3}_{\omega\omega}-
(C_{1}+2a_{1}a)\varphi^{3}_{\omega}+
(a_{1}^{2}-a_{1}\varphi^{1})\varphi^{3}=0.
\end{equation}

\mbox{A.} If $\varphi^{1}$ is determined by (\ref{e6.27}), it
follows that
\begin{displaymath}
\begin{array}{ll}
\varphi^{3}=\!\!\!\!&\exp
\bigl(\frac{1}{2}(1+a^{2})^{-1}(C_{1}+2a_{1}a)\omega\bigr)\cdot\\
\\
&\left\{
\begin{array}{ll}
C_{5}\exp(\nu^{1/2}\omega)+C_{6}\exp(-\nu^{1/2}\omega),&
\nu>0\\
C_{5}\cos((-\nu)^{1/2}\omega)+C_{6}\sin((-\nu)^{1/2}\omega),&
\nu<0\\
C_{5}+C_{6}\omega,&
\nu=0\\
\end{array}
\right\},
\end{array}
\end{displaymath}
where $\:\nu=\frac{1}{4}(1+a^{2})^{-2}(C_{1}+2a_{1}a)^{2}-
(1+a^{2})^{-1}(a_{1}^{2}-a_{1}\varphi^{1})$.

\vspace{2ex}

\mbox{B.} If $\:C_{1}=4a,\:$ that is, $\varphi^{1}$ is determined by
(\ref{e6.27}), we obtain
\begin{displaymath}
\varphi^{3}=\exp\bigl(a(a_{1}+2)(1+a^{2})^{-1}\omega\bigr)
\theta(\tau),
\end{displaymath}
where $\:\tau=(1+a^{2})^{-1/2}\omega+C_{4}$. Here the function
$\theta=\theta(\tau)$ is the general solution of of the following
Lame equation (\cite{kamke}):
\begin{displaymath}
\theta_{\tau\tau}+\bigl(6a_{1}\wp(\tau)+a_{1}^{2}+2a_{1}-
a^{2}(2+a_{1})^{2}(1+a^{2})^{-1}\bigr)\theta=0
\end{displaymath}
with the Weierstrass function
\begin{displaymath}
\begin{array}{l}
\wp(\tau)=\wp\Bigl(\tau,\frac{1}{3}\bigl(4-(1+a^{2})^{-1}
(C_{1}^{2}+2C_{2})\bigr),C_{3}\Bigr).
\end{array}
\end{displaymath}

Consider the particular case when $\:C_{2}=2-6a^{2}\:$ and $\:C_{3}=0\:$
additionally, i.e., $\varphi^{1}$ can be given in form
(\ref{e6.29}). Depending on the values of $a$ and $a_{1}$, we
obtain the following expression for $\varphi^{3}$:

\vspace{1ex}

Case 1. $a_{1}\not=-2$, $\:a_{1}\not=2a^{2}$:
\begin{displaymath}
\varphi^{3}= \vert\omega\vert^{1/2}
exp\left(\frac{a(2+a_{1})}{1+a^{2}}\omega\right)
Z_{\nu}\left(\frac{\bigl((2+a_{1})(a_{1}-2a^{2})\bigr)^{1/2}}
{1+a^{2}}\omega\right),
\end{displaymath}
where $\:\nu=(\frac{1}{4}-6a_{1})^{1/2}$.

\vspace{1ex}

Case 2. $a_{1}=-2$: \quad
$\varphi^{3}=C_{5}\omega^{4}+C_{6}\omega^{-3}$.

\vspace{1ex}

Case 3. $a_{1}=2a^{2}$:

\vspace{1ex}

\quad Case 3.1. $48a^{2}<1$: \quad
$\varphi^{3}=\vert\omega\vert^{1/2}e^{2a\omega}\bigl(
C_{5}\omega^{\sigma}+C_{6}\omega^{-\sigma}\bigr)$, \\
where $\:\sigma=\frac{1}{2}\sqrt{1-48a^{2}}$.

\vspace{1ex}

\quad Case 3.2. $48a^{2}=1$, that is,
$a=\pm\frac{1}{12}\sqrt{3}$: \quad
$\varphi^{3}=\vert\omega\vert^{1/2}(C_{5}+C_{6}\ln\omega)$.

\vspace{1ex}

\quad Case 3.3. $48a^{2}>1$:\quad
$\varphi^{3}=\vert\omega\vert^{1/2}e^{2a\omega}\bigl(
C_{5}\cos(\gamma\ln\omega)+C_{6}\sin(\gamma\ln\omega)\bigr)$,\\
where $\:\gamma=\frac{1}{2}\sqrt{48a^{2}-1}$.

\vspace{2ex}

C. Let the conditions
\begin{displaymath}
C_{1}\not=4a, \quad
(1+a^{2})^{-1}(C_{1}^{2}+2C_{2})-4=-9\mu^{4}
\end{displaymath}
be satisfied, that is, let
$\varphi^{1}$ be determined by (\ref{e6.30}). Transforming the
variables in equation (\ref{e6.33}) by the formulas:
\begin{displaymath}
\begin{array}{l}
\varphi^{3}=\tau^{-1/2}
\exp\Bigl(\frac{1}{2}(C_{1}+2aa_{1})(1+a^{2})^{-1}\omega\Bigr)
\theta(\tau),
\end{array}
\end{displaymath}
\begin{displaymath}
\tau=\exp\bigl(-\mu(1+a^{2})^{-1/2}\omega\bigr),
\end{displaymath}
we obtain the following equation in the function
$\theta=\theta(\tau)$:
\begin{equation}\label{e6.34}
\tau^{2}\theta_{\tau\tau}+\bigl(6a_{1}\tau^{2}
\wp(\tau+C_{4},0,C_{3})+\sigma\bigr)\theta=0,
\end{equation}
where $\sigma=\mu^{-2}\bigl(a_{1}^{2}+2a_{1}-
\frac{1}{4}(1+a^{2})^{-1}(C_{1}^{2}+2aa_{1})^{2}\bigr)-3a_{1}+
\frac{1}{4}$.
If $\sigma=0$, equation (\ref{e6.34}) is a Lame equation.

In the particular case when $\varphi^{1}$ is determined by
(\ref{e6.32}), equation (\ref{e6.34}) has the form:
\begin{equation}\label{e6.35}
\tau^{2}(\tau+C_{4})^{2}\theta_{\tau\tau}+\bigl(6a_{1}\tau^{2}+
\sigma(\tau+C_{4})^{2}\bigr)\theta=0.
\end{equation}
By means of the following transformation of variables:
\begin{displaymath}
\theta=\vert\xi\vert^{\nu_{1}}\vert\xi-1\vert^{\nu_{2}}\psi(\xi),
\quad \xi=-C_{4}^{-1}\tau,
\end{displaymath}
where $\:\nu_{1}(\nu_{1}-1)+\sigma=0\:$ and
$\:\nu_{2}(\nu_{2}-1)+6a_{1}=0,\:$
equation (\ref{e6.35}) is reduced to a hypergeometric
equation of the form (see \cite{kamke}):
\begin{displaymath}
\xi(\xi-1)\psi_{\xi\xi}+\bigl(2(\nu_{1}+\nu_{2})\xi-2\nu_{1})
\psi_{\xi}+2\nu_{1}\nu_{2}\psi=0.
\end{displaymath}
If $\:\sigma=0,\:$ equation (\ref{e6.35}) implies that
\begin{displaymath}
(\tau+C_{4})^{2}\theta_{\tau\tau}+6a_{1}\theta=0.
\end{displaymath}
Therefore,
\begin{displaymath}
\theta=C_{5}\vert\tau+C_{4}\vert^{1/2-\nu}+
C_{6}\vert\tau+C_{4}\vert^{1/2+\nu}
\end{displaymath}
if $\:a_{1}<\frac{1}{24}\:,\:$ where $\:\nu=(\frac{1}{4}-6a_{1})^{1/2}\:$,
\begin{displaymath}
\theta=\vert\tau+C_{4}\vert^{1/2}\bigl(
C_{5}+C_{6}\ln\vert\tau+C_{4}\vert\bigr)
\end{displaymath}
if $\:a_{1}=\frac{1}{24}\:,\:$ and
\begin{displaymath}
\theta=\vert\tau+C_{4}\vert^{1/2}\bigl(
C_{5}\cos(\nu\ln\vert\tau+C_{4}\vert)+
C_{6}\sin(\nu\ln\vert\tau+C_{4}\vert)\bigr)
\end{displaymath}
if $\:a_{1}>\frac{1}{24}\:,\:$ where $\:\nu=(6a_{1}-\frac{1}{4})^{1/2}$.

\setcounter{equation}{0}

\section{
Exact solutions of system (\protect\ref{e2.9})}
\label{sec7}

Among the reduced systems from Sec.~\ref{sec2}, only particular
cases of system (\ref{e2.9}) have Lie symmetry operators that are
not induced by elements from $A(NS)$. Therefore, Lie reductions
of the other systems from Sec.~\ref{sec2} give only solutions
that can be obtained by means of reducing the NSEs with two- and
three-dimensional subalgebras of $A(NS)$.

Here we consider system (\ref{e2.9}) with $\rho^{i}$ vanishing.
As mentioned in Note \ref{n2.5}, in this case the
vector-function $\vec m$ has the form $\vec m=\eta(t)\vec e$,
where $\vec e=\mbox{\rm const}$, $\vert\vec e\vert=1$, and
\mbox{$\eta=\eta(t)=\vert\vec m(t)\vert\not=0$}. Without loss of
generality we can assume that $\vec e=(0,0,1)$, i.e.,
\begin{displaymath}
\vec m=(0,0,\eta(t)).
\end{displaymath}
For such vector $\vec m$, conditions (\ref{e2.5}) are satisfied
by the following vector $\vec n^{i}$:
\begin{displaymath}
\vec n^{1}=(1,0,0), \quad \vec n^{2}=(0,1,0).
\end{displaymath}
Therefore, ansatz (\ref{e2.4}) and system (\ref{e2.9}) can be
written, respectively, in the forms:

\begin{equation}\label{e7.1}
\begin{array}{l}
u^{1}=v^{1}, \quad u^{2}=v^{2}, \quad
u^{3}=\bigl(\eta(t)\bigr)^{-1}
\bigl(v^{3}+\eta_{t}(t)x_{3}\bigr),
\\[0.8em]
p=q-\frac{1}{2}\eta_{tt}(t)\bigl(\eta(t)\bigr)^{-1}x_{3}^{2},
\end{array}
\end{equation}
where $\:v=v(y_{1},y_{2},y_{3}),\:$ $\:q=q(y_{1},y_{2},y_{3}),\:$
$\:y_{i}=x_{i},\:$ $\:y_{3}=t\:$, and

\begin{equation}\label{e7.2}
\begin{array}{l}
v^{i}_{t}+v^{j}v^{i}_{j}-v^{i}_{jj}+q_{i}=0,
\\[0.8em]
v^{3}_{t}+v^{j}v^{3}_{j}-v^{3}_{jj}=0,
\\[0.8em]
v^{i}_{i}+\rho^{3}=0,
\end{array}
\end{equation}
where $\rho^{3}=\rho^{3}(t)=\eta_{t}/\eta$.

It was shown in Note \ref{n2.8} that there exists a local
transformation which make $\rho^{3}$ vanish. Therefore, we can
consider system (\ref{e7.2}) only with $\rho^{3}$ vanishing and
extend the obtained results in the case $\rho^{3}\not=0$ by means of
transformation (\ref{e2.12}).
However it will be sometimes convenient to investigate, at once,
system (\ref{e7.2}) with an arbitrary function $\rho^{3}$.

The MIA of (\ref{e7.2}) with $\rho^{3}=0$ is given by the algebra
\begin{displaymath}
B=<R_{3}(\bar\psi),\: Z^{1}(\lambda),\: D^{1}_{3},\:
\partial_{t},\: J^{1}_{12},\: \partial_{v^{3}},\:
v^{3}\partial_{v^{3}}>
\end{displaymath}
(see notations in Subsec. \ref{subsec2.1}). We construct complete
sets of inequivalent one-dimensional subalgebras of $B$ and choose
such algebras, among these subalgebras, that can be used to reduce
system (\ref{e7.2}) and do not lie in the linear span of the
operators
$\:R_{3}(\bar\psi),\:$ $\:Z^{1}(\lambda),\:$
$J^{1}_{12},\:$  i.e., the operators which are induced by operators from
$A(NS)$ for arbitrary $\rho^{3}$. As a result we obtain the
following algebras (more exactly, the following classes of algebras):

The one-dimentional subalgebras:

\vspace{1ex}

1. $B^{1}_{1}=<D^{1}_{3}+2\kappa J^{1}_{12}+2\gamma v^{3}\partial_{v^{3}}+
2\beta\partial_{v^{3}}>,\:$ where
$\:\gamma\beta=0.$

\vspace{1ex}

2. $B^{1}_{2}=<\partial_{t}+\kappa J^{1}_{12}+\gamma v^{3}\partial_{v^{3}}+
\beta\partial_{v^{3}}>,\:$
where
$\:\gamma\beta=0,$ $\:\kappa\in\{0;1\}.$

\vspace{1ex}

3. $B^{1}_{3}=<J^{1}_{12}+\gamma v^{3}\partial_{v^{3}}+Z^{1}(\lambda(t))>,
 \:$ where
 $\:\gamma\not=0,$
$\: \lambda\in C^{\infty}((t_{0},t_{1}),\R).$

\vspace{1ex}

4. $B^{1}_{4}=<R_{3}(\bar\psi(t))+\gamma v^{3}\partial_{v^{3}}>,\:$
 where \vspace{1ex}
$\:\gamma\not=0,$
\mbox{$\:\bar\psi(t)=(\psi^{1}(t),\psi^{2}(t))\not=(0,0)$}
\mbox{$\forall t\in(t_{0},t_{1}),$}

\noindent
\hspace*{10mm}
$\:\psi^{i}\in C^{\infty}((t_{0},t_{1}),\R).$

\vspace{1ex}

The two-dimentional subalgebras:

\vspace{1ex}

1. $B^{2}_{1}=<\partial_{t}+\beta_{2}\partial_{v^{3}}, \;
D^{1}_{3}+\kappa J^{1}_{12}+\gamma v^{3}\partial_{v^{3}}+
\beta_{1}\partial_{v^{3}}>,\:$
\vspace{1ex}
where
$\:\gamma\beta_{1}=0,$
$\:(\gamma-2)\beta_{2}=0.$

\vspace{1ex}

2.\vspace{1ex} $B^{2}_{2}=<D^{1}_{3}+2\gamma_{1} v^{3}\partial_{v^{3}}+
2\beta_{1}\partial_{v^{3}}, \quad
J^{1}_{12}+\gamma_{2} v^{3}\partial_{v^{3}}+
\beta_{2}\partial_{v^{3}}+Z^{1}(\varepsilon |t|^{-1})>,\;$\quad
where

\noindent
\hspace*{10mm}
$\:\gamma_{1}\beta_{1}=0,$ $\:\gamma_{2}\beta_{2}=0,$
$\:\gamma_{1}\beta_{2}-\gamma_{2}\beta_{1}=0.$

\vspace{1ex}

3.
\vspace{1ex}
$B^{2}_{3}=<D^{1}_{3}+2\kappa J^{1}_{12}+2\gamma_{1} v^{3}\partial_{v^{3}}+
2\beta_{1}\partial_{v^{3}}, \quad
R_{3}(|t|^{\sigma+1/2}\cos\tau,
|t|^{\sigma+1/2}\sin\tau)+
\gamma_{2} v^{3}\partial_{v^{3}}+$

\noindent
\hspace*{10mm}
$\beta_{2}\partial_{v^{3}}+
Z^{1}(\varepsilon |t|^{\sigma-1})>,\:$
\vspace{1ex}
where
$\:\tau=\kappa\ln|t|,$
$\:(\gamma_{1}+\sigma)\beta_{1}-\gamma_{2}\beta_{1}=0,$
$\:\sigma\gamma_{2}=0,$
$\:\varepsilon\sigma=0.$

\vspace{1ex}

4.~$B^{2}_{4}=<\partial_{t}+\gamma_{1} v^{3}\partial_{v^{3}}+
\beta_{1}\partial_{v^{3}}, \;
J^{1}_{12}+\gamma_{2} v^{3}\partial_{v^{3}}+
\beta_{2}\partial_{v^{3}}+Z^{1}(\varepsilon)>,\:$
\vspace{1ex}
where
$\:\gamma_{1}\beta_{1}=0,$ 

\noindent
\hspace*{10mm}
$\:\gamma_{2}\beta_{2}=0,$
$\:\gamma_{1}\beta_{2}-\gamma_{2}\beta_{1}=0.$

\vspace{1ex}

5.
\vspace{1ex}
$B^{2}_{5}=<\partial_{t}+\kappa J^{1}_{12}+
\gamma_{1} v^{3}\partial_{v^{3}}+
\beta_{1}\partial_{v^{3}}, \;
R_{3}(e^{\sigma t}\cos\kappa t,
e^{\sigma t}\sin\kappa t)
+Z^{1}(\varepsilon e^{\sigma t})+
\gamma_{2} v^{3}\partial_{v^{3}}+$

\noindent
\hspace*{10mm}
$\beta_{2}\partial_{v^{3}}>,\:$
where
$\:(\gamma_{1}+\sigma)\beta_{1}-\gamma_{2}\beta_{1}=0,$
$\:\sigma\gamma_{2}=0,$
$\:\varepsilon\sigma=0.$

\vspace{1ex}

6.~$B^{2}_{6}=<R_{3}(\bar\psi^{1})+\gamma v^{3}\partial_{v^{3}}, \;
R_{3}(\bar\psi^{2})>,\:$
\vspace{1ex}
where
\mbox{$\:\bar\psi^{i}=(\psi^{i1}(t),\psi^{i2}(t))\not=(0,0)$}
$\forall t\in(t_{0},t_{1}),$

\noindent
\hspace*{10mm}
$\:\psi^{ij}\in C^{\infty}((t_{0},t_{1}),\R),$
$\:\bar\psi^{1}_{tt}\cdot\bar\psi^{2}-
\bar\psi^{1}\cdot\bar\psi^{2}_{tt}=0,$
$\:\gamma\not=0.\:$
\vspace{1ex}
Hereafter $\:\bar\psi^{1}\cdot\bar\psi^{2}:=\psi^{1i}\psi^{2i}.$

\vspace{1ex}

Let us reduce system (\ref{e7.2}) to systems of PDEs in two
independent variables. With the algebras $B^{1}_{1}$--$B^{1}_{4}$
we can construct the following complete set of Lie ansatzes of
codimension 1 for system (\ref{e7.2}) with $\rho^{3}=0$:

\begin{equation}\label{e7.3}
\!\!\!\!\!\!\!\!\!\!\!\!\!
\begin{array}{llcl}
\mbox{1.}
& v^{1} \!\!\!\!& = \!\!\!\!&
|t|^{-1/2}(w^{1}\cos\tau-w^{2}\sin\tau)+\frac{1}{2}y_{1}t^{-1}-
\kappa y_{2}t^{-1},
\\[0.8em]
& v^{2} \!\!\!\!& = \!\!\!\!&
|t|^{-1/2}(w^{1}\sin\tau+w^{2}\cos\tau)+\frac{1}{2}y_{2}t^{-1}+
\kappa y_{1}t^{-1},
\\[0.8em]
& v^{3} \!\!\!\!& = \!\!\!\!&
|t|^{\gamma}w^{3}+\beta\ln|t|,
\\[0.8em]
& q     \!\!\!\!& = \!\!\!\!&
|t|^{-1}s + \frac{1}{2}(\kappa^{2}+\frac{1}{4})t^{-2}r^{2},
\end{array}
\end{equation}
where
$\:\tau = \kappa\ln|t|,$ $\:\gamma\beta=0,$
\begin{displaymath}
z_{1}=|t|^{-1/2}( y_{1}\cos\tau+y_{2}\sin\tau),
\quad
z_{2}=|t|^{-1/2}(-y_{1}\sin\tau+y_{2}\cos\tau).
\quad
\end{displaymath}
Here and below $\:w^{a} = w^{a}(z_{1},z_{2})$,
$\:s = s(z_{1},z_{2})$,
$\:r = ( y^{2}_{1} + y^{2}_{2} )^{1/2}$.

\begin{equation}\label{e7.4}
\!\!\!\!\!\!\!\!\!\!\!\!\!
\begin{array}{llcl}
\mbox{2.}
& v^{1} \!\!\!\!& = \!\!\!\!&
w^{1}\cos\kappa t-w^{2}\sin\kappa t-\kappa y_{2},
\\[0.8em]
& v^{2} \!\!\!\!& = \!\!\!\!&
w^{1}\sin\kappa t+w^{2}\cos\kappa t+\kappa y_{1},
\\[0.8em]
& v^{3} \!\!\!\!& = \!\!\!\!&
w^{3}e^{\gamma t}+\beta t,
\\[0.8em]
& q     \!\!\!\!& = \!\!\!\!&
s+\frac{1}{2}\kappa^{2}r^{2},
\end{array}
\end{equation}
where
$\:\kappa\in\{0;1\},$ $\:\gamma\beta=0,$
\begin{displaymath}
z_{1}=y_{1}\cos\kappa t+y_{2}\sin\kappa t,
 \quad
z_{2}=-y_{1}\sin\kappa t+y_{2}\cos\kappa t.
\end{displaymath}

\begin{equation}\label{e7.5}
\!\!\!\!\!\!\!\!\!\!\!\!\!
\begin{array}{llcl}
\mbox{3.}
& v^{1}\!\!\!\!& = \!\!\!\!&
y_{1}r^{-1}w^{3} - y_{2}r^{-2}w^{1}-\gamma y_{2}r^{-2},
\\[0.8em]
& v^{2} \!\!\!\!& = \!\!\!\!&
y_{2}r^{-1}w^{3} + y_{1}r^{-2}w^{1}+\gamma y_{1}r^{-2},
\\[0.8em]
& v^{3} \!\!\!\!& = \!\!\!\!&
w^{2}e^{\gamma\arctan y_{2}/y_{1}},
\\[0.8em]
& q     \!\!\!\!& = \!\!\!\!&
s+\lambda(t)\arctan y_{2}/y_{1},
\end{array}
\end{equation}
where
$\;z_{1}=t,$ $\:z_{2}=r,$ $\:\gamma\not=0,$
$\:\lambda\in C^{\infty}((t_{0},t_{1}),\R).$

\begin{equation}\label{e7.6}
\!\!\!\!\!\!\!\!\!\!\!\!\!
\begin{array}{llcl}
\mbox{4.}
& \bar v \!\!\!\!& = \!\!\!\!&
(\bar\psi\cdot\bar\psi)^{-1}\Bigl((w^{1}+\gamma)\bar\psi+w^{3}\bar\theta+
(\bar\psi\cdot\bar y)\bar\psi_{t}-z_{2}\bar\theta_{t}\Bigr)
\\[0.8em]
& v^{3} \!\!\!\!& = \!\!\!\!&
w^{2}\exp\bigl(\gamma(\bar\psi\cdot\bar\psi)^{-1}
(\bar\psi\cdot\bar y)\bigr)
\\[0.8em]
& q     \!\!\!\!& = \!\!\!\!&
s-(\bar\psi\cdot\bar\psi)^{-1}(\bar\psi_{tt}\cdot\bar y)
(\bar\psi\cdot\bar y)+\frac{1}{2}
(\bar\psi\cdot\bar\psi)^{-2}(\bar\psi_{tt}\cdot\bar\psi)
(\bar\psi\cdot\bar y)^{2},
\end{array}
\end{equation}
where
$\;z_{1}=t,$ $\:z_{2}=(\bar\theta\cdot\bar y),$ $\:\gamma\not=0,$
$\:\bar v=(v^{1},v^{2}),$ $\:\bar y=(y_{1},y_{2}),$
\mbox{$\:\psi^{i}\in C^{\infty}((t_{0},t_{1}),\R),$}
$\:\bar\theta=(-\psi^{2},\psi^{1}).$

\vspace{1ex}

Substituting ansatzes (\ref{e7.3}) and (\ref{e7.4}) into system
(\ref{e7.2}) with $\rho^{3}=0$, we obtain a reduced system
of the form (\ref{e6.1}), where
\[
\!\!\!\!
\begin{array}{l}
\alpha_{1}=0, \quad \alpha_{2}=-1, \quad \alpha_{3}=-2\kappa, \quad
\alpha_{4}=\gamma, \quad \alpha_{5}=\beta \quad \mbox{if} \quad t>0 \quad
\mbox{and}
\\[0.7em]
\alpha_{1}=0, \quad \alpha_{2}=1, \quad \alpha_{3}=2\kappa, \quad
\alpha_{4}=-\gamma, \quad \alpha_{5}=-\beta \quad \mbox{if} \quad t<0
\end{array}
\]
for ansatz (\ref{e7.3}) and
\[
\alpha_{1}=0, \quad \alpha_{2}=0, \quad \alpha_{3}=-2\kappa, \quad
\alpha_{4}=\gamma, \quad \alpha_{5}=\beta
\]
for ansatz (\ref{e7.4}). System (\ref{e6.1}) is investigated in
Sec.~\ref{sec6} in detail.

Because the form of ansatzes (\ref{e7.3}) is not changed after
transformation (\ref{e2.12}), it is convinient to substitute their
into a system of form (\ref{e7.2}) with an arbitrary function
$\rho^{3}$. As a result of substituting, we obtain the following
reduced systems:

\begin{equation}\label{e7.7}
\!\!\!\!\!\!\!\!\!\!\!\!\!
\begin{array}{ll}
\mbox{3.}
& w^{3}_{1}+w^{3}w^{3}_{2}-z_{2}^{-3}(w^{1}+\gamma)^{2}-
(w^{3}_{22}+z_{2}^{-1}w^{3}_{2}-z_{2}^{-2}w^{3})+s_{2}=0,
\\[0.8em]
& w^{1}_{1}+w^{3}w^{1}_{2}-w^{1}_{22}+z_{2}^{-1}w^{1}_{2}+\lambda=0,
\\[0.8em]
& w^{2}_{1}+w^{3}w^{2}_{2}-w^{2}_{22}-z_{2}^{-1}w^{2}_{2}+
\gamma z_{2}^{-2}w^{1}w^{2}=0,
\\[0.8em]
& w^{3}_{2}+z_{2}^{-1}w^{3}=-\eta_{1}/\eta.
\end{array}
\end{equation}

\begin{equation}\label{e7.8}
\!\!\!\!\!\!\!\!\!\!\!\!\!
\begin{array}{ll}
\mbox{4.}
&w^{1}_{1}+w^{3}w^{1}_{2}-(\bar\psi\cdot\bar\psi)w^{1}_{22}=0,
\\[0.8em]
&w^{3}_{1}+w^{3}w^{3}_{2}-(\bar\psi\cdot\bar\psi)w^{3}_{22}+
(\bar\psi\cdot\bar\psi)s_{2}+
2(w^{1}+\gamma)(\bar\psi\cdot\bar\theta)(\bar\psi\cdot\bar\psi)^{-1}-
\\[0.8em]
&\quad 2(\bar\psi_{t}\cdot\bar\psi)(\bar\psi\cdot\bar\psi)^{-1}w^{3}+
(2\bar\psi_{t}\cdot\bar\psi_{t}-\bar\psi_{tt}\cdot\bar\psi)
(\bar\psi\cdot\bar\psi)^{-1}z_{2}=0,
\\[0.8em]
&w^{2}_{1}+w^{3}w^{2}_{2}-(\bar\psi\cdot\bar\psi)w^{2}_{22}+
\gamma(\bar\psi\cdot\bar\psi)^{-1}\bigl(
w^{1}+(\bar\psi_{t}\cdot\bar\theta)(\bar\psi\cdot\bar\psi)^{-1}z_{2}
\bigr)w^{2}=0,
\\[0.8em]
&w^{3}_{2}+\eta_{t}/\eta=0.
\end{array}
\end{equation}
Unlike systems 8 and 9 from Subsec.~\ref{subsec3.2}, systems (\ref{e7.7})
and (\ref{e7.8}) are not reduced to linear systems of PDEs.

Let us investigate system (\ref{e7.7}). The last equation of (\ref{e7.7})
immediately gives

\begin{equation}\label{e7.9}
\!\!\!\!
\begin{array}{l}
(w^{3}_{2}+z_{2}^{-1}w^{3})_{2}=
w^{3}_{22}+z_{2}^{-1}w^{3}_{2}-z_{2}^{-2}w^{3}=0, \quad
w^{3}=(\chi-1)z_{2}^{-1}-\frac{1}{2}\eta_{t}\eta^{-1}z_{2},
\end{array}
\end{equation}
where $\:\chi=\chi(t)\:$ is an arbitrary differentiable function of
$t=z_{2}$. Then it follows from the first equation of (\ref{e7.7})
that
\[
\!\!\!\!
\begin{array}{l}
s=\int z_{2}^{-3}(w^{1}+\gamma)^{2}dz_{2}-
\frac{1}{2}(\chi-1)^{2}z_{2}^{-2}+
\frac{1}{4}z_{2}^{2}
\Bigl((\eta_{t}/\eta)_{t}-\frac{1}{2}(\eta_{t}/\eta)^{2}\Bigr)-
\chi_{t}\ln|z_{2}|.
\end{array}
\]
Substituting (\ref{e7.9}) into the remaining equations of (\ref{e7.7}),
we get

\begin{equation}\label{e7.10}
\!\!\!\!
\begin{array}{l}
w^{1}_{1}-w^{1}_{22}+
\bigl(\chi z_{2}^{-1}-\frac{1}{2}\eta_{t}\eta^{-1}z_{2}\bigr)
w^{1}_{2}+\lambda=0,
\\[0.8em]
w^{2}_{1}-w^{2}_{22}+
\bigl((\chi-2)z_{2}^{-1}-\frac{1}{2}\eta_{t}\eta^{-1}z_{2}\bigr)
w^{2}_{2}+\gamma z_{2}^{-2}w^{1}w^{2}=0.
\end{array}
\end{equation}
By means of changing the independent variables

\begin{equation}\label{e7.11}
\begin{array}{l}
\tau=\int|\eta(t)|dt, \quad z=|\eta(t)|^{1/2}z_{2},
\end{array}
\end{equation}
system (\ref{e7.10}) can be transformed to a system of a simpler form:

\begin{equation}\label{e7.12}
\!\!\!\!
\begin{array}{l}
w^{1}_{\tau}-w^{1}_{zz}+
\hat\chi z^{-1}w^{2}_{z}+\hat\lambda|\hat\eta|^{-1}=0,
\\[0.8em]
w^{2}_{\tau}-w^{2}_{zz}+(\hat\chi-2)z^{-1}w^{2}_{z}
+\gamma z^{-2}w^{1}w^{2}=0,
\end{array}
\end{equation}
where $\hat\eta(\tau)=\eta(t)$, $\hat\chi(\tau)=\chi(t)$,
and $\hat\lambda(\tau)=\lambda(t)$.

If $\:\lambda(t)=-2C\eta(t)(\chi(t)-1)\:$ for some fixed constant
$C$, particular solutions of (\ref{e7.10}) are functions
\[
\!\!\!\!
\begin{array}{l}
w^{1}=C\eta(t)z_{2}^{2}, \quad
w^{2}=f(z_{1},z_{2})\exp\bigl(\gamma C\int\eta(t)dt\bigr),
\end{array}
\]
where $f$ is an arbitrary solution of the following equation

\begin{equation}\label{e7.13}
\!\!\!\!
\begin{array}{l}
f_{1}-f_{22}+
\bigl((\chi-2)z_{2}^{-1}-\frac{1}{2}\eta_{t}\eta^{-1}z_{2}\bigr)
f_{2}=0.
\end{array}
\end{equation}
In the variables from (\ref{e7.11}), equation (\ref{e7.13}) has form
(\ref{e5.22}) with \mbox{$\:\tilde\eta(\tau)=\chi(t)-2$}.

In the case
$\:\lambda(t)=8C(\chi(t)-1)\eta(t)\int\eta(t)(\chi(t)-3)dt$
\mbox{($C=\mbox{\rm const}$)}, particular solutions of (\ref{e7.10})
are functions
\[
\!\!\!\!
\begin{array}{l}
w^{1}=C\Bigl((\eta(t))^{2}z_{2}^{4}-
4z_{2}^{2}\eta(t)\int\eta(t)(\chi(t)-3)dt\Bigr),
\\[0.8em]
w^{2}=f(z_{1},z_{2})\exp
\bigl(\frac{1}{2}(\gamma C)^{1/2}\eta(t)z_{2}^{2}+\xi(t)\bigr),
\end{array}
\]
where $\:\xi(t)=-(\gamma C)^{1/2}\int\eta(t)(\chi(t)-3)dt+
4\gamma C\int\eta(t)\bigl(\int\eta(t)(\chi(t)-3)dt\bigr)dt$
and $f$ is an arbitrary solution of the following equation

\begin{equation}\label{e7.14}
\!\!\!\!
\begin{array}{l}
f_{1}-f_{22}+
\bigl((\chi-2)z_{2}^{-1}-(\frac{1}{2}\eta_{t}\eta^{-1}+2(\gamma C)^{1/2})
z_{2}\bigr)f_{2}=0.
\end{array}
\end{equation}
After the change of the independent variables
\[
\!\!\!\!
\begin{array}{l}
\tau=\int|\eta(t)|\exp\bigl(4(\gamma C)^{1/2}\int\eta(t)dt\bigr)dt,
\quad
z=|\eta(t)|^{1/2}\exp\bigl(2(\gamma C)^{1/2}\int\eta(t)dt\bigr)z_{2}
\end{array}
\]
in (\ref{e7.14}), we obtain equation (\ref{e5.22}) with
\mbox{$\:\tilde\eta(\tau)=\chi(t)-2$} again.

Let us continue to system (\ref{e7.8}). The last equation of
(\ref{e7.8}) integrates with respect to $z_{2}$ to the following
expression:
$\:w^{3}=-\eta_{t}\eta^{-1}z_{2}+\chi.\:$
Here $\chi=\chi(t)$ is an differentiable function of
\mbox{$z_{1}=y_{3}=t$}. Let us make the transformation from the
symmetry group of (\ref{e7.2}):
\[
\!\!\!\!
\begin{array}{l}
\bar{\tilde v}(t,\bar y)=\bar v(t,\bar y-\bar\xi(t))+\bar\xi_{t}(t),
\quad
\tilde v^{3}=v^{3},
\quad
\tilde q(t,\bar y)=q(t,\bar y-\bar\xi(t))-\bar\xi_{tt}(t)\cdot\bar y,
\end{array}
\]
where $\:\bar\xi_{tt}\cdot\bar\psi-\bar\xi\cdot\bar\psi_{tt}=0\:$ and
\[
\bar\xi_{t}\cdot\bar\theta+\chi+
\eta_{t}\eta^{-1}(\bar\xi\cdot\bar\theta)-
|\bar\psi|^{-2}(\bar\xi\cdot\bar\psi)(\bar\psi_{t}\cdot\bar\theta)+
|\bar\psi|^{-2}(\bar\xi\cdot\bar\theta)(\bar\theta_{t}\cdot\bar\theta)=0.
\]
Hereafter $\:|\bar\psi|^{2}=\bar\psi\cdot\bar\psi.\:$
This transformation does not modify ansatz (\ref{e7.6}), but it makes

the function $\chi$ vanish, i.e.,
\mbox{$\tilde w^{3}=-\eta_{t}\eta^{-1}z_{2}$}. Therefore, without loss
of generality we may assume, at once, that
\mbox{$w^{3}=-\eta_{t}\eta^{-1}z_{2}$}.

Substituting the expression for $w^{3}$ in the other equations of
(\ref{e7.8}), we obtain that
\[
\!\!\!\!
\begin{array}{ll}
s=\!\!\!\!&z_{2}^{2}|\bar\psi|^{-2}\Bigl(\bigl(
\frac{1}{2}\bar\psi_{tt}\cdot\bar\psi-\bar\psi_{t}\cdot\bar\psi_{t}-
(\bar\psi_{t}\cdot\bar\psi)\eta_{t}\eta^{-1}\bigr)|\bar\psi|^{-2}+
\frac{1}{2}\eta_{tt}\eta^{-1}-(\eta_{t})^{2}\eta^{-2}\Bigr)-
\\[0.9em]
\!\!\!\!&2(\bar\psi_{t}\cdot\bar\theta)|\bar\psi|^{-2}
\int w^{1}(z_{1},z_{2})dz_{2},
\end{array}
\]
\begin{equation}\label{e7.15}
\!\!\!\!
\begin{array}{l}
w^{1}_{1}-\eta_{1}\eta^{-1}z_{2}w^{1}_{2}-|\bar\psi|^{2}w^{1}_{22}=0,
\\[0.8em]
w^{2}_{1}-\eta_{1}\eta^{-1}z_{2}w^{2}_{2}-|\bar\psi|^{2}w^{2}_{22}+
\gamma|\bar\psi|^{-2}\bigl(2(\bar\psi_{t}\cdot\bar\theta)
|\bar\psi|^{-2}z_{2}+w^{1}\bigr)w^{2}=0.
\end{array}
\end{equation}
The change of the independent variables
\[
\!\!\!\!
\begin{array}{l}
\tau=\int(\eta(t)|\bar\psi|)^{2}dt, \quad
z=\eta(t)z_{2}
\end{array}
\]
reduces system (\ref{e7.15}) to the following form:
\begin{equation}\label{e7.16}
\!\!\!\!
\begin{array}{l}
w^{1}_{\tau}-w^{1}_{zz}=0,
\\[0.8em]
w^{2}_{\tau}-w^{2}_{zz}+
\gamma|\bar{\hat\psi}|^{-4}\hat\eta^{-2}
\bigl(2(\bar{\hat\psi}_{t}\cdot\bar{\hat\theta})\hat\eta z
+w^{1}\bigr)w^{2}=0,
\end{array}
\end{equation}
where $\:\bar{\hat\psi}(\tau)=\bar\psi(t),\:$
$\:\bar{\hat\theta}(\tau)=\bar\theta(t),\:$
$\:\hat\eta(\tau)=\eta(t)$.

Particular solutions of (\ref{e7.15}) are the functions
\[
\!\!\!\!
\begin{array}{l}
w^{1}=C_{1}+C_{2}\eta(t)z_{2}+C_{3}
\bigl(\frac{1}{2}(\eta(t)z_{2})^{2}+\int(\eta(t)|\bar\psi|)^{2}dt\bigr),
\\[0.8em]
w^{2}=f(t,z_{2})\exp\bigl(\xi^{2}(t)z_{2}^{2}+\xi^{1}(t)z_{2}+
\xi^{0}(t)\bigr),
\end{array}
\]
where $\:(\xi^{2}(t),\xi^{1}(t),\xi^{0}(t))\:$ is a particular
solution of the system of ODEs:
\[
\!\!\!\!
\begin{array}{l}
\xi^{2}_{t}-2\eta_{t}\eta^{-1}\xi^{2}-4|\bar\psi|^{2}(\xi^{2})^{2}+
\frac{1}{2}C_{3}\gamma \eta^{2}|\bar\psi|^{-2}=0,
\\[0.8em]
\xi^{1}_{t}-\eta_{t}\eta^{-1}\xi^{1}-4|\bar\psi|^{2}\xi^{2}\xi^{1}+
2\gamma(\bar\psi_{t}\cdot\bar\theta)|\bar\psi|^{-4}+
C_{2}\gamma \eta|\bar\psi|^{-2}=0,
\\[0.8em]
\xi^{0}_{t}-2|\bar\psi|^{2}\xi^{2}-|\bar\psi|^{2}(\xi^{1})^{2}+
\gamma\bigl(C_{1}+C_{3}\int(\eta(t)|\bar\psi|)^{2}dt\bigr)
|\bar\psi|^{-2}=0,
\end{array}
\]
and $f$ is an arbitrary solution of the following equation

\begin{equation}\label{e7.17}
\!\!\!\!
\begin{array}{l}
f_{1}-|\bar\psi|^{2}f_{22}+
\bigl((\eta_{t}\eta^{-1}+4|\bar\psi|^{2}\xi^{2})z_{2}+
2|\bar\psi|^{2}\xi^{1}\bigr)f_{2}=0.
\end{array}
\end{equation}
Equation (\ref{e7.17}) is reduced by means of a local transformation
of the independent variables to the heat equation.

Consider the Lie reductions of system (\ref{e7.2}) to systems of ODEs.
The second basis operator of the each algebra $\:B^{2}_{k}$,
$k=\overline{1,5}\:$ induces, for the reduced system obtained from
system (\ref{e7.2}) by means of the first basis operator, either
a Lie symmetry operator from Table 2 or a operator giving
a ansatz of form (\ref{e6.5}). Therefore, the Lie reduction of system
(\ref{e7.2}) with the algebras \mbox{$B^{2}_{1}-B^{2}_{5}$} gives only
solutions that can be constructed for system (\ref{e7.2}) by means of
reducing with the algebras $B^{1}_{1}$ and $B^{1}_{2}$ to system
(\ref{e6.1}).

With the algebra $B^{2}_{6}$ we obtain an ansatz and a reduced system
of the following forms:
\begin{equation}\label{e7.18}
\!\!\!\!
\begin{array}{l}
\bar v=\bar\phi+\lambda^{-1}(\bar\theta^{i}\cdot\bar y)\bar\psi^{i}_{t},
\quad
v^{3}=\phi^{3}\exp\bigl(\gamma\lambda(\bar\theta^{1}\cdot\bar y)\bigr),
\\[0.8ex]
s=h-\frac{1}{2}\lambda^{-1}(\bar\psi^{i}_{tt}\cdot\bar y)
(\bar\theta^{i}\cdot\bar y),
\end{array}
\end{equation}
where $\:\phi^{a}=\phi^{a}(\omega),\:$ $\:h=h(\omega),\:$ $\:\omega=t,\:$
$\:\lambda=\psi^{11}\psi^{22}-\psi^{12}\psi^{21}=
\bar\psi^{1}\cdot\bar\theta^{1}=\bar\psi^{2}\cdot\bar\theta^{2},\:$
$\:\bar\theta^{1}=(\psi^{22},-\psi^{21}),\:$
$\:\bar\theta^{2}=(-\psi^{12},\psi^{11}),\:$
and

\begin{equation}\label{e7.19}
\!\!\!\!
\begin{array}{l}
\bar\phi_{t}+\lambda^{-1}(\bar\theta^{i}\cdot\bar\phi)\bar\psi^{i}_{t}=0,
\quad
\phi^{3}_{t}+\bigl(\gamma\lambda^{-1}(\bar\theta^{1}\cdot\bar\phi)-
\gamma^{2}\lambda^{-2}(\bar\theta^{1}\cdot\bar\theta^{1})\bigr)\phi^{3}=0,
\\[0.8em]
\lambda^{-1}(\bar\theta^{i}\cdot\bar\psi^{i}_{t})+\eta_{t}\eta^{-1}=0.
\end{array}
\end{equation}
Let us make the transformation from the symmetry group of system
(\ref{e7.2}):
\[
\bar{\tilde v}(t,\bar y)=\bar v(t,\bar y-\bar\xi)+\bar\xi_{t},
\quad
\tilde v^{3}(t,\bar y)=v^{3}(t,\bar y-\bar\xi),
\quad
\tilde s(t,\bar y)=s(t,\bar y-\bar\xi)-\bar\xi_{tt}\cdot\bar y,
\]
where

\begin{equation}\label{e7.20}
\bar\xi_{t}+\lambda^{-1}(\bar\theta^{i}\cdot\bar\xi)\bar\psi^{i}_{t}
+\bar\phi=0.
\end{equation}
It follows from (7.20) that
$\:\bar\xi_{tt}=\lambda^{-1}(\bar\theta^{i}\cdot\bar\xi)\bar\psi^{i}_{tt},
\:$ i.e.,
$\:\bar\theta^{i}_{tt}\cdot\bar\xi-
\bar\theta^{i}\cdot\bar\xi_{tt}=0.\:$ Therefore, this trasformation
does not modify ansatz (\ref{e7.18}), but it
makes the functions $\phi^{i}$ vanish.
And without loss of
generality we may assume, at once, that $\:\phi^{i}\equiv 0.\:$
Then
\[
\!\!\!\!
\begin{array}{l}
\phi^{3}=C\exp\Bigl(\int\bigl(\gamma\lambda^{-1}|\theta|\bigr)^{2}dt\Bigr),
\quad
C=\mbox{const}.
\end{array}
\]
The last equation of system (\ref{e7.19}) is the compatibility condition
of system (\ref{e7.2}) and ansatz (\ref{e7.18}).

\setcounter{equation}{0}

\section{
Conclusion}
\label{sec8}

In this article we reduced the NSEs to systems of PDEs in 
three and two independent variables and systems of ODEs by 
means of the Lie method. Then, we investigated symmetry 
properties of the reduced systems of PDEs and made Lie 
reductions of systems which admitted non-trivial symmetry 
operators, i.e., operators that are not induced by operators
from $A(NS)$. Some of the systems in two independent variables
were reduced to linear systems of either two one-dimensional heat
equations or two translational equations. We also managed to find 
exact solutions for most of the reduced systems of ODEs.

Now, let us give some remaining problems. Firstly, we 
failed, for the present, to describe the non-Lie ansatzes of form 
\ref{e1.6} that reduce the NSEs. (These ansatzes include, for 
example, the well-known ansatzes for the Karman swirling flows (see 
bibliography in \cite{goldshtik}). One can also consider non-local 
ansatzes for the Navier-Stokes field, i.e., ansatzes containing 
derivatives of new unknown functions. 
 
Second problem is to study non-Lie (i.e., non-local, conditional, 
and Q-conditional) symmetries of the NSEs \cite{fshs}.

And finally, it would be interesting to investigate compatability 
and to construct exact solutions of overdetermined systems that are
obtained from the NSEs by means of different additional conditions.
Usually one uses the condition where the nonlinearity has a simple 
form, for example, the potential form (see review \cite{wang}), 
i.e., $\mbox{\rm rot}((\vec u\cdot\vec\nabla)\vec u)=\vec 0$ (the NS 
fields satisfying this condition is called the generalized Beltrami 
flows). We managed to describe the general solution of the NSEs 
with the additional condition where the convective terms vanish
\cite{nonlinearity_vanish,my_phdthesis}.
But one can give other conditions, for example, 
\begin{displaymath}
\triangle\vec u=\vec 0, \quad 
\vec u_{t}+(\vec u\cdot\vec\nabla)\vec u=\vec 0,
\end{displaymath}
and so on.

We will consider the problems above elsewhere.

\setcounter{equation}{0}

\appendix
%\part*{Appendix}
%\addcontentsline{toc}{part}{Appendix}
\section*{Appendix}
\addcontentsline{toc}{section}{Appendix}

\section{
Inequivalent one-, two-, and  three-dimensional subal-\\ gebras of
$A(NS)$} \label{sec_a}

To find complete sets of inequivalent subalgebras of $A(NS)$, we
use the method given, for example, in \cite{olver,ovsiannikov}.
Let us describe it briefly.

1. We find the commutation relations between the basis elements
of $A(NS)$.

2. For arbitrary basis elements $V$, $W^{0}$ of $A(NS)$ and each
$\varepsilon\in\R$ we calculate the adjoint action
\begin{displaymath}
W(\varepsilon)=\mbox{Ad}(\varepsilon V)W^{0}=
\mbox{Ad}(\exp(\varepsilon V))W^{0}
\end{displaymath}
of the element $\exp(\varepsilon V)$ from the one-parameter group
generated by the operator $V$ on $W^{0}$. This calculation can be
made in two ways: either by means of summing the Lie series

\begin{equation}\label{ea.1}
W(\varepsilon)=\sum_{n=0}^{\infty}\frac{\varepsilon^{n}}{n!}
\{V^{n},W^{0}\}
=W^{0}+\frac{\varepsilon}{1!}[V,W^{0}]+
\frac{\varepsilon^{2}}{2!}[V,[V,W^{0}]]+\ldots\, ,
\end{equation}
% \begin{equation}\label{eaa.1}
% W(\varepsilon)=\sum_{n=0}^{\infty}\frac{\varepsilon^{n}}{n!}
% \{V^{n},W^{0}\}=
% W^{0}\!+\!\frac{\varepsilon}{1!}[V,W^{0}]\!+\!
% \frac{\varepsilon^{2}}{2!}[V,[V,W^{0}]]\!+\ldots ,
% \end{equation}
where $\:\{V^{0},W^{0}\}=W^{0}$,
$\:\{V^{n},W^{0}\}=[V,\{V^{n-1},W^{0}\}],\:$
or directly by means of solving the initial value problem
\begin{equation}\label{ea.2}
\frac{dW(\varepsilon)}{d\varepsilon}=[V,W(\varepsilon)],
\quad
W(0)=W^{0}.
\end{equation}

3. We take a subalgebra of a general form with a fixed dimension.
Taking into account that the subalgebra is closed under the Lie
bracket, we try to simplify it by means of adjoint actions as
much as possible.

\subsection{
The commutation relations and the adjoint representation of
the algebra $A(NS)$}\label{subsec_a.1}

Basis elements (\ref{e1.2}) of $A(NS)$ satisfy the following
commutation relations:
\begin{equation}\label{ea.3}
\!\!\!\begin{array}{l}
[J_{12},J_{23}]=-J_{31},\quad
[J_{23},J_{31}]=-J_{12},\quad
[J_{31},J_{12}]=-J_{23},\quad \\
\\ {}
 [\partial_{t},J_{ab}]=[D,J_{ab}]=0, \quad
 [\partial_{t},D]=2\partial_{t},\\
\\ {}
 [\partial_{t},R(\vec m)]=R(\vec m_{t}),\quad
 [D,R(\vec m)]=R(2t\vec m_{t}-\vec m),\\
\\ {}
 [\partial_{t},Z(\chi)]=Z(\chi_{t}),\quad
 [D,Z(\chi)]=Z(2t\chi_{t}+2\chi),\\
\\ {}
 [R(\vec m),R(\vec n)]=
Z(\vec m_{tt}\cdot\vec n-\vec m\cdot\vec n_{tt}),
\quad
 [J_{ab},R(\vec m)]=R(\vec{\tilde m}),
\\
\\ {}
 [J_{ab},Z(\chi)]=[Z(\chi),R(\vec m)]=[Z(\chi),Z(\eta)]=0,
\end{array}
\end{equation}
where $\tilde m^{a}=m^{b}$,\quad $\tilde m^{b}=-m^{a}$, \quad
$\tilde m^{c}=0$,\quad $a\not=b\not=c\not=a$.

\begin{note}\label{na.1}
Relations (\ref{ea.3}) imply that the set of operators
(\ref{e1.2}) generates an algebra when, for example, the
parameter-functions $m^{a}$ and $\chi$ belong to
$C^{\infty}((t_{0},t_{1}),\R)$
($C^{\infty}_{0}((t_{0},t_{1}),\R)$,
$A((t_{0},t_{1}),\R)$), i.e., the set of
infinite-differentiable (infinite-differentiable finite, real
analytic) functions from $(t_{0},t_{1})$ in $\R$,
where $-\infty\le t_{0}<t_{1}\le+\infty$. But the NSEs (\ref{e1.1})
admit operators (\ref{e1.3}) and (\ref{e1.4}) with
parameter-functions of a less degree of smoothness. Moreover,
the minimal degree of their smoothness depends on the smoothness
that is demanded for the solutions of the NSEs (\ref{e1.1}). Thus, if
$u^{a}\in C^{2}((t_{0},t_{1})\times\Omega,\R)$ and
$p\in C^{1}((t_{0},t_{1})\times\Omega,\R)$, where
$\Omega$ is a domain in $\R^{3}$, then it is
sufficient that
$m^{a}\in C^{3}((t_{0},t_{1}),\R)$ and
$\chi\in C^{1}((t_{0},t_{1}),\R)$.
Therefore, one can consider the "pseudoalgebra" generated by
operators (\ref{e1.2}). The prefix "pseudo-" means that in this
set of operators the commutation operation is not determined for all
pairs of its elements, and the algebra axioms are
satisfied only by elements, where they are defined. It is
better to indicate the functional classes that are sets of values
for the parameters $m^{a}$ and $\chi$ in the notation of the
algebra $A(NS)$. But below, for simplicity, we fix these classes,
taking
$m^{a},\chi\in C^{\infty}((t_{0},t_{1}),\R)$,
and keep the notation of the algebra generated by operators
(\ref{e1.2}) in the form $A(NS)$. However, all calculations will
be made in such a way that they can be translated for the case of a less
degree of smoothness.
\end{note}

Most of the adjoint actions are calculated simply as sums of
their Lie series. Thus,
\begin{equation}\label{ea.4}
\!\!\!\begin{array}{l}
\mbox{Ad}(\varepsilon\partial_{t})D=D+2\varepsilon\partial_{t},
\quad
\mbox{Ad}(\varepsilon D)\partial_{t}=e^{-2\varepsilon}\partial_{t},
\\ \\
\mbox{Ad}(\varepsilon Z(\chi))\partial_{t}=\partial_{t}-
\varepsilon Z(\chi_{t}),
\quad
\mbox{Ad}(\varepsilon Z(\chi))D=D-\varepsilon
Z(2t\chi_{t}+2\chi),
\\ \\
\mbox{Ad}(\varepsilon R(\vec m))\partial_{t}=\partial_{t}-
\varepsilon R(\vec m_{t})-\frac{1}{2}\varepsilon^{2}
Z(\vec m_{t}\cdot\vec m_{tt}-\vec m\cdot\vec m_{ttt}),
\\ \\
\mbox{Ad}(\varepsilon R(\vec m))D=D-
\varepsilon R(2t\vec m_{t}-\vec m)-
\\[2ex]\quad\qquad\qquad\qquad
\frac{1}{2}\varepsilon^{2}
Z(2t\vec m_{t}\cdot\vec m_{tt}-2t\vec m\cdot\vec m_{ttt}-
4\vec m\cdot\vec m_{tt}),
\\ \\
\mbox{Ad}(\varepsilon R(\vec m))J_{ab}=J_{ab}-
\varepsilon R(\vec{\tilde m})+\varepsilon^{2}
Z(m^{a}m^{b}_{tt}-m^{a}_{tt}m^{b}),
\\ \\
\mbox{Ad}(\varepsilon R(\vec m))R(\vec n)=R(\vec n)+\varepsilon
Z(\vec m_{tt}\cdot\vec n-\vec m\cdot\vec n_{tt}),
\quad
\mbox{Ad}(\varepsilon J_{ab})R(\vec m)=R(\vec{\hat m}),
\\ \\
\mbox{Ad}(\varepsilon J_{ab})J_{cd}=J_{cd}\cos\varepsilon+
[J_{ab},J_{cd}]\sin\varepsilon \quad
\bigl((a,b)\not=(c,d)\not=(b,a)\bigr),
\end{array}
\end{equation}
where
\begin{displaymath}
\tilde m^{a}=m^{b},\quad \tilde m^{b}=-m^{a},\quad
\tilde m^{c}=0, \quad a\not=b\not=c\not=a,
\end{displaymath}
\begin{displaymath}
\hat m^{d}=m^{d}\cos\varepsilon+\tilde m^{d}\sin\varepsilon, \quad
\hat m^{c}=m^{c}, \quad a\not=b\not=c\not=a,\quad d\in\{a;b\}.
\end{displaymath}
Four adjoint actions are better found by means of integrating a
system of form (\ref{ea.2}). As a result we obtain that
\begin{equation}\label{ea.5}
\!\!\!\begin{array}{llllll}
\mbox{Ad}(\varepsilon \partial_{t})Z(\chi(t))\!\!\!\!\!&=\!\!\!\!\!&
Z(\chi(t+\varepsilon)),
&\:
\mbox{Ad}(\varepsilon D)Z(\chi(t))\!\!\!\!\!&=\!\!\!\!\!&
Z(e^{2\varepsilon}\chi(te^{2\varepsilon})),
\\[0.8em]
\mbox{Ad}(\varepsilon \partial_{t})R(\vec m(t))\!\!\!\!\!&=\!\!\!\!\!&
R(\vec m(t+\varepsilon)),
&\:
\mbox{Ad}(\varepsilon D)R(\vec m(t))\!\!\!\!\!&=\!\!\!\!\!&
R(e^{-\varepsilon}\vec m(te^{2\varepsilon})).
\end{array}
\end{equation}
Cases where adjoint actions coincide with the identical mapping
are omitted.

\begin{note}\label{na.2}
If $\:Z(\chi(t))\!\in\!A(NS)[C^{\infty}((t_{0},t_{1}),\R)]\:$
with $\:-\infty<t_{0}$ or $t_{1}<+\infty$, the adjoint
representation $\mbox{Ad}(\varepsilon \partial_{t})$
 ($\mbox{Ad}(\varepsilon D)$) gives an equivalence relation
between the operators $Z(\chi(t))$ and $Z(\chi(t+\varepsilon))$
 ($Z(\chi(t))$  and $Z(e^{2\varepsilon}\chi(te^{2\varepsilon}))$)
that belong to the different algebras
\begin{displaymath}
\begin{array}{l}
A(NS)[C^{\infty}((t_{0},t_{1}),\R)]\quad \mbox{and}\quad
A(NS)[C^{\infty}((t_{0}-\varepsilon,t_{1}-\varepsilon),\R)]
\\[1ex]
(A(NS)[C^{\infty}((t_{0},t_{1}),\R)] \quad \mbox{and}\quad
A(NS)[C^{\infty}((t_{0}e^{-2\varepsilon},t_{1}e^{-2\varepsilon}),
\R)] )
\end{array}
\end{displaymath}
respectively. An analogous statement is true for
the operator $R(\vec m)$. Equivalence of subalgebras in Theorems
\ref{ta.1} and \ref{ta.2} is also meant in this sense.
\end{note}
\begin{note}\label{na.3}
Besides the adjoint representations of operators (\ref{e1.2}) we
make use of discrete transformation (\ref{e1.6}) for classifying
the subalgebras of $A(NS)$,
\end{note}

To prove the theorem of this section, the following obvious lemma
is used.

\begin{lemma}\label{la.1}
Let $N\in\N$.

\vspace{1ex}

A. If $\chi\in C^{N}((t_{0},t_{1}),\R)$, then
$\exists\eta\in C^{N}((t_{0},t_{1}),\R)$:
$2t\eta_{t}+2\eta=\chi$.

\vspace{1ex}

B. If $\chi\in C^{N}((t_{0},t_{1}),\R)$, then
$\exists\eta\in C^{N}((t_{0},t_{1}),\R)$:
$2t\eta_{t}-\eta=\chi$.

\vspace{1ex}

C. If $m^{i}\in C^{N}((t_{0},t_{1}),\R)$ and $a\in\R$,
then
$\exists l^{i}\in C^{N}((t_{0},t_{1}),\R)$:
\begin{displaymath}
2tl^{1}_{t}-l^{1}+al^{2}=m^{1}, \quad
2tl^{2}_{t}-l^{2}-al^{1}=m^{2}.
\end{displaymath}
\end{lemma}

\subsection{
One-dimensional subalgebras}\label{subsec_a.2}

\begin{theorem}\label{ta.1}
A complete set of $A(NS)$-inequivalent one-dimensional
subalgebras of $A(NS)$ is exhausted by the following algebras:

\vspace{2ex}

1. $A^{1}_{1}(\kappa)=<D+2\kappa J_{12}>,\:$ where $\:\kappa\ge0$.

\vspace{2ex}

2. $A^{1}_{2}(\kappa)=<\partial_{t}+\kappa J_{12}>,\:$ where
$\:\kappa\in\{0;1\}$.

\vspace{2ex}

3. $A^{1}_{3}(\eta,\chi)=<J_{12}+R(0,0,\eta(t))+Z(\chi(t))>\:$
with smooth functions $\eta$ and $\chi$.
Algebras $\:A^{1}_{3}(\eta,\chi)\:$ and
$\:A^{1}_{3}(\tilde\eta,\tilde\chi)\:$ are equivalent if
\mbox{$\:\exists\varepsilon,\delta\in\R$},
\mbox{$\:\exists\lambda\in C^{\infty}((t_{0},t_{1}),\R)$}:

\begin{equation}\label{ea.9}
\tilde\eta(\tilde t)=e^{-\varepsilon}\eta(t), \quad
\tilde\chi(\tilde t)=e^{2\varepsilon}
(\chi(t)+\lambda_{tt}(t)\eta(t)-\lambda(t)\eta_{tt}(t)),
\end{equation}
where $\:\tilde t=te^{-2\varepsilon}+\delta$.

\vspace{2ex}

4. $A^{1}_{4}(\vec m,\chi)=<R(\vec m(t))+Z(\chi(t))>\:$ with
smooth functions $\vec m$ and $\chi$:
$\:(\vec m,\chi)\not\equiv(\vec 0,0).\:$
Algebras $\:A^{1}_{4}(\vec m,\chi)\:$ and
$\:A^{1}_{4}(\vec{\tilde m},\tilde\chi)\:$ are equivalent if
\mbox{$\:\exists\varepsilon,\delta\in\R$}, \quad
\mbox{$\exists C\not=0$},
\mbox{$\:\exists B\in O(3)$},
\mbox{$\:\exists\vec l\in C^{\infty}((t_{0},t_{1}),\R^{3})$}:

\begin{equation}\label{ea.10}
\begin{array}{l}
\vec{\tilde m}(\tilde t)=Ce^{-\varepsilon}B\vec m(t),
\quad
\tilde\chi(\tilde t)=Ce^{2\varepsilon}\bigl(\chi(t)+
\vec l_{tt}(t)\cdot\vec m(t)-\vec m_{tt}(t)\cdot\vec l(t)\bigr),
\end{array}
\end{equation}
where $\:\tilde t=te^{-2\varepsilon}+\delta$.
\end{theorem}

{\bf P r o o f}\  \ \ \ Consider an arbitrary one-dimensional subalgebra generated
by
\begin{displaymath}
V=a_{1}D+a_{2}\partial_{t}+a_{3}J_{12}+a_{4}J_{23}+a_{5}J_{31}+
R(\vec m)+Z(\chi).
\end{displaymath}
The coefficients $a_{4}$ and $a_{5}$ are omitted below since
they always can be made to vanish by means of the
adjoint representations $\mbox{Ad}(\varepsilon_{1}J_{12})$ and
$\mbox{Ad}(\varepsilon_{2}J_{31})$.

If $a_{1}\not=0$ we get $\tilde a_{1}=1$ by means of a change of
basis. Next, step-by-step we make $a_{2}$, $\vec m$, and $\chi$
vanish by means of the adjoint representations
$\mbox{Ad}(-\frac{1}{2}a_{2}a_{1}^{-1}\partial_{t})$,
$\mbox{Ad}(R(\vec l))$, and $\mbox{Ad}(Z(\chi))$, where
\begin{displaymath}
\begin{array}{l}
\vec l\!\in\!C^{\infty}((t_{0}+\frac{1}{2}a_{2}a_{1}^{-1},
t_{1}+\frac{1}{2}a_{2}a_{1}^{-1}),\R^{3}),
\quad
\eta\!\in\!C^{\infty}((t_{0}+\frac{1}{2}a_{2}a_{1}^{-1},
t_{1}+\frac{1}{2}a_{2}a_{1}^{-1}),\R),
\end{array}
\end{displaymath}
and $\vec l$, $\eta$ are solutions of the equations
\begin{displaymath}
2t\vec l_{t}-\vec l+a_{3}a_{1}^{-1}(l^{2},-l^{1},0)^{T}=
\vec{\hat m}, \quad
\begin{array}{l}
2t\eta_{t}+2\eta=\hat\chi+\frac{1}{2}
(\vec l_{tt}\cdot\vec{\hat m}-\vec l\cdot\vec{\hat m}_{tt})
\end{array}
\end{displaymath}
with
$\vec{\hat m}(t)=a_{1}^{-1}\vec m(t-\frac{1}{2}a_{2}a_{1}^{-1})$
and $\hat\chi(t)=a_{1}^{-1}\chi(t-\frac{1}{2}a_{2}a_{1}^{-1})$.
Such $\vec l$ and $\eta$ exist in virtue of Lemma \ref{la.1}. As
a result we obtain the algebra $A_{1}^{1}(\kappa)$, where
$2\kappa=a_{3}a_{1}^{-1}$. In case $\kappa<0$ additionally one
has to apply transformation (\ref{e1.6}) with $b=1$.

If $a_{1}=0$ and $a_{2}\not=0$, we make $\tilde a_{2}=1$ by means
of a change of basis. Next, step-by-step we make $\vec m$ and $\chi$
vanish by means of the adjoint representations
$\mbox{Ad}(R(\vec l))$ and $\mbox{Ad}(Z(\chi))$, where
$\vec l\in C^{\infty}((t_{0},t_{1}),\R^{3})$,
$\eta\in C^{\infty}((t_{0},t_{1}),\R)$, and
\begin{displaymath}
a_{2}\vec l_{t}+a_{3}(l^{2},-l^{1},0)^{T}=\vec m, \quad
\begin{array}{l}
a_{2}\eta_{t}=\chi+\frac{1}{2}
(\vec l_{tt}\cdot\vec m-\vec l\cdot\vec m_{tt}).
\end{array}
\end{displaymath}
If $a_{3}=0$ we obtain the algebra $A^{1}_{2}(0)$ at once. If
$a_{3}\not=0$, using the adjoint representation
$\mbox{Ad}(\varepsilon D)$ and transformation (\ref{e1.6})
(in case of need), we obtain the algebra $A^{1}_{2}(1)$.

If $a_{1}=a_{2}=0$ and $a_{3}\not=0$, after a change of basis and
applying the adjoint representation
$\mbox{Ad}(R(-a_{3}^{-1}m^{2}, a_{3}^{-1}m^{1},0))$ we get the
algebra $A^{1}_{3}(\eta,\tilde\chi)$, where
$\eta=a_{3}^{-1}m^{3}$ and
$\tilde\chi=a_{3}^{-1}\chi+a_{3}^{-2}(m^{1}_{tt}m^{2}-m^{1}m^{2}_{tt})$.
Equivalence relation (\ref{ea.9}) is generated by the adjoint
representations $\mbox{Ad}(\varepsilon D)$,
$\mbox{Ad}(\delta\partial_{t})$, and
$\mbox{Ad}(R(0,0,\lambda))$.

If $a_{1}=a_{2}=a_{3}=0$, at once we get the algebra
$A^{1}_{4}(\vec m,\chi)$. Equivalence relation (\ref{ea.10})
is generated by the adjoint representations
$\mbox{Ad}(\varepsilon D)$,
$\mbox{Ad}(\delta\partial_{t})$,
$\mbox{Ad}(R(\vec l))$, and $\mbox{Ad}(\varepsilon_{ab}J_{ab})$.

\subsection{
Two-dimensional subalgebras}\label{subsec_a.3}

\begin{theorem}\label{ta.2}
A complete set of $A(NS)$-inequivalent two-dimensional
subalgebras of $A(NS)$ is exhausted by the following algebras:

\vspace{2ex}

1. $A^{2}_{1}(\kappa)=<\partial_{t},\: D+\kappa J_{12}>,\:$
where $\:\kappa\ge0$.

\vspace{2ex}

2. $A^{2}_{2}(\kappa,\varepsilon)=<D,\:
J_{12}+R(0,0,\kappa\vert t \vert^{1/2})+Z(\varepsilon t^{-1})>,\:$
where $\:\kappa\ge 0$, $\:\varepsilon\ge 0$.

\vspace{2ex}

3.\vspace{1ex} $A^{2}_{3}(\kappa,\varepsilon)=<\partial_{t}, \:
J_{12}+R(0,0,\kappa)+Z(\varepsilon)>\:$,
where $\:\kappa\in\{0;1\}$,
$\:\varepsilon\ge0\:$ if $\:\kappa=1\:$ and

\noindent
\hspace*{10mm}
$\:\varepsilon\in\{0;1\}\:$ if $\:\kappa=0$.

\vspace{2ex}

4. \vspace{1ex} $A^{2}_{4}(\sigma,\kappa,\mu,\nu,\varepsilon)=
<D+2\kappa J_{12},\:
R\bigl(\vert t \vert^{\sigma+1/2}(\nu\cos\tau,\,\nu\sin\tau,\,\mu)\bigr)+
Z(\varepsilon\vert t \vert^{\sigma-1})>,\:$
where

\noindent
\hspace*{10mm}
$\:\tau=\kappa\ln\vert t \vert$,
$\:\kappa>0$, $\:\mu\ge0$,  $\:\nu\ge0$,
$\:\mu^{2}+\nu^{2}=1$,
$\:\varepsilon\sigma=0,\:$ and $\:\varepsilon\ge0$.

\vspace{2ex}

5. $A^{2}_{5}(\sigma,\varepsilon)=<D, \:
R(0,0,\vert t \vert^{\sigma+1/2})+
Z(\varepsilon\vert t \vert^{\sigma-1})>,\:$
where
$\:\varepsilon\sigma=0\:$ and $\:\varepsilon\ge0$.

\vspace{2ex}

6. \vspace{1ex} $A^{2}_{6}(\sigma,\mu,\nu,\varepsilon)=
<\partial_{t}+J_{12},\:
R(\nu e^{\sigma t}\cos t, \nu e^{\sigma t}\sin t,
\mu e^{\sigma t})+Z(\varepsilon e^{\sigma t})>$,
where
$\:\mu\ge0$,

\noindent
\hspace*{10mm}
$\:\nu\ge0$,  $\:\mu^{2}+\nu^{2}=1$,
$\:\varepsilon\sigma=0,\:$ and $\:\varepsilon\ge0$.

\vspace{2ex}

7. $A^{2}_{7}(\sigma,\varepsilon)=<\partial_{t},\:
R(0,0,e^{\sigma t})+Z(\varepsilon e^{\sigma t})>,\:$
where  $\:\sigma\in\{-1;0;1\}$,
$\:\varepsilon\sigma=0,\:$ and $\:\varepsilon\ge0$.

\vspace{2ex}

8. $A^{2}_{8}(\lambda,\psi^{1},\rho,\psi^{2})=<
J_{12}+R(0,0,\lambda)+Z(\psi^{1}), \: R(0,0,\rho)+Z(\psi^{2})>$ with
smooth functions (of $t$) $\lambda$, $\rho$, and $\psi^{i}$:
\mbox{$\:(\rho,\psi^{2})\not\equiv(0,0)\:$} and
\mbox{$\:\lambda_{tt}\rho-\lambda\rho_{tt}\equiv0.$}
Algebras $\:A^{2}_{8}(\lambda,\psi^{1},\rho,\psi^{2})\:$ and
$\:A^{2}_{8}(\tilde\lambda,\tilde\psi^{1},\tilde\rho,\tilde\psi^{2})\:$
are equivalent if  $\:\exists C_{1}\not=0$,
$\:\exists \varepsilon,\delta,C_{2}\in\R$,
\mbox{$\:\exists\theta\in C^{\infty}((t_{0},t_{1}),\R)$}:

\begin{equation}\label{ea.12}
\begin{array}{l}
\tilde\lambda(\tilde t)=e^{\varepsilon}
(\lambda(t)+C_{2}\rho(t)),
\quad
\tilde\rho(\tilde t)=C_{1}e^{-\varepsilon}\rho(t),\\
\\
\tilde\psi^{1}(\tilde t)=e^{2\varepsilon}\bigl(
\psi^{1}(t)+\theta_{tt}(t)\lambda(t)-\theta(t)\lambda_{tt}(t)+
\\[1ex]
\quad\quad\quad\quad\quad\quad\quad\quad
+C_{2}(\psi^{2}(t)+\theta_{tt}(t)\rho(t)-\theta(t)\rho_{tt}(t))
\bigr),\\
\\
\tilde\psi^{2}(\tilde t)=C_{1}e^{2\varepsilon}
(\psi^{2}(t)+\theta_{tt}(t)\rho(t)-\theta(t)\rho_{tt}(t)),
\end{array}
\end{equation}
where $\:\tilde t=te^{-2\varepsilon}+\delta$.

\vspace{2ex}

9. $A^{2}_{9}(\vec m^{1},\chi^{1},\vec m^{2},\chi^{2})=
<R(\vec m^{1}(t))+Z(\chi^{1}(t)),\:
R(\vec m^{2}(t))+Z(\chi^{2}(t))>$
with smooth functions $\vec m^{i}$ and $\chi^{i}$:
\begin{displaymath}
\vec m^{1}_{tt}\cdot\vec m^{2}-\vec m^{1}\cdot\vec m^{2}_{tt}=0,
\quad
\mbox{\rm rank}\bigl((\vec m^{1},\chi^{1}),(\vec m^{2},\chi^{2})\bigr)=2.
\end{displaymath}
Algebras
$\:A^{2}_{9}(\vec m^{1},\chi^{1},\vec m^{2},\chi^{2})\:$
and $\:A^{2}_{9}(\vec{\tilde m^{1}},\tilde\chi^{1},
\vec{\tilde m^{2}},\tilde\chi^{2})\:$ are equivalent if
$\:\:\exists\varepsilon,\delta\!\in\!\R,$
$\:\exists\{a_{ij}\}_{i,j=1,2}:\det\{a_{ij}\}\not=0,\:$
$\:\exists B\in O(3),\:$
$\:\exists\vec l\in C^{\infty}((t_{0},t_{1}),\R^{3})$:

\begin{equation}\label{ea.13}
\begin{array}{l}
\vec{\tilde m^{i}}(\tilde t)=e^{-\varepsilon}a_{ij}B
\vec m^{j}(t),\\
\\
\tilde\chi^{i}(\tilde t)=e^{2\varepsilon}a_{ij}\bigl(\chi^{j}(t)+
\vec l_{tt}(t)\cdot\vec m^{j}(t)-\vec l(t)\cdot\vec m^{j}_{tt}(t)
\bigr),
\end{array}
\end{equation}
where $\:\tilde t= te^{-2\varepsilon}+\delta$.

\vspace{2ex}

10. $A^{2}_{10}(\kappa,\sigma)=<D+\kappa J_{12},\:
Z(\vert t \vert^{\sigma})>,\:$
where $\:\kappa\ge0$, $\:\sigma\in\R$.

\vspace{2ex}

11. $A^{2}_{11}(\sigma)=<\partial_{t}+J_{12}, \:
Z(e^{\sigma t})>,\:$ where $\:\sigma\in\R$.

\vspace{2ex}

12. $A^{2}_{12}(\sigma)=<\partial_{t}, \:
Z(e^{\sigma t})>\:$ where $\:\sigma\in\{-1;0;1\}$.
\end{theorem}

The proof of Theorem \ref{ta.2} is analogous to that of Theorem
\ref{ta.1}. Let us take an arbitrary two-dimensional subalgebra
generated by two linearly independent operators of the form
\begin{displaymath}
V^{i}=a^{i}_{1}D+a^{i}_{2}\partial_{t}
+a^{i}_{3}J_{12}+a^{i}_{4}J_{23}+a^{i}_{5}J_{31}+
R(\vec m^{i})+Z(\chi^{i}),
\end{displaymath}
where $\:a^{i}_{n}=\mbox{const}\:(n=\overline{1,5})\:$
and $\:[V^{1},V^{2}]\in<V^{1},V^{2}>.\:$
Considering the different possible cases we try to simplify
$V^{i}$ by means of adjoint representation as much as possible.
Here we do not present the proof of Theorem \ref{ta.2} as it is
too cumbersome.

\subsection{
Three-dimensional subalgebras}\label{subsec_a.4}

We also constructed a complete set of $A(NS)$-inequivalent
three-dimensional subalgebras. It contains 52 classes of algebras.
By means of 22 classes from this set one can obtain ansatzes of
codimension three for the Navier-Stokes field. Here we only give
8 superclasses that arise from unification of some of these
classes:

\vspace{2ex}

1. $A^{3}_{1}=<D,\: \partial_{t}, \: J_{12}>$.

\vspace{2ex}

2. $A^{3}_{2}=<D+\kappa J_{12},\: \partial_{t},\: R(0,0,1)>,\:$
where $\:\kappa\ge0$.\\
Here and below $\kappa$, $\sigma$, $\varepsilon_{1}$,
$\varepsilon_{2}$, $\mu$, $\nu$, and $a_{ij}$ are real constants.

\vspace{2ex}

3. \vspace{1ex} $A^{3}_{3}(\sigma,\nu,\varepsilon_{1},\varepsilon_{2})=
<D,\: J_{12}+\nu\bigl(R(0,0,\vert t \vert^{1/2}\ln\vert t \vert)+
Z(\varepsilon_{2}\vert t \vert^{-1}\ln\vert t \vert)\bigr)
+Z(\varepsilon_{1}\vert t \vert^{-1}),$
$R(0,0,\vert t \vert^{\sigma+1/2})+
Z(\varepsilon_{2}\vert t \vert^{\sigma-1})>$,
where
$\:\nu\sigma=0$, $\:\varepsilon_{1}\ge0$, $\:\nu\ge0,\:$
and  $\:\sigma\varepsilon_{2}=0$.

\vspace{2ex}

4. $A^{3}_{4}(\sigma,\nu,\varepsilon_{1},\varepsilon_{2})=
<\partial_{t}, \:
J_{12}+Z(\varepsilon_{1})+\nu\bigl(R(0,0,t)+
Z(\varepsilon_{2}t)\bigr),
\:
R(0,0,e^{\sigma t})+Z(\varepsilon_{2}e^{\sigma t})>$,
\\[1ex]
where $\:\nu\sigma=0,\:$ $\:\sigma\varepsilon_{2}=0,\:$
and, if $\:\sigma=0,\:$
the constants $\nu$, $\varepsilon_{1}$, and $\varepsilon_{2}$ satisfy
one of the following conditions:
\begin{displaymath}
\nu=1,\:\varepsilon_{1}\ge0;\quad
\nu=0,\:\varepsilon_{1}=1,\:\varepsilon_{2}\ge0;\quad
\nu=\varepsilon_{1}=0,\:\varepsilon_{2}\in\{0;1\}.
\end{displaymath}

\vspace{2ex}

5. $A^{3}_{5}(\kappa,\vec m^{1},\vec m^{2},\chi^{1},\chi^{2})=
<D+2\kappa J_{12}, \: R(\vec m^{1})+Z(\chi^{1}), \:
R(\vec m^{2})+Z(\chi^{2})>$,
\\[1ex]
where $\:\kappa\ge0,\:$ $\:\mbox{rank}(\vec m^{1},\vec m^{2})=2$,
\begin{displaymath}
\!\!\!\begin{array}{l}
t\vec m^{i}_{t}-\frac{1}{2}\vec m^{i}+
\kappa(m^{i2},-m^{i1},0)^{T}=a_{ij}\vec m^{j},
\end{array}
\end{displaymath}
\begin{displaymath}
t\chi^{i}_{t}+\chi^{i}=a_{ij}\chi^{j}, \quad a_{ij}=\mbox{const},
\end{displaymath}

\begin{equation}\label{ea.14}
\!\!\!\begin{array}{l}
(a_{11}+a_{22})\bigl(a_{21}\vec m^{1}\cdot\vec m^{1}+
(a_{22}-a_{11})\vec m^{1}\cdot\vec m^{2}-
a_{12}\vec m^{2}\cdot\vec m^{2}+
\\[1ex]
+2\kappa(m^{12}m^{21}-m^{11}m^{22})\bigr)=0.
\end{array}
\end{equation}
This superclass contains eight inequivalent classes of subalgebras
that can be obtained from it by means of a change of basis and the
adjoint actions
\begin{displaymath}
\mbox{Ad}(\delta_{1} D),\quad \mbox{Ad}(\delta_{2}J_{12}), \quad
\mbox{Ad}(R(\vec n)+Z(\eta))
\end{displaymath}
\begin{displaymath}
\bigl( \mbox{Ad}(\delta D),\quad
\mbox{Ad}(\varepsilon_{ab}J_{ab}), \quad
\mbox{Ad}(R(\vec n)+Z(\eta)) \bigr)
\end{displaymath}
if $\:\kappa>0\:$ $\:(\kappa=0)\:$ respectively. Here the functions
$\vec n$ and $\eta$ satisfy the following equations:
\begin{displaymath}
\!\!\!\begin{array}{l}
t\vec n_{t}-
\frac{1}{2}\vec n+\kappa(n^{2},-n^{1},0)^{T}=b_{i}\vec m^{i},
\end{array}
\end{displaymath}
\begin{displaymath}
\!\!\!\begin{array}{l}
t\eta_{t}+\eta=b_{i}\chi_{i}+\frac{1}{2}t(\vec n_{ttt}\cdot\vec n
-\vec n_{tt}\cdot\vec n_{t})+\vec n_{tt}\cdot\vec n+
\kappa(n^{1}n^{2}_{tt}-n^{1}_{tt}n^{2}).
\end{array}
\end{displaymath}

\vspace{2ex}

6. $A^{3}_{6}(\kappa,\vec m^{1},\vec m^{2},\chi^{1},\chi^{2})=
<\partial_{t}+\kappa J_{12},\: R(\vec m^{1})+Z(\chi^{1}),
\: R(\vec m^{2})+Z(\chi^{2})>$,
\\[1ex]
where $\:\kappa\in\{0;1\},\:$
$\:\mbox{rank}(\vec m^{1},\vec m^{2})=2,$
\begin{displaymath}
\vec m^{i}_{t}-\kappa(m^{i2},-m^{i1},0)^{T}=a_{ij}\vec m^{j},
\quad
t\chi^{i}_{t}=a_{ij}\chi^{j},
\end{displaymath}
and $a_{ij}$ are constants satisfying (\ref{ea.14}).
This superclass contains eight inequivalent classes of subalgebras
that can be obtained from it by means of a change of basis and the
adjoint actions
\begin{displaymath}
\mbox{Ad}(\delta_{1}\partial_{t}),\quad
\mbox{Ad}(\delta_{2}J_{12}), \quad
\mbox{Ad}(R(\vec n)+Z(\eta))
\end{displaymath}
\begin{displaymath}
\bigl(\mbox{Ad}(\delta_{1}\partial_{t}),\quad
\mbox{Ad}(\delta_{2} D),\quad
\mbox{Ad}(\varepsilon_{ab}J_{ab}), \quad
\mbox{Ad}(R(\vec n)+Z(\eta))\bigr)
\end{displaymath}
if $\:\kappa=1\:$ $\:(\kappa=0)\:$ respectively. Here the functions
$\vec n$ and $\eta$ satisfy the following equations:
\begin{displaymath}
\begin{array}{l}
\vec n_{t}+\kappa(n^{2},-n^{1},0)^{T}=b_{i}\vec m^{i},
\end{array}
\end{displaymath}
\begin{displaymath}
\begin{array}{l}
\eta_{t}=b_{i}\chi_{i}+\frac{1}{2}(\vec n_{ttt}\cdot\vec n
-\vec n_{tt}\cdot\vec n_{t})+
\kappa(n^{1}n^{2}_{tt}-n^{1}_{tt}n^{2}).
\end{array}
\end{displaymath}

\vspace{2ex}

7. $A^{3}_{7}(\eta^{1},\eta^{2},\eta^{3},\chi)=
< J_{12}+R(0,0,\eta^{3}), \: R(\eta^{1},\eta^{2},0), \:
R(-\eta^{2},\eta^{1},0) >$, \quad where
\[
\eta^{a}\in C^{\infty}((t_{0},t_{1}),\R),\quad
\eta^{1}_{tt}\eta^{2}-\eta^{1}\eta^{2}_{tt}\equiv 0,\quad
\eta^{i}\eta^{i}\not\equiv0, \quad \eta^{3}\not=0.
\]
Algebras $\:A^{3}_{7}(\eta^{1},\eta^{2},\eta^{3})\:$ and
$\:A^{3}_{7}(\tilde\eta^{1},\tilde\eta^{2},\tilde\eta^{3})\:$
are equivalent if $\:\:\exists\delta_{a}\in\R,\:$
\mbox{$\exists\delta_{4}\not=0$}:

\begin{equation}\label{ea.15}
\!\!\!\!
\begin{array}{ll}
\tilde\eta^{1}(\tilde t)\!\!\!\!&=\delta_{4}(\eta^{1}(t)\cos\delta_{3}-
\eta^{2}(t)\sin\delta_{3}),
\\[1ex]
\tilde\eta^{2}(\tilde t)\!\!\!\!&=\delta_{4}(\eta^{1}(t)\sin\delta_{3}+
\eta^{2}(t)\cos\delta_{3}),
\\[1ex]
\tilde\eta^{3}(\tilde t)\!\!\!\!&=e^{-\delta_{1}}\eta^{3}(t),
\end{array}
\end{equation}
where $\:\tilde t=te^{-2\delta_{1}}+\delta_{2}$.

\vspace{2ex}

8. \mbox{$A^{3}_{8}(\vec m^{1},\vec m^{2},\vec m^{3})=
<R(\vec m^{1}),\: R(\vec m^{2}), \: R(\vec m^{3})>$}, \quad
 where
\[
\mbox{$\vec m^{a}\in C^{\infty}((t_{0},t_{1}),\R^{3})$},
\quad\,
\mbox{$\mbox{rank}(\vec m^{1},\vec m^{2},\vec m^{3})=3$},
\quad\,
\mbox{$\vec m^{a}_{tt}\cdot\vec m^{b}-
\vec m^{a}\cdot\vec m^{b}_{tt}=0$}.
\]
Algebras
\mbox{$\:A^{3}_{8}(\vec m^{1},\vec m^{2},\vec m^{3})\:$} and
\mbox{$\:A^{3}_{8}(\vec{\tilde m}^{1},\vec{\tilde m}^{2},
\vec{\tilde m}^{3})\:$} are equivalent if
\mbox{$\:\exists\delta_{i}\in\R^{3}$},
\mbox{$\:\:\exists B\in O(3)$},
\mbox{$\:\:\exists\{d_{ab}\}:\det\{d_{ab}\}\not=0\:\:$} such that

\begin{equation}\label{ea.16}
\vec{\tilde m^{a}}(\tilde t)=d_{ab}B\vec m^{b}(t),
\end{equation}
where $\:\tilde t=te^{-2\delta_{1}}+\delta_{2}$.

\setcounter{equation}{0}
\section{
On construction of ansatzes for the Navier-Stokes field by means 
of the Lie method}
\label{sec_b}

The general method for constructing a complete set of inequivalent 
Lie ansatzes of a system of PDEs are well known and described, for 
examle, in \cite{olver,ovsiannikov}. However, in some cases when 
the symmetry operators of the system have a special form, this 
method can be modified \cite{f_ansatz}. Thus, in the case of the NSEs, 
coefficients of an arbitrary operator 
\[
Q=\xi^{0}\partial_{t}+\xi^{a}\partial_{a}+
\eta^{a}\partial_{u^{a}}+\eta^{0}\partial_{p}
\]
from $A(NS)$ satisfy the following conditions:
\begin{equation}\label{eb.1}
\!\!\!\!
\begin{array}{l}
\xi^{0}=\xi^{0}(t,\vec x), \quad 
\xi^{a}=\xi^{a}(t,\vec x), \quad 
\eta^{a}=\eta^{ab}(t,\vec x)u^{b}+\eta^{a0}(t,\vec x), \quad 
\\[0.8em] 
\eta^{0}=\eta^{01}(t,\vec x)p+\eta^{00}(t,\vec x).
\end{array}
\end{equation}
(The coefficients $\xi^{a}$, $\xi^{0}$, $\eta^{a}$, and $\eta^{0}$ 
also satisfy stronger conditions than (\ref{eb.1}). For example if 
$\:Q\in A(NS),\:$ then 
$\:\xi^{0}=\xi^{0}(t),\:$
$\:\eta^{ab}=\mbox{\rm const}\:$, 
and so on. But conditions (\ref{eb.1}) are sufficient to simplify 
the general method.) Therefore, ansatzes for the Navier-Stokes 
field can be constructing in the following way:

1. We fix a $M$-dimensional subalgebra of  $A(NS)$ with the basis 
elements
\begin{equation}\label{eb.2}
Q^{m}=\xi^{m0}\partial_{t}+\xi^{ma}\partial_{a}+
(\eta^{mab}u^{b}+\eta^{ma0})\partial_{u^{a}}+
(\eta^{m01}p+\eta^{m00})\partial_{p},
\end{equation}
where $\:M\in\{1;2;3\},\:$ $\:m=\overline{1,M},\:$ and
\begin{equation}\label{eb.3}
\mbox{\rm rank}\{(\xi^{m0},\xi^{m1},\xi^{m2},\xi^{m3}),\:
m=\overline{1,M}\}=M.
\end{equation}
To construct a complete set of inequivalent Lie ansatzes of 
codimension $M$ for the Navier-Stokes field, we have to use the set 
of $M$-dimensional subalgebras from Sec.~\ref{sec_a}. Condition (\ref{eb.3}) 
is neeeded for the existance of ansatzes connected with this subalgebra.

2. We find the invariant independent variables 
$\:\omega_{n}=\omega_{n}(t,\vec x), \: n=\overline{1,N},\:$ where
$\:N=4-M,\:$ as a set of functionally independent solutions of the 
following system:
\begin{equation}\label{eb.4}
L^{m}\omega=Q^{m}\omega=\xi^{m0}\partial_{t}\omega+
\xi^{ma}\partial_{a}\omega=0, \quad m=\overline{1,M},
\end{equation}
where $\:L^{m}:=\xi^{m0}\partial_{t}+\xi^{ma}\partial_{a}.\:$

3. We present the Navier-Stokes field in the form:
\begin{equation}\label{eb.5}
u^{a}=f^{ab}(t,\vec x)v^{b}(\bar\omega)+g^{a}(t,\vec x), \quad 
p=f^{0}(t,\vec x)q(\bar\omega)+g^{0}(t,\vec x), 
\end{equation}
where $v^{a}$ and $q$ are new unknown functions of 
$\:\bar\omega=\{\omega_{n},\:n=\overline{1,N}\}$.
Acting on representation (\ref{eb.5}) with the operators $Q^{m}$, 
we obtain the following equations on functions $f^{ab}$, $g^{a}$, 
$f^{0}$, and $g^{0}$:
\begin{equation}\label{eb.6}
\!\!\!\!
\begin{array}{ll}
L^{m}f^{ab}=\eta^{mac}f^{cb},
&
L^{m}g^{a}=\eta^{mab}g^{b}+\eta^{ma0}, \quad c=\overline{1,3},
\\[0.8em]
L^{m}f^{0}=\eta^{m01}f^{0},
&
L^{m}g^{0}=\eta^{m01}g^{0}+\eta^{m00}.
\end{array}
\end{equation}
If the set of functions $f^{ab}$, $f^{0}$, $g^{a}$, and $g^{0}$ is 
a particular solution of (\ref{eb.6}) and satisfies the conditions 
$\:\mbox{\rm rank}\{(f^{1b},f^{2,b},f^{3b}),\:b=\overline{1,3}\}=3\:$ 
and $\:f^{0}\not=0,\:$ formulas (\ref{eb.5}) give an ansatz for 
the Navier-Stokes field.

The ansatz connected with the fixed subalgebra is not determined in 
an unique manner. Thus, if 
\begin{equation}\label{eb.7}
\!\!\!\!
\begin{array}{l}
\tilde\omega_{l}=\tilde\omega_{l}(\bar\omega), \quad 
{\displaystyle \det\left\{\frac{\partial\tilde\omega_{l}}{\partial\omega_{n}}
\right\}_{l,n=\overline{1,N}}\not=0,}
\\[1.4em]
\!\!\!\!
\begin{array}{ll}
\tilde f^{ab}(t,\vec x)=f^{ac}(t,\vec x)F^{cb}(\bar\omega),
&
\tilde g^{a}(t,\vec x)=g^{a}(t,\vec x)+f^{ac}(t,\vec x)G^{c}(\bar\omega),
\\[0.8em]
\tilde f^{0}(t,\vec x)=f^{0}(t,\vec x)F^{0}(\bar\omega),
&
\tilde g^{0}(t,\vec x)=g^{0}(t,\vec x)+f^{0}(t,\vec x)G^{0}(\bar\omega),
\end{array}
\end{array}
\end{equation}
the formulas 
\begin{equation}\label{eb.8}
u^{a}=\tilde f^{ab}(t,\vec x)\tilde v^{b}(\bar{\tilde\omega})+
\tilde g^{a}(t,\vec x), \quad 
p=\tilde f^{0}(t,\vec x)q(\bar{\tilde \omega})+\tilde g^{0}(t,\vec x)
\end{equation}
give an ansatz which is equivalent to ansatz (\ref{eb.5}). The reduced 
system of PDEs on the functions $\tilde v^{a}$ and $\tilde q$ is 
obtained from the system on $v^{a}$ and $q$ by means of a local 
transformation. Our problem is to find or "to guess", at once, such 
an ansatz that the corresponding reduced system has a simple 
and convenient form for our investigation. Otherwise, we can obtain a 
very complicated reduced system which will be not convenient for 
investigation and we can not simplify it.

Consider a simple example. 

Let $\:M=1\:$ and let us give the algebra 
$\:<\partial_{t}+\kappa J_{12}>,\:$ where $\:\kappa\in\{0;1\}\:$. 
For this algebra, the invariant independent variables 
$\:y_{a}=y_{a}(t,\vec x)\:$ are functionally independent solutions 
of the equation $\:Ly=0\:$ (see (\ref{eb.4})), where 
\begin{equation}\label{eb.9}
L:=\partial_{t}+\kappa(x_{1}\partial_{x_{2}}-
x_{2}\partial_{x_{1}}).
\end{equation}
There exists an infinite set of choices for the variables 
$y_{a}$. For example, we can give the following expressions for 
$y_{a}$:
\[
y_{1}=\arctan\frac{x_{1}}{x_{2}}-\kappa t, \quad 
y_{2}=(x_{1}^{2}+x_{2}^{2})^{1/2}, \quad 
y_{3}=x_{3}.
\]
However choosing $y_{a}$ in such a way, for $\:\kappa\not=0\:$ we 
obtain a reduced system which strongly differs from the "natural" 
reduced system for $\:\kappa=0\:$ (the NSEs for steady flows of a 
viscous fluid in Cartesian coordinates). It is better to choose the 
following variables $y_{a}$:
\[
y_{1}=x_{1}\cos\kappa t+x_{2}\sin\kappa t, \quad 
y_{2}=-x_{1}\sin\kappa t+x_{2}\cos\kappa t, \quad 
y_{3}=x_{3}.
\]
The vector-functions $\:\vec f\,^{b}=(f^{1b},f^{2b},f^{3b})$, 
$b=\overline{1,3},\:$ 
should be linearly independent solutions of the system
\[
Lf^{1}=-\kappa f^{2}, \quad Lf^{2}=\kappa f^{1}, \quad Lf^{3}=0
\]
and the function $f^{0}$ should satisfy the equation $\:Lf^{0}=0\:$ and 
the condition $\:f^{0}\not=0$. Here the operator $L$ is defined by 
(\ref{eb.9}). We give the following values of these functions:
\[
\vec f\,^{1}=(\cos\kappa t,\sin\kappa t,0), \; 
\vec f\,^{2}=(-\sin\kappa t,\cos\kappa t,0), \; 
\vec f\,^{3}=(0,0,1), \; f^{0}=1.
\]
The functions $g^{a}$ and $g^{0}$ are solutions of the equations
\[
Lg^{1}=-\kappa g^{2}, \quad Lg^{2}=\kappa g^{1}, \quad Lg^{3}=0,
\quad Lg^{0}=0.
\]
We can make, for example, $g^{a}$ and $g^{0}$ vanish. Then the 
corresponding ansatz has the form:
\begin{equation}\label{eb.10}
\!\!\!\!
\begin{array}{l}
u^{1}=\tilde v^{1}\cos\kappa t-\tilde v^{2}\sin\kappa t, \quad
u^{2}=\tilde v^{1}\sin\kappa t+\tilde v^{2}\cos\kappa t, 
\quad 
u^{3}=\tilde v^{3}, \quad p=\tilde q,
\end{array}
\end{equation}
where 
$\:\tilde v^{a}=\tilde v^{a}(y_{1},y_{2},y_{3})\:$ and 
$\:\tilde q=\tilde q(y_{1},y_{2},y_{3})\:$ 
are the new unknown functions. Substituting ansatz (\ref{eb.10}) 
into the NSEs, we obtain the following reduced system:
\begin{equation}\label{eb.11}
\!\!\!\!
\begin{array}{l}
\tilde v^{a}\tilde v^{1}_{a}-\tilde v^{1}_{aa}+\tilde q_{1}+
\kappa y_{2}\tilde v^{1}_{1}-\kappa y_{1}\tilde v^{1}_{2}-
\kappa\tilde v^{2}=0,
\\[0.8em]
\tilde v^{a}\tilde v^{2}_{a}-\tilde v^{2}_{aa}+\tilde q_{2}+
\kappa y_{2}\tilde v^{2}_{1}-\kappa y_{1}\tilde v^{2}_{2}+
\kappa\tilde v^{1}=0,
\\[0.8em]
\tilde v^{a}\tilde v^{3}_{a}-\tilde v^{3}_{aa}+\tilde q_{3}+
\kappa y_{2}\tilde v^{3}_{1}-\kappa y_{1}\tilde v^{3}_{2}=0,
\\[0.8em]
\tilde v^{a}_{a}=0.
\end{array}
\end{equation}
Here subscripts 1,2, and 3 of functions in (\ref{eb.11}) 
denote differentiation with 
respect to $y_{1}$, $y_{2}$, and $y_{3}$ accordingly. System 
(\ref{eb.11}), having variable coefficients, can be simplified by 
means of the local transformation 
\begin{equation}\label{eb.12}
\tilde v^{1}=v^{1}-\kappa y_{2}, \quad 
\tilde v^{2}=v^{2}+\kappa y_{1}, \quad 
\tilde v^{3}=v^{3}, \quad 
\tilde q=q+\!\!\!\begin{array}{l}\frac{1}{2}\end{array}\!\!
(y_{1}^{2}+y_{2}^{2}).
\end{equation}
Ansatz (\ref{eb.10}) and system (\ref{eb.11}) are transformed under 
(\ref{eb.12}) into ansatz (\ref{e2.2}) and system (\ref{e2.7}), 
where 
\begin{equation}\label{eb.13}
g^{1}=-\kappa x_{2},\quad g^{2}=\kappa x_{1},\quad g_{3}=0,\quad 
g^{0}=\!\!\!\begin{array}{l}\frac{1}{2}\end{array}\!\!
\kappa^{2}(x_{1}^{2}+x_{2}^{2}),
\end{equation} 
$\gamma_{1}=-2\kappa,\:$ and $\:\gamma_{2}=0$. Therefore, we can 
give the values of $g^{a}$ and $g^{0}$ from (\ref{eb.13}) and 
obtain ansatz (\ref{e2.2}) and system (\ref{e2.7}) at once.

The above is a good example how a reduced system can be 
simplified by means of modifying (complicating) an ansatz 
corresponding to it. Thus, system (\ref{e2.7}) is simpler than 
system (\ref{eb.11}) and ansatz (\ref{e2.2}) is more complicated 
than ansatz (\ref{eb.10}).

Finally, let us make several short notes about constructing other 
ansatzes for the Navier-Stokes field.

Ansatz corresponding to the algebra $\:A^{1}_{4}(\vec m,\chi)\:$ 
(see Subsec.~\ref{subsec_a.2}) can be constructed 
only for such $t$ that $\:\vec m(t)\not=\vec 0$. 
For these values of $t$, the parameter-function $\chi$ can be made 
to vanish by means of equivalence transformations (\ref{ea.10}). 

Ansatz corresponding to the algebra 
$\:A^{2}_{8}(\lambda,\psi^{1},\rho,\psi^{2})\:$ 
(see Subsec.~\ref{subsec_a.3}) can be constructed 
only for such $t$ that $\:\rho(t)\not=0$. 
For these values of $t$, the parameter-function $\psi^{2}$ can be made 
to vanish by means of equivalence transformations (\ref{ea.12}). 
Moreover, it can be considered that 
$\:\lambda_{t}\rho-\lambda\rho_{t}\in\{0;1\}$. The algebra obtained 
finally is denoted by $\:A^{2}_{8}(\lambda,\chi,\rho,0)$.

Ansatz corresponding to the algebra 
$\:A^{2}_{9}(\vec m^{1},\chi^{1},\vec m^{2},\chi^{2})\:$ 
(see Subsec.~\ref{subsec_a.3}) can be constructed 
only for such $t$ that $\:\mbox{\rm rank}(\vec m^{1},\vec m^{2})=2$. 
For these values of $t$, the parameter-functions $\chi^{i}$ can be made 
to vanish by means of equivalence transformations (\ref{ea.13}). 

The algebras $\:A^{2}_{10}(\kappa,\sigma)$, 
$\:A^{2}_{11}(\sigma),\:$ and $\:A^{2}_{12}(\sigma)$ can not be  
used to construct ansatzes by means of the Lie algorithm.

In view of equivalence transformation (\ref{ea.15}), the 
functions $\eta^{i}$ in the algebra 
$\:A^{3}_{7}(\eta^{1},\eta^{2},\eta^{3})\:$ (see Subsec.~\ref{subsec_a.4})
can be considered to satisfy the following condition:
\[
\eta^{1}_{t}\eta^{2}-\eta^{1}\eta^{2}_{t}\in\{0;
\!\!\!\begin{array}{l}\frac{1}{2}\end{array}\!\!\}.
\]

\setcounter{equation}{0}

\end{document}